\documentclass[acmsmall,screen]{acmart}

\usepackage{listings}
\usepackage{tikz}
\usepackage[dvipsnames]{xcolor}
\usepackage{xspace}
\usepackage{caption}
\usepackage{subcaption}
\usepackage{enumitem}
\usepackage[normalem]{ulem}
\usepackage{pifont}
\usepackage{array}
\usepackage{fp}

\newcounter{codeline}
\newcommand{\codeline}[1][]{%
  \refstepcounter{codeline}%
  \qquad\makebox[0pt][r]{\thecodeline\quad}#1%
}

\newenvironment{numberedtext}
  {\setcounter{codeline}{0}}
  {}

\usepackage{siunitx}
\definecolor{GrayCodeBlock}{RGB}{241,241,241}
\definecolor{BlackText}{RGB}{110,107,94}
\definecolor{RedTypename}{RGB}{182,86,17}
\definecolor{GreenString}{RGB}{96,172,57}
\definecolor{PurpleKeyword}{RGB}{184,84,212}
\definecolor{GrayComment}{RGB}{100,100,100}
\definecolor{GoldDocumentation}{RGB}{180,165,45}
\lstdefinelanguage{CustomRust}
{
    columns=fullflexible,
    keepspaces=true,
    showstringspaces=false,
    frame=single,
    framesep=0pt,
    framerule=0pt,
    framexleftmargin=4pt,
    framexrightmargin=4pt,
    framextopmargin=5pt,
    framexbottommargin=3pt,
    xleftmargin=4pt,
    xrightmargin=4pt,
    backgroundcolor=\color{GrayCodeBlock},
    basicstyle=\ttfamily\color{BlackText},
    keywords={
        true,false,
        unsafe,async,await,move,
        use,pub,crate,super,self,mod,
        struct,enum,fn,const,static,let,mut,ref,type,impl,dyn,trait,where,as,
        break,continue,if,else,while,for,loop,match,return,yield,in
    },
    keywordstyle=\color{RedTypename},
    ndkeywords={
        bool,u8,u16,u32,u64,u128,i8,i16,i32,i64,i128,char,str,usize,
        Self,Option,Some,None,Result,Ok,Err,String,Box,Vec,Rc,Arc,Cell,RefCell,HashMap,Vec,Weak,Ref,RefMut,
        macro_rules
    },
    ndkeywordstyle=\color{RedTypename},
    comment=[l][\color{GrayComment}\slshape]{//},
    morecomment=[s][\color{GrayComment}\slshape]{/*}{*/},
    morecomment=[l][\color{GoldDocumentation}\slshape]{///},
    morecomment=[s][\color{GoldDocumentation}\slshape]{/*!}{*/},
    morecomment=[l][\color{GoldDocumentation}\slshape]{//!},
    morecomment=[s][\color{RedTypename}]{\#![}{]},
    morecomment=[s][\color{RedTypename}]{\#[}{]},
    stringstyle=\color{GreenString},
    string=[b]",
    escapeinside={(@}{@)},
    mathescape=true,
}

\lstdefinestyle{GrayRegion}{
  keywordstyle=\color{gray},
  ndkeywordstyle=\color{gray},
  commentstyle=\itshape\color{gray},
  stringstyle=\color{gray},
  identifierstyle=\color{gray},
}

\lstdefinelanguage{LLVM}{language=}

\makeatletter
\let\oldttfamily\ttfamily
\renewcommand{\ttfamily}{%
  \ifx\f@family\ttdefault
    \oldttfamily
  \else
    \oldttfamily
    \ifdim\f@size pt>9.5pt
      \small
    \else\ifdim\f@size pt>8.5pt
      \footnotesize
    \else
      \fontsize{\f@size}{\f@baselineskip}\selectfont
    \fi\fi
  \fi
}
\makeatother

\makeatletter
\renewcommand{\p@subfigure}{\thefigure}

\makeatother

\newcommand\grumbler[3]{{\color{#1}#2 says: #3}}
\newcommand\mike[1]{\grumbler{blue}{Mike}{#1}}
\newcommand\victor[1]{\grumbler{red}{Victor}{#1}}

\makeatletter
\newcommand\later[1]{\@bsphack\@esphack} 
\newcommand\savespace[1]{\@bsphack\@esphack} 
\newcommand\recentsavespace[1]{\@bsphack\@esphack} 
\makeatother

\newtoggle{includes-appendix}
\toggletrue{includes-appendix}

\iftoggle{includes-appendix}
  {\newcommand\supplement[1]{#1}
   \newcommand\Supplement[1]{\supplement{#1}}}
  {\newcommand\supplement[1]{our extended arXiv version~\cite{refinator-extended-arxiv}}
   \newcommand\Supplement[1]{Our extended arXiv version~\cite{refinator-extended-arxiv}}}

\newcommand\irule[4]{\begin{aligned}&\text{[\scshape#1]}\\&\frac{#2}{#3}\ \text{#4}\end{aligned}}
\newcommand{\lifetime}[1]{\texttt{\textquotesingle#1}}
\renewcommand{\textrm}[1]{\textsf{#1}}

\newcommand\refinator{\&\textsc{inator}\xspace}
\newcommand\Refinator{\refinator}
\newcommand\trmcolor{}
\newcommand\trm[1]{{\trmcolor#1}}
\newcommand\nt[1]{\textbf{\textsf{#1}}}

\newcommand\kw[1]{\texttt{\textbf{#1}}}
\newcommand\sty[1]{\texttt{\color[rgb]{0.75, 0, 0}#1}}
\newcommand\rty[1]{\texttt{\color[rgb]{0, 0.5, 0}#1}}
\newcommand\ofRefCell{\ensuremath{^\rty{Cell}}}

\newcommand\labeled[2]{\ensuremath{[\texttt{#2}]^{#1}}}

\newcommand\NN{\mathbb{N}}

\newcommand\CONST{\textsf{\textbf{Const}}\xspace}
\newcommand\IDENT{\textsf{\textbf{Ident}}\xspace}

\newcommand\TYPELAB{\textsf{\textbf{TypeLab}}\xspace}

\newcommand\STYPE{\textsf{\textbf{SType}}\xspace}
\newcommand\STYPEFRAG{\textsf{\textbf{STypeFrag}}\xspace}

\newcommand\VAR{\textsf{\textbf{Var}}\xspace}
\newcommand\STRUCT{\textsf{\textbf{Struct}}\xspace}
\newcommand\FUNCTION{\textsf{\textbf{Function}}\xspace}
\newcommand\INSTRLAB{\textsf{\textbf{InstrLab}}\xspace}
\newcommand\POINT{\textsf{\textbf{Point}}\xspace}

\newcommand\BASETYPE{\textsf{\textbf{BaseType}}\xspace}

\newcommand\RTYPEFRAG{\textsf{\textbf{RTypeFrag}}\xspace}
\newcommand\PTRMOD{\textsf{\textbf{Qual}}\xspace}

\newcommand\LIFETIMEVAR{\textsf{\textbf{LifeVar}}\xspace}
\newcommand\LIFETIMELIST{\textsf{\textbf{LifeList}}\xspace}

\newcommand\PATH{\textsf{\textbf{Path}}\xspace}
\newcommand\LOAN{\textsf{\textbf{Loan}}\xspace}

\newcommand\refbind{\textsf{RefBind}\xspace}
\newcommand\structbind{\textsf{StructBind}\xspace}

\newcommand\BaseType{\textsf{BaseType}\xspace}
\newcommand\Array{\textsf{Array}\xspace}
\newcommand\Cell{\textsf{Cell}\xspace}
\newcommand\RefLifetime{\textsf{RefLT}\xspace}
\newcommand\StructLifetime{\textsf{StructLT}\xspace}

\newcommand\RustType{\textsf{RustType}\xspace}

\newcommand\StructGenerics{\textsf{StructGenerics}\xspace}
\newcommand\StructRank{\textsf{StructRank}\xspace}

\newcommand\Lifetime{\textsf{LT}\xspace}
\newcommand\LifetimeEnd{\textsf{LE}\xspace}

\newcommand\Loans{\textsf{Loans}\xspace}
\newcommand\gen{\textsf{gen}\xspace}
\newcommand\dfkill{\textsf{kill}\xspace}
\newcommand\borrowpath{\textsf{\detokenize{bor}}\xspace}
\newcommand\deeppaths{\textsf{\detokenize{deep}}\xspace}

\newcommand\cost{\textrm{cost}\xspace}

\newcommand\bench[1]{\textsf{#1}\xspace}
\newcommand\mc[3]{\multicolumn{#1}{#2}{#3}}
\newcommand\sigfigs[2]{\num[round-mode=figures,round-precision=#1,round-minimum=1,round-pad=false]{#2}}
\newcommand\tsec[1]{%
  \FPiflt{#1}{1}%
    \num[round-mode=places,round-precision=1]{#1}%
  \else
    \sigfigs{2}{#1}%
  \fi\,s%
}

\newcommand\assertions[1]{\sigfigs{3}{#1}}
\newcommand\stat[1]{\num{#1}}

\setcopyright{cc}
\setcctype{by}
\acmDOI{10.1145/3808270}
\acmYear{2026}
\acmJournal{PACMPL}
\acmVolume{10}
\acmNumber{PLDI}
\acmArticle{192}
\acmMonth{6}
\acmSubmissionID{pldi26main-p110-p}
\received{2025-11-13}
\received[accepted]{2026-04-03}



\begin{document}

\title{\Refinator: Correct, Precise C-to-Rust Interface Translation}

\newcommand\mytitle{\Refinator: Correct, Precise C-to-Rust Interface Translation}
\title{\mytitle\iftoggle{includes-appendix}{
    \framebox{\normalsize\normalfont This extended version of a PLDI 2026 paper adds an appendix with additional material.}\title{\mytitle}%
}{}}

\author{Victor Chen}
\orcid{0009-0001-9395-4086}
\affiliation{%
  \institution{The Ohio State University}
  \country{USA}
}
\email{chen.10604@buckeyemail.osu.edu}

\author{Ayden Coughlin}
\orcid{0009-0004-6024-4856}
\affiliation{%
  \institution{The Ohio State University}
  \country{USA}
}
\email{coughlin.166@buckeyemail.osu.edu}

\author{Michael D. Bond}
\orcid{0000-0002-8971-4944}
\affiliation{%
  \institution{The Ohio State University}
  \country{USA}
}
\email{mikebond@cse.ohio-state.edu}

\begin{abstract}
Automatically translating system software from C to Rust is an appealing but challenging problem,
as it requires whole-program reasoning to satisfy Rust's ownership and borrowing discipline.
A key enabling step in whole-program translation is \emph{interface translation}, which produces Rust declarations for the C program's top-level declarations (i.e., structs and function signatures), enabling modular and incremental code translation.

This paper introduces correct, precise C-to-Rust interface translation, called \emph{\refinator}.
\Refinator employs a novel constraint-based formulation of semantic equivalence and type correctness including borrow-checking rules to produce a Rust interface that is correct (i.e., the interface admits a semantics-preserving implementation in safe Rust) and precise (i.e., it uses the simplest, least costly types).
Our results show \refinator produces correct, precise Rust interfaces for real C programs, but support for certain C features and scaling to large
programs are challenges left for future work.
This work advances the state of the art by being the first correct, precise approach to C-to-Rust interface translation.
\end{abstract}

\maketitle

\section{Introduction}
\label{sec:intro}

The C programming language's lack of memory safety has led to countless bugs and vulnerabilities, costing billions of dollars annually and imperiling human well-being and security~\cite{cisq-report-2022}.
There is growing support among researchers, practitioners, and government for migrating safety- and security-critical system software from C to memory-safe languages, with (the safe subset of) Rust as the exemplar, owing to its combination of safety and bare-metal performance~\cite{rust-taking-over-tech,memory-safety-fact-sheet,darpa-tractor}.

However, translating the world's system software from C to Rust requires more than localized syntactic and semantic conversion~\cite{code-quality-c-to-rust-2026,c-to-rust-user-study}.
The central obstacle is Rust's strict ownership and borrowing discipline, which ensures safety but is tough to adhere to.
Large language models (LLMs) are good at converting C code into \emph{idiomatic} Rust code, but not code that satisfies Rust's ownership and borrowing rules while also ensuring semantic equivalence~\cite{tymcrat,fluorine,C2SaferRust,sactor}---a challenge that requires deep, program-wide reasoning.
Mechanical (non-LLM) C-to-Rust translators emit large amounts of \emph{unsafe} Rust code~\cite{c2rust,laertes-2021,laertes-2023,crown,CRustS} or cannot handle common aliasing patterns~\cite{mechanical-c-to-safe-rust}.

\subsubsection*{Key problem and motivation}

\begin{figure}
\centering
\includegraphics[page=1,scale=0.48]{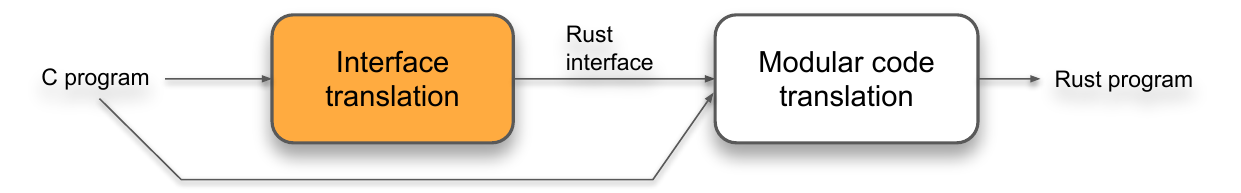}


\caption{C-to-Rust interface translation enables modular code translation, which could be be LLM-based, mechanical, or manual.}
\label{fig:interface-translation}
\end{figure}

A key enabling step of C-to-Rust translation is \emph{interface translation}: choosing Rust types for a C program's top-level declarations (e.g., struct fields, function parameters, and return values).
While not technically required as an explicit step,
interface translation is the foundation for modular code translation: Once correct interfaces are chosen, function bodies can be translated one at time, with confidence that the pieces will fit together.
By contrast, a mistaken type choice in the interface cascades into widespread translation failures~\cite{tymcrat,fluorine}.
\autoref{fig:interface-translation} shows how interface translation fits into the C-to-Rust translation workflow.
Furthermore, interface translation is compatible with a translation paradigm that 
preserves a direct mapping of functions and structs between the C and Rust programs, making it compatible with incremental translation (i.e., some functions in idiomatic Rust and others still in C or unidiomatic or unsafe Rust).
\recentsavespace{Performing interface translation as an explicit step allows for using a mechanical (i.e., non-LLM) for correct, precise interface translation and an LLM approach for idiomatic code translation.}

Prior work recognizes the importance of interface translation. Tymcrat performs LLM-based interface translation, with the authors stating that
``[t]his step is crucial because directly requesting the LLM to translate the function may not migrate types correctly''~\cite{tymcrat}.
The authors of Fluorine, a Go-to-Rust translator, similarly report that a mistaken interface type cascades into widespread translation failures~\cite{fluorine}. CRUST-Bench, a C-to-Rust translation benchmark suite, distributes hand-translated interfaces so that code translators can even be evaluated~\cite{CRUST-Bench}.
\savespace{Collectively, these efforts suggest that interface translation is widely acknowledged as essential, yet remains unsolved.}

\subsubsection*{Interface translation is challenging}

Interface translation is hard for both humans and machines.
The difficulty lies in simultaneously satisfying Rust's ownership and borrowing rules \emph{and} preserving semantic equivalence with the C program. This requires whole-program reasoning. A successful translation must have the following properties:
\begin{itemize}
  \item \emph{Correct}: The chosen types admit a safe Rust implementation that preserves the semantics of the original program.
  \item \emph{Precise}: Among correct types, the types should be the least costly (according to some cost function).
\end{itemize}
For example, a C function parameter of type \lstinline{T*} might plausibly map to Rust types \lstinline{&T}, \lstinline{T}, \lstinline{&[T]}, \lstinline{Vec<T>}, \lstinline{Box<T>}, \lstinline{Rc<RefCell}\allowbreak\lstinline{<T>>}, and others. Which are correct, and which is best?

Consider the C program in \autoref{fig:running-example:c} and a corresponding Rust interface in \autoref{fig:running-example:rust}. The Rust interface consists of struct and function declarations. The figure shows semantics-preserving function bodies (in gray) to demonstrate that the interface is viable. Even in this small example, subtle challenges arise.
For example, the most precise correct type for \lstinline{replace}'s parameter \lstinline{dst} is \lstinline{&mut i32}. The types \lstinline{&i32} and \lstinline{i32} are incorrect, while \lstinline{Rc<RefCell<i32>>} is correct but imprecise.

\begin{figure}
\centering
\begin{subfigure}[]{0.49\linewidth}
\begin{lstlisting}[language=c,breaklines=true,basicstyle=\ttfamily\footnotesize]
struct Node {
  int data;
  struct Node *next;
};
struct Node *new() {
  struct Node *n = malloc(sizeof(Node));
  n->next = NULL;
  return n;
}
void push(struct Node *list, int x) {
  struct Node *n = malloc(sizeof(Node));
  n->data = x;
  n->next = list->next;
  list->next = n;
}
void replace(int *dst, int x) {
  *dst = x;
}
void replace_first(struct node *list, int x) {
  replace(&list->next->data, x);
}
int first(struct node *list) {
  return list->next->data;
}
int main() {
  struct Node *list = new();
  push(list, 1);
  struct Node *sublist = list->next;
  push(sublist, 2);
  replace_first(list, 3);
  replace_first(sublist, 4);
}
\end{lstlisting}
\caption{An example C program.\label{fig:running-example:c}}
\end{subfigure}
\hfill
\begin{subfigure}[]{0.47\linewidth}
\lstset{breaklines=true}
\begin{lstlisting}[basicstyle=\ttfamily\footnotesize]
struct Node {
  data: i32,
  next: Option<Rc<RefCell<Node>>>,
}
fn new() -> Node {
  @Node { data: i32::default(), next: None }@
}
fn push(list: &mut Node, x: i32) {
  @let n = Rc::new(RefCell::new(Node {
    data: x,
    next: list.next.take(),
  }));
  list.next = Some(n);@
}
fn replace(dst: &mut i32, x: i32) {
  @*dst = x;@
}
fn replace_first(list: &Node, x: i32) {
  @replace(&mut list.next.as_ref().unwrap().borrow_mut().data, x);@
}
fn first(list: &Node) -> i32 {
  @list.next.unwrap().borrow().data@
}
fn main() {
  @let mut list = new();
  push(&mut list, 1);
  let sublist = list.next.as_ref().unwrap();
  push(&mut sublist.borrow_mut(), 2);
  replace_first(&list, 3);
  replace_first(&sublist.borrow(), 4);@
}
\end{lstlisting}
\caption{A translated Rust interface for the C program.\label{fig:running-example:rust}}
\end{subfigure}
\caption{An example C program and translated Rust interface. The Rust interface shows grayed-out Rust function bodies that are compatible with the interface. Note that \lstinline{replace_first} and \lstinline{first} are so named because they remove the first element \emph{after the dummy head node}.}
\label{fig:running-example}
\end{figure}

As another example, the \lstinline{Node::next} field must be an owning type instead of a borrowed reference (e.g., \lstinline{Node} or \lstinline{Rc<Node>} instead of \lstinline{&Node}) because a pointer escapes from \lstinline{push} into a \lstinline{Node}.
The function \lstinline{replace_first} must take an immutable reference to avoid borrowing conflicts in \lstinline{main} even though it mutates its \lstinline{list} parameter, which forces \lstinline{Node::next} to be interior mutable (i.e., to use \lstinline{RefCell}).

Making such determinations requires a global understanding of the C program's semantics and Rust's ownership and borrowing rules. For example, the analysis must recognize that \lstinline{list.next.}\allowbreak\lstinline{as_ref().unwrap()} and \lstinline{*list.next.take()} are available to transform Rust values in a way that obeys Rust's borrowing rules and preserves the semantics of the C program.

Prior work has not solved C-to-Rust interface translation (\autoref{sec:related}). \emph{Tymcrat} attempts LLM-based interface translation but frequently fails to produce correct, precise interfaces~\cite{tymcrat}. \emph{Scylla's} mechanical translation  approach includes interface translation implicitly~\cite{mechanical-c-to-safe-rust}, but it cannot handle aliasing, lacks dynamic borrowing support, and fails to choose precise types because it forgoes precise global reasoning. \emph{CRUST-Bench} allows translation approaches to sidestep the difficulty entirely, by distributing \emph{hand-translated} interfaces~\cite{CRUST-Bench}.
\later{\textcolor{red}{These manually translated interfaces are often incorrect or imprecise, we found and the CRUST-Bench authors confirmed (\autoref{sec:eval-interfaces}), further motivating the difficulty of interface translation.} \mike{We likely don't want to make this point or include this evaluation, especially assuming we don't evaluate all of the three programs for which we evaluated the manual interfaces. It doesn't really matter that manual interface translation isn't high quality; the main issue is that it's manual!}}


\subsubsection*{Our approach}

This paper takes the first steps in developing interface translation that is correct and precise.
The paper's novel interface translation approach, called \emph{\refinator},\footnote{Pronounced REF-inator.} encodes type correctness and semantic equivalence as logical constraints.
Key insights include (1) a novel method for mapping ``type fragments'' between C and Rust to make the type space bounded and solvable and (2) a novel formulation of Rust's ownership and borrowing rules, including support for legal transformations between Rust types that enable precise translation.

Conceptually, \refinator's approach can be viewed as inferring an ownership and borrowing discipline for C code and expressing it using Rust's type system. This perspective connects our work to prior efforts to infer safety-oriented types for C~\cite{ccured,cyclone,vault,checked-c-by-3c} (\autoref{sec:related:type-inference}).

\Refinator currently has four main limitations.
First, \refinator's interface translation is based on a strict interpretation of the aliasing in the input C program. As a result, the translated interface may be less precise than what could be produced by considering semantics-preserving reordering that can impact aliasing.
Second, \refinator currently cannot handle certain C features, including function pointers and polymorphic pointers.
Third, for some Rust types, \refinator infers a set of possible types rather than an exact type.
Fourth, \refinator's running time currently scales superlinearly with program size, and it takes several hours to solve constraints for C programs with a few thousand lines of code.
Addressing these limitations
is left for future work.
\recentsavespace{This work demonstrates the potential for correct, precise C-to-Rust interface translation.}

We implemented \refinator to take LLVM IR as input and generate SMT constraints solved by Z3.
We evaluated \refinator using three metrics: correctness and precision of translated interfaces and run-time performance of interface translation.
The evaluation uses several programs from CRUST-Bench~\cite{CRUST-Bench}. The results show that \refinator chooses correct, precise interface types nearly universally. Constraint solving takes several hours for C programs with a few thousand lines of code.

This paper contributes a novel formulation of cross-language type inference under a safety discipline, demonstrated empirically on real C programs. It shows, for the first time, the potential for correct, precise C-to-Rust interface translation.

\section{Background and Related Work}
\label{sec:related}

No prior work solves the interface translation problem.
\autoref{tab:qualitative-comparison} compares C-to-Rust interface translation approaches with \refinator. Note that several of the approaches perform interface translation implicitly as part of whole-program translation, rather than as an explicit step.
An approach \emph{supports aliasing} if it handles C programs with arbitrary aliasing.
An approach is \emph{modular} if it does not require whole-program analysis.
\Refinator requires whole-program analysis, but its analysis results enable modular code translation.

\begin{table}
\caption{Qualitative comparison of automatic C-to-Rust interface translation approaches (plus one Go-to-Rust approach~\cite{fluorine}), including approaches that translate interfaces implicitly as part of whole-program translation.}
\label{tab:qualitative-comparison}

\small
\newcommand{\cmark}{\textcolor{ForestGreen}{\ding{51}}}
\newcommand{\xmark}{\textcolor{red}{\ding{55}}}
\newcommand{\Yes}{\cmark}
\newcommand{\No}{\xmark}
\newcommand{\Maybe}[1]{\textcolor{BurntOrange}{\bf #1}}
\newcolumntype{H}{>{\setbox0=\hbox\bgroup}c<{\egroup}@{}}
\begin{tabular}{lHcccc}
\toprule
& matic & Supports aliasing & Correct & Precise & Modular \\
\midrule
Mechanical translation to safe Rust~\cite{mechanical-c-to-safe-rust} & \Yes & \No & \Yes & \No & \Yes \\
Mechanical translation to unsafe Rust~\cite{c2rust,crown,CRustS,laertes-2021,laertes-2023}  & \Yes & \Yes & \No & N/A & \Yes \\
LLM-based translation~\cite{tymcrat,C2SaferRust,sactor,fluorine} & \Yes & \Yes & \No & \No & \No \\
\midrule
\Refinator & \Yes & \Yes & \Yes & \Yes & \No \\
\bottomrule
\end{tabular}
\end{table}

\subsubsection*{Mechanical translation to safe Rust}

Fromherz and Protzenko introduce Scylla, a mechanical approach for whole-program C-to-Rust translation~\cite{mechanical-c-to-safe-rust}.
However, Scylla does not support aliasing in general nor dynamic borrowing, making it infeasible to refactor some C programs to be eligible for translation.
Furthermore, Scylla fails to choose precise types because it forgoes global reasoning.

\subsubsection*{Mechanical translation to unsafe Rust}

Several approaches translate C to a mix of safe and unsafe Rust, which we consider incorrect solutions to the problem of producing safe Rust interfaces (\autoref{tab:qualitative-comparison}).

C2Rust is a mature project that mechanically translates C programs to Rust~\cite{c2rust}. However, most of the generated code is unsafe
(i.e., uses operations marked \lstinline{unsafe}) and may violate memory safety, particularly dereferences of raw pointers.

Several approaches attempt to convert the output of C2Rust to use less unsafe code, with limited success.
Laertes uses heavyweight pointer analysis to make C2Rust's output safer, but it cannot change many raw pointers to safe types~\cite{laertes-2021, laertes-2023}.

In a similar spirit, Crown rewrites Rust code produced by C2Rust to use fewer raw pointers~\cite{crown}. Crown avoids Laertes's heavyweight pointer analysis, instead encoding ownership rules as SMT constraints that encode a conservative version of Rust's ownership and borrowing rules.
While Crown bears some resemblance to \refinator, there are significant differences.
Crown's analysis is mostly \emph{intra-function}: It does not modify function signatures to eliminate raw pointers and unsafe types, except for output parameters. (Crown however uses global constraints to choose new types for struct fields.) In contrast, \refinator translates function signatures, using global constraints that capture dependencies across function call boundaries and between multiple variables' types. Furthermore, Crown determines only whether types should be owning, while \refinator determines complete type information, enabling interface translation. Finally,
Crown fails to remove many raw pointers, while \refinator removes raw pointers, even in the presence of aliasing.

CRustS performs template-based translation on C2Rust-translated code to use less unsafe code, but it primarily encapsulates unsafe code without eliminating it~\cite{CRustS}.

\subsubsection*{LLM-based translation}

Tymcrat performs LLM-based \emph{interface} translation as its first step,
generating multiple candidate interfaces and trying each of them in LLM-based \emph{code} translation~\cite{tymcrat}. There are no guarantees about the correctness or precision of the interfaces, which are often incorrect~\cite{tymcrat}, and code translation treats the interfaces as a hint rather than a mandate.

In addition to Tymcrat~\cite{tymcrat}, recent work uses LLMs to translate C to Rust, but generally fails to produce compilable, correct translations for nontrivial inputs.
LLM-based translation is not modular, and one can expect its correctness and precision to degrade with increasing program size and complexity due to LLM reasoning capacity limits.
\Refinator is not modular, but its correctness and precision do not degrade with increasing program size and complexity, and it can enable modular LLM-based code translation.

C2SaferRust uses LLMs to transform (mechanically translated) unsafe Rust to more safe Rust~\cite{C2SaferRust}, but it still leaves much code unsafe.
SACTOR likewise translates unsafe Rust to safe Rust, using an LLM guided by hints from static program analysis~\cite{sactor}. \savespace{SACTOR's static analysis provides dependencies between program elements and mostly intraprocedural pointer and aliasing semantics to the LLM.} Unlike \refinator's constraint generation--based approach, SACTOR's static information is insufficient for correctly translating types, but instead is intended to help the LLM choose better types more often.
Fluorine shows the utility of ``feature mapping'' to help LLMs translate across languages and ``type compatibility'' to ensure compatibility of types across languages~\cite{fluorine}.
However, neither technique takes into account the semantics of ownership and borrowing in Rust.


Outside the context of translation,
M\"{u}ndler et al.\ constrain LLM token generation to produce type-correct code~\cite{type-constrained-llm}, but their approach does not support ownership or borrowing.

\subsubsection*{Type inference}
\label{sec:related:type-inference}

Rust interface translation can be viewed as a form of global type inference.
Edwards and Li explore global \emph{interface} inference in the context of translation~\cite{determining-interfaces-2004}, and other approaches perform type inference on languages with memory safety features~\cite{ccured,cyclone,vault,checked-c-by-3c}. None of these approaches infer types for languages with statically enforced ownership and borrowing.


\section{\Refinator Overview}

Given an input C program, \refinator infers Rust types for interface variables: struct fields, function parameters and return values, and global variables.

\subsection{Guarantees and Requirements}
\label{sec:refinator-overview:guarantees-requirements}

\Refinator's inferred interface is \emph{correct} if there exists a safe Rust translation using the interface that preserves the C program's behavior, except for potential dynamic borrowing conflicts and memory leaks mentioned below.
For a C program without undefined behavior, \refinator infers a correct interface (modulo the limitations below).
For a C program with undefined behavior, \refinator still infers an interface, which we expect in practice to admit a safe Rust translation with the intended behavior of the C program---a translation that likely panics in cases where the C program would dereference a \lstinline[language=C]{NULL} pointer or read past the end of a buffer, for example.

\Refinator always infers a correct interface (i.e., its constraints are always satisfiable) because it is possible to implement any C interface in safe Rust by using \lstinline{Rc<RefCell<T>>} for every pointer.

\Refinator requires whole-program analysis to perform correct and precise interface translation, impacting its scalability. In addition, the entire C program needs to be available as input to \refinator. For an incomplete program such as a library, \refinator requires a client test suite that uses the library. The same requirement applies to other translation approaches.

\subsection{Limitations}
\label{sec:refinator-overview:limitations}

\Refinator infers types based on a strict interpretation of aliasing in the input C program. As a result, it cannot infer an interface that relies on code reordering that impacts aliasing.

\Refinator may not guarantee that there exists a \emph{panic-free} safe Rust translation that uses the inferred interface and preserves the C program's behavior---this is because of the possibility of dynamic borrowing conflicts caused by conflicting \lstinline{RefCell::borrow} and \lstinline{RefCell::borrow_mut} calls.\footnote{Note that expressions such as \lstinline{*p1 = *p2} where the pointers may alias do not cause dynamic borrowing conflicts in the Rust translations admitted by \refinator. This is because the analysis operates on the LLVM IR--based source language, and such assignments are lowered to a read followed by a write through a temporary (e.g., \lstinline{tmp = *p2; *p1 = tmp}), yielding legal Rust code of the form \lstinline{let tmp = *p2.borrow(); *p1.borrow_mut() = tmp}.}
We are most concerned about cases in which \refinator chooses an interface variable type of \lstinline{std::cell::Ref} or \lstinline{std::cell::RefMut} (the return types of \lstinline{RefCell::borrow} and \lstinline{RefCell::borrow_mut}, respectively). This could lead to a function that takes a \lstinline{RefMut} as input and yet requires access to the same \lstinline{RefCell}, for example. This concern could be addressed by disallowing \lstinline{Ref} and \lstinline{RefMut} as interface types.
Future work should determine the extent of this issue and how to address it.

\Refinator does not differentiate between some types, which we discuss in \autoref{sec:undifferentiated-types}.
\Refinator cannot guarantee that the inferred interface admits a \emph{memory leak free} safe Rust translation, even when the C program is memory leak free, since the interface may require that any safe Rust implementation uses \lstinline{Rc} cycles.
\Refinator does not currently support multi-threaded programs or libraries.

\Refinator currently provides partial or no support for certain C program features: function pointers, polymorphic pointers, \lstinline{union} types, variadic functions, and the ternary operator.
Supported type casts include coercions on integral types and \lstinline{void} pointers cast to a single non-\lstinline{void} pointer type.
Unsupported type casts include \lstinline{void} pointers cast to more than one other pointer type and struct pointers cast to pointers to the struct's first field.
\autoref{sec:impl} provides details on these limitations.

\subsection{How \Refinator Works}

\begin{figure}
\centering
\hspace*{-0.5em}\includegraphics[page=4,scale=0.48]{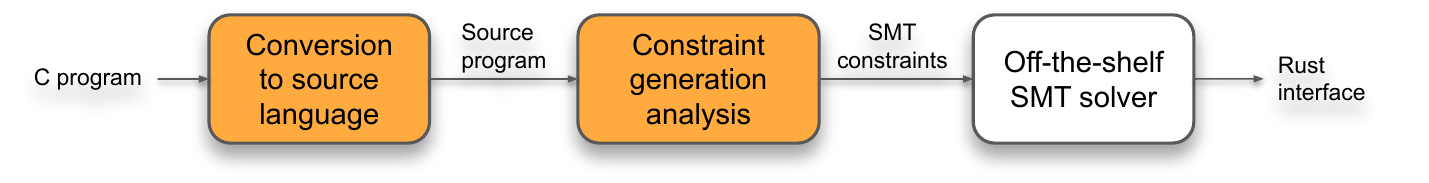}


\caption{\Refinator performs interface translation (see \autoref{fig:interface-translation}) using three top-level components.}
\label{fig:refinator-overview}
\end{figure}

\Refinator performs the following steps as shown in \autoref{fig:refinator-overview}:
\begin{enumerate}
    \item Convert the C program to \refinator's source language (\autoref{sec:source}).
    \item Generate SMT constraints on the inferred Rust types that ensure type correctness, semantic equivalence, and precision (\autoref{sec:constraints-overview}--\autoref{sec:precision}).
    \item Run an off-the-shelf SMT solver to solve the constraints, yielding correct, precise Rust types.
\end{enumerate}

\section{Source Language for Constraint Generation Analysis}
\label{sec:source}

\Refinator analyzes programs written in a language with C-like syntax that resembles a three-address-code intermediate representation (IR).
This \emph{source language} is one-to-one with LLVM IR~\cite{llvm}, but it represents more precise types than LLVM IR, called \emph{source types}; is explicit about when variables are dropped; and tags parts of its syntax with \emph{type labels}.

\begin{figure}
    \small
    \begin{gather*}
        \begin{aligned}
            \nt{Program} \ni M \ &{::=} \ \{ \Delta_g \mid \Delta_s \mid \Delta_f \}^\ast \\
            \nt{GlobDef} \ni \Delta_g \ &{::=} \ \texttt{$\vec{\tau}$ $v$ \trm{=} $c$\trm{;}} \\
            \nt{StructDef} \ni \Delta_s \ &{::=} \ \texttt{\kw{struct} $s$ \trm{\{}$\{\texttt{$\vec{\tau}$ $f$\trm{;}}\}^\ast$\trm{\};}} \\
            \nt{FuncDef} \ni \Delta_f \ &{::=} \ \texttt{$\vec{\tau}$ $F$\trm{(}$\{\texttt{$\vec{\tau}$ $v$\trm{,}}\}^\ast$) \trm{\{} $\{ \{ L\trm{:} \}^? \  I\}^\ast$ \trm{\}}} \\
            \nt{Instr} \ni I \ &{::=} \ \texttt{$\vec{\tau}$ $v$ \trm{=} \kw{alloca}\trm{($c$);}}
                \mid \texttt{$\vec{\tau}$ $v$ \trm{=} \labeled{\ell}{\trm{*}$v$}\trm{;}}
                \mid \texttt{\trm{*}$v$ \trm{=} \labeled{\ell}{$e$}\trm{;}}
                \mid \texttt{$\vec{\tau}$ $v$ \trm{=} \labeled{\ell}{\trm{\&*(*}$v$\trm{).}$f$}\trm{;}} \\
              & \mid \texttt{$\vec{\tau}$ $v$ \trm{=} \labeled{\ell}{\trm{\&}$v$\trm{[}$e$\trm{]}}\trm{;}}
                \mid \texttt{$\{\texttt{$\vec{\tau}$ $v$ \trm{=}}\}^?$ $F$\trm{(}$\{\texttt{\labeled{\ell}{$e$}\trm{,}}\}^\ast$\trm{);}}
                \mid \texttt{$\vec{\tau}$ $v$ \trm{=} \kw{phi}($\{ \texttt{\labeled{\ell}{$e$}\trm{:} $L$\trm{,}} \}^+$\trm{);}} \\
              & \mid \texttt{$\{\texttt{$\vec{\tau}$ $v$ \trm{=}}\}^?$ \kw{use}\trm{(}$\{\texttt{\labeled{\ell}{$v$}\trm{,}}\}^+$\trm{);}}
                \mid \texttt{\kw{goto} $L$\trm{;}}
                \mid \texttt{\kw{if} \trm{(}\labeled{\ell}{$e$}\trm{)} \kw{goto} $L$\trm{;} \kw{else} \kw{goto} $L$\trm{;}} \\
              & \mid \texttt{\kw{return} \labeled{\ell}{$e$}\trm{;}}
                \mid \texttt{\kw{drop}\trm{(}{$\{v\trm{,}\}^+$}\trm{);}} \\
            \STYPE \ni \vec{\tau} \ &{::=} \ \texttt{$\langle \{\labeled{\ell}{$\tau$},\}^+ \rangle$} \\
            \STYPEFRAG \ni \tau \ &{::=} \
                \sty{unknown}
                \mid \sty{void}
                \mid \sty{bool}
                \mid \sty{i8}
                \mid \sty{i16}
                \mid \sty{i32}
                \mid \sty{i64}
                \mid \sty{f32}
                \mid \sty{f64}
                \mid \texttt{\sty{struct}\trm{(}$s$\trm{)}}
                \mid \texttt{\sty{ptr}} \\
        \end{aligned} \\
        e \in \IDENT \cup \CONST \quad F, L, f, s, v \in \IDENT \quad c \in \CONST \quad \ell \in \NN
    \end{gather*}

    \caption{\label{fig:source-lang}Context-free grammar for the source language of \refinator's constraint generation analysis.}
\end{figure}

\begin{figure}
    \begin{minipage}[c]{\linewidth}
    \ttfamily
    \footnotesize
    \begin{numberedtext}
    \codeline \kw{struct} \detokenize{Node} \{$\langle \labeled{0}{\sty{ptr}}, \labeled{1}{\sty{i32}} \rangle$ \detokenize{data};
        $\langle \labeled{2}{\sty{ptr}}, \labeled{3}{\sty{ptr}}, \labeled{4}{\sty{struct}(\detokenize{Node})} \rangle$ \detokenize{next};
    \};\\
    \codeline $\langle \labeled{5}{\sty{ptr}}, \labeled{6}{\sty{struct}(\detokenize{Node})} \rangle$ \detokenize{new}() \{
    \textcolor{gray}{/-- snip --/}
    \}\\
    \codeline $\langle \labeled{7}{\sty{void}} \rangle$ \detokenize{push}($\langle \labeled{8}{\sty{ptr}}, \labeled{9}{\sty{struct}(\detokenize{Node})} \rangle$ \detokenize{list},
        $\langle \labeled{10}{\sty{i32}} \rangle$ \detokenize{x},
    ) \{\\
    \codeline \hspace*{1em}$\langle \labeled{11}{\sty{ptr}}, \labeled{12}{\sty{struct}(\detokenize{Node})} \rangle$ \detokenize{n} = \detokenize{malloc}(\labeled{13}{$c_1$}, );\\
    \codeline \hspace*{1em}\textcolor{gray}{/-- snip --/}\\
    \codeline \hspace*{1em}$\langle \labeled{14}{\sty{ptr}}, \labeled{15}{\sty{ptr}}, \labeled{16}{\sty{struct}(\detokenize{Node})} \rangle$ \detokenize{next} = \labeled{17}{\&*(*\detokenize{list}).next};\\
    \codeline \hspace*{1em}*\detokenize{next} = \labeled{18}{\detokenize{n}};\\
    \codeline \hspace*{1em}\kw{drop}(\detokenize{next},
        \textcolor{gray}{/-- snip --/}
        \detokenize{n},
        \detokenize{x},
        \detokenize{list},
    );\\
    \codeline \}\\
    \codeline $\langle \labeled{19}{\sty{void}} \rangle$ \detokenize{replace}($\langle \labeled{20}{\sty{ptr}}, \labeled{21}{\sty{i32}} \rangle$ \detokenize{dst},
        $\langle \labeled{22}{\sty{i32}} \rangle$ \detokenize{x},
    ) \{
    \textcolor{gray}{/-- snip --/}
    \}\\
    \codeline $\langle \labeled{23}{\sty{void}} \rangle$ \detokenize{replace_first}($\langle \labeled{24}{\sty{ptr}}, \labeled{25}{\sty{struct}(\detokenize{Node})} \rangle$ \detokenize{list},
        $\langle \labeled{26}{\sty{i32}} \rangle$ \detokenize{x},
    ) \{
    \textcolor{gray}{/-- snip --/}
    \}\\
    \codeline $\langle \labeled{27}{\sty{i32}} \rangle$ \detokenize{first}($\langle \labeled{28}{\sty{ptr}}, \labeled{29}{\sty{struct}(\detokenize{Node})} \rangle$ \detokenize{list},
    ) \{
    \textcolor{gray}{/-- snip --/}
    \}\\
    \codeline $\langle \labeled{30}{\sty{i32}} \rangle$ \detokenize{main}() \{\\
    \codeline \hspace*{1em}$\langle \labeled{31}{\sty{ptr}}, \labeled{32}{\sty{struct}(\detokenize{Node})} \rangle$ \detokenize{list} = \detokenize{new}();\\
    \codeline \hspace*{1em}\detokenize{push}(\labeled{33}{\detokenize{list}}, \labeled{34}{$c_2$}, );\\
    \codeline \hspace*{1em}$\langle \labeled{35}{\sty{ptr}}, \labeled{36}{\sty{ptr}}, \labeled{37}{\sty{struct}(\detokenize{Node})} \rangle$ \detokenize{next} = \labeled{38}{\&*(*\detokenize{list}).next};\\
    \codeline \hspace*{1em}$\langle \labeled{39}{\sty{ptr}}, \labeled{40}{\sty{struct}(\detokenize{Node})} \rangle$ \detokenize{sublist} = \labeled{41}{*\detokenize{next}};\\
    \codeline \hspace*{1em}\detokenize{push}(\labeled{42}{\detokenize{sublist}}, \labeled{43}{$c_3$}, );\\
    \codeline \hspace*{1em}\detokenize{replace_first}(\labeled{44}{\detokenize{list}}, \labeled{45}{$c_4$}, );\label{line:source-snippet:main-calls-replace-first}\\
    \codeline \hspace*{1em}\detokenize{replace_first}(\labeled{46}{\detokenize{sublist}}, \labeled{47}{$c_5$}, );\\
    \codeline \hspace*{1em}\kw{drop}(\detokenize{sublist},
        \detokenize{next},
        \detokenize{list},
    );\\
    \codeline \}
    \end{numberedtext}
    \end{minipage}

    \caption{A program snippet in the source language of \refinator's constraint generation analysis, derived from the C program in \autoref{fig:running-example:c}. For brevity the snippet omits about two-thirds of the code (i.e., \texttt{\textcolor{gray}{/-- snip --/}}).}
    \label{fig:source-snippet}
\end{figure}

\subsubsection*{Syntax and semantics}

The source language captures struct definitions, global variable definitions, and function definitions including function bodies.
\autoref{fig:source-lang} shows a grammar for the source language.
\autoref{fig:source-snippet} shows (an abridged version of) the program in \autoref{fig:running-example:c} expressed in the source language.
The semantics of each instruction in the source language is the same as its counterpart in LLVM IR and the C statement that it resembles.
Instructions that begin with \texttt{$\vec{\tau}$ $v$} declare a local variable $v$ with source type $\vec{\tau}$.
More explicitly,
\begin{enumerate}
    \item \texttt{$\vec{\tau}$ $v$ = \kw{alloca}($c$);} allocates $c$ bytes on the stack and assigns a pointer to the allocation to $v$.
    \item \texttt{$\vec{\tau}$ $v_1$ = \labeled{\ell}{*$v_2$};} dereferences $v_2$ and copies the result to $v_1$.
    \item \texttt{*$v$ = \labeled{\ell}{$e$};} stores a copy of the value of $e$ at the address referenced by the pointer $v$.
    \item \texttt{$\vec{\tau}$ $v_1$ = \labeled{\ell}{\&*(*$v_2$).$f$};} creates an alias of the pointer \texttt{(*$v_2$).$f$}, and assigns it into $v_1$.
    \item \texttt{$\vec{\tau}$ $v_1$ = \labeled{\ell}{\&$v_2$[$e$]};} assigns \texttt{$v_2$ + $e$}, computed according to C pointer arithmetic, where $v_2$ is a pointer and $e$ is an integer, into $v_1$.
    \item \texttt{$\vec{\tau}$ $v$ = $F$(\labeled{\ell_1}{$e_1$}, $...$, \labeled{\ell_n}{$e_n$});} calls the function $F$ with arguments $e_1, ..., e_n$, and assigns the return value into $v$.
    \item \texttt{$\vec{\tau}$ $v$ = \kw{phi}(\labeled{\ell_1}{$e_1$}: $L_1$, $...$, \labeled{\ell_n}{$e_n$}: $L_n$);} is the SSA $\phi$-function, and it assigns into $v$ the value of $e_i$, where control was previously transferred from the basic block starting at $L_i$.
    \item \texttt{$\vec{\tau}$ $v$ = \kw{use}(\labeled{\ell_1}{$e_1$}, $...$, \labeled{\ell_n}{$e_n$});} represents expressions that do not involve pointers or control flow, e.g., arithmetic operations and type casts between numeric types.
    \item Instructions with \texttt{\kw{goto}} and \texttt{\kw{return}} are standard control-flow and return instructions.
    \item \texttt{\kw{drop}($v_1$, $...$, $v_n$);} deallocates variables $v_1, ..., v_n$. \Refinator's conversion to the source language inserts a drop of $v$ at the start of the immediate postdominator of the nearest common ancestor in the postdominator tree of all basic blocks where $v$ is live. If the nearest common ancestor is the exit node, then a drop is inserted at the exit node.
\end{enumerate}

\subsubsection*{Source types}

In the source language, each variable is declared with a \emph{source type} corresponding to its type in the original C program, represented as \STYPE in the grammar.
A source type is a list of \emph{source type fragments}, represented as \STYPEFRAG in the grammar.
For example, in \autoref{fig:source-snippet}, the parameter \texttt{\detokenize{list}} of \texttt{\detokenize{push}} is declared with source type $\langle \labeled{8}{\sty{ptr}}, \labeled{9}{\texttt{\sty{struct}(Node)}} \rangle$ (i.e., the type consists of two source type fragments), which corresponds to its type in \autoref{fig:running-example:c}, \lstinline[language=C]{struct Node*}.
\Supplement{\autoref{appendix:computing-source-types}} presents the details of how \refinator infers source types from LLVM IR.

\subsubsection*{Type labels}

In the source language, each source type fragment and non-nested rvalue expression, excluding operands of \kw{drop},
is tagged with a type label, denoted by $\ell$ in \autoref{fig:source-lang}.
A type label is a natural number used by \refinator's constraints to uniquely identify an item in the program for which \refinator must infer a \emph{Rust type fragment}, discussed in \autoref{sec:rtypes}.
For example, in \autoref{fig:source-snippet}, the \sty{ptr} and \texttt{\sty{struct}(\detokenize{Node})} type fragments of the parameter \lstinline{list} of \lstinline{push} have type labels 8 and 9, respectively.
As we shall see, tagging each \emph{fragment} of each source type in the program instead of entire source types with type labels bounds the Rust type search space, and tagging each non-nested rvalue expression with a type label enables \refinator to choose more precise types.

\section{Inferred Rust Types}
\label{sec:rtypes}
\label{sec:undifferentiated-types}

\savespace{For an input program in the source language, \refinator infers Rust types for each interface variable (struct field, global variable, function parameter, or return value), local variable, and rvalue expression such that there exists a safe Rust program that uses the inferred Rust types, is functionally equivalent to the input program, and adheres to Rust's borrow-checking rules.}
\Refinator performs interface translation by inferring a \emph{Rust type fragment} for each type label.
Rust type fragments are defined in \autoref{fig:rtypes-frag}, and each Rust type fragment corresponds to a set of concrete Rust types, shown in \autoref{tab:rust-type-classes}.
In \autoref{fig:rtypes-snippet}, we show the program in \autoref{fig:source-snippet} annotated with Rust type fragments instead of source type fragments.

\begin{figure}
  \small
  \begin{gather*}
  \begin{aligned}
      \RTYPEFRAG \ni T \ &{::=} \
          \rty{unknown}
          \mid \rty{void}
          \mid \rty{bool}
          \mid \rty{i8}
          \mid \rty{i16}
          \mid \rty{i32}
          \mid \rty{i64}
          \mid \rty{f32}
          \mid \rty{f64}
          \mid \texttt{\rty{struct}($s$, $\vec{\rho}$)} \\
         &\mid \rty{ghost}^\mu
          \mid \rty{Box}^\mu
          \mid \rty{Rc}^\mu
          \mid \texttt{$\rty{shared}^\mu$($\rho$)}
          \mid \texttt{$\rty{mut}^\mu$($\rho$)} \\
  \end{aligned} \\
  \PTRMOD \ni \mu \ {::=} \
      \epsilon
      \mid \rty{Cell}
      \mid \texttt{[]}
      \mid \texttt{[\rty{Cell}]} \quad
  \LIFETIMELIST \ni \vec{\rho} \ {::=} \ \langle \{ \rho \texttt{,} \}^* \rangle \quad
  \LIFETIMEVAR \ni \rho \ {::=} \ \lifetime{a} \mid \lifetime{b} \mid \cdots
  \end{gather*}
  \caption{\label{fig:rtypes-frag}Grammar for Rust type fragments (\RTYPEFRAG).}
\end{figure}

\begin{table}[t]
\caption{Corresponding concrete Rust types for each Rust type fragment without pointer qualifiers, and corresponding concrete Rust types for \rty{Box} with qualifiers excluding \lstinline|Option|-wrapped types. The concrete Rust types for the other pointer Rust type fragments with qualifiers are similar.}
\label{tab:rust-type-classes}
\footnotesize
\begin{tabular}{l|l}
\bf Rust type fragment & \bf Concrete Rust type \\\hline
\rty{unknown} & Any Rust type \\
\rty{void} & \lstinline|()| \\
\rty{bool} & \lstinline|bool| \\
\rty{i}$N$ & \lstinline[morendkeywords={u}]|u|$N$, \lstinline[morendkeywords={i}]|i|$N$ \\
\rty{f}$N$ & \lstinline[morendkeywords={f}]|f|$N$ \\
\texttt{\rty{struct}($s$, $\langle \rho_1\texttt{,} ...\texttt{,} \rho_n \rangle$)} & \lstinline[mathescape]|$s$<$\rho_1$, $...$, $\rho_n$>| \\
\rty{ghost} & \lstinline|T|, \lstinline|Option<T>| \\
\rty{Box} & \lstinline|Box<T>|, \lstinline|Option<Box<T>>| \\
\rty{Rc} & \lstinline|Rc<T>|, \lstinline|Option<Rc<T>>|, \lstinline|Weak<T>|, \lstinline|Option<Weak<T>>| \\
\texttt{\rty{shared}($\rho$)} & \lstinline[mathescape]|&$\rho$ T|, \lstinline[mathescape]|Option<&$\rho$ T>|, \lstinline[mathescape]|Ref<$\rho$, T>|, \lstinline[mathescape]|Option<Ref<$\rho$, T>>| \\
\texttt{\rty{mut}($\rho$)} & \lstinline[mathescape]|&$\rho$ mut T|, \lstinline[mathescape]|Option<&$\rho$ mut T>|, \lstinline[mathescape]|RefMut<$\rho$, T>|, \lstinline[mathescape]|Option<RefMut<$\rho$, T>>| \\\hline
$\rty{Box}^\texttt{[]}$ & \lstinline|Box<[T]>|, \lstinline|(Box<[T]>, usize)|, \\
$\rty{Box}^\rty{Cell}$ & \lstinline|Box<Cell<T>>|, \lstinline|Box<RefCell<T>>|, \\
$\rty{Box}^\texttt{[\rty{Cell}]}$ & \lstinline|Box<[Cell<T>]>|, \lstinline|Box<[RefCell<T>]>|, \lstinline|(Box<[Cell<T>]>, usize)|, \\
& \lstinline|(Box<[RefCell<T>]>, usize)| \\
\end{tabular}
\end{table}

Reference and struct Rust type fragments include \emph{lifetime inference variables}.
The Rust type fragments \texttt{$\rty{shared}^\texttt{\textunderscore}$($\rho$)} and \texttt{$\rty{mut}^\texttt{\textunderscore}$($\rho$)} include a lifetime variable $\rho \in \LIFETIMEVAR$,\footnote{We write \texttt{\textunderscore} in place of specific qualifiers or lifetime inference variables to mean any qualifier or lifetime inference variable(s).} and the Rust type fragment \texttt{\rty{struct}($s$, $\vec{\rho}$)} corresponds to a struct type $s$ with lifetimes $\vec{\rho} \in \LIFETIMELIST$.
For example, the Rust type fragment \texttt{\rty{struct}(Node, $\langle \lifetime{a} \rangle$)} corresponds to the concrete Rust type \lstinline{Node<'a>}.
The concrete Rust type that corresponds to a \rty{ghost} Rust type fragment is the type of its referent instead of a pointer to the type of its referent, which allows \refinator to remove unnecessary indirection.
A \emph{pointer} Rust type fragment (\rty{ghost}, \rty{Box}, \rty{Rc}, \rty{shared}, or \rty{mut})
may be annotated with a combination of \rty{Cell} and \texttt{[]} qualifiers to indicate that its referent is wrapped in a \lstinline{Cell} or slice type.
\lstinline{Cell} and slice types are special since their presence is determined during constraint solving, instead of during source language conversion as for other types (primitives, pointers, and structs).
Only pointer types need \lstinline{Cell} and slice qualifiers: \lstinline{Cell} is necessary only when mutating the referent of a pointer, and all arrays are accessed via pointers in the source language.

\Refinator is only concerned with inferring correct pointer types with respect to ownership, borrowing, and mutability.
Whether a pointer must be wrapped in an \lstinline{Option} depends only on whether the pointer may be \lstinline[language=C]{NULL}, and does not depend on nor affect correctness with respect to ownership, borrowing, and mutability.
As a result, \refinator's pointer Rust type fragments do not differentiate between \lstinline{Option}-wrapped and non-wrapped pointer types, and are correct with respect to ownership, borrowing, and mutability regardless of \lstinline[language=C]{NULL} pointers.
In a conservative interface translation, all pointer types may be wrapped in an \lstinline{Option}.
Determining more precise interfaces with respect to \lstinline{Option} is a subject of future work.
Additionally, whether to use \lstinline{Rc<T>} or \lstinline{Weak<T>} is difficult to determine in a fully automated manner, but a conservative translation may use \lstinline{Rc<T>} for all \rty{Rc} Rust type fragments, which
may introduce memory leaks, but does not affect correctness with respect to ownership and borrowing.
Similarly, a conservative interface translation can use \lstinline{RefCell<T>} for all \rty{Cell}-qualified Rust type fragments, but additional analysis can determine whether the simpler \lstinline{Cell<T>} can be used.

\begin{figure}
    \begin{minipage}[t]{\linewidth}
    \ttfamily
    \footnotesize
    \begin{numberedtext}
    \codeline \kw{struct} \detokenize{Node} \{
        $\langle \labeled{0}{\rty{ghost}}, \labeled{1}{\rty{i32}} \rangle$ \detokenize{data};
        $\langle \labeled{2}{\rty{ghost}}, \labeled{3}{$\rty{Rc}\ofRefCell$}, \labeled{4}{\rty{struct}(\detokenize{Node}, $\langle\rangle$)} \rangle$ \detokenize{next};
    \};\\
    \codeline $\langle \labeled{5}{\rty{ghost}}, \labeled{6}{\rty{struct}(\detokenize{Node}, $\langle\rangle$)} \rangle$ \detokenize{new}() \{
    \textcolor{gray}{/-- snip --/}
    \}\\
    \codeline $\langle \labeled{7}{\rty{void}} \rangle$ \detokenize{push}($\langle \labeled{8}{\rty{mut}(\lifetime{a})}, \labeled{9}{\rty{struct}(\detokenize{Node}, $\langle\rangle$)} \rangle$ \detokenize{list},
        $\langle \labeled{10}{\rty{i32}} \rangle$ \detokenize{x},
    ) \{\\
    \codeline \hspace*{1em}$\langle \labeled{11}{\rty{ghost}}, \labeled{12}{\rty{struct}(\detokenize{Node}, $\langle\rangle$)} \rangle$ \detokenize{n} = \detokenize{malloc}(\labeled{13}{$c_1$}{\color{gray}: \rty{i64}}, );\\
    \codeline \hspace*{1em}\textcolor{gray}{/-- snip --/}\\
    \codeline \hspace*{1em}$\langle \labeled{14}{\rty{mut}(\lifetime{b})}, \labeled{15}{\rty{Rc}\ofRefCell}, \labeled{16}{\rty{struct}(\detokenize{Node}, $\langle\rangle$)} \rangle$ \detokenize{next} = \labeled{17}{\&(*\detokenize{list}).next}{\color{gray}: \rty{mut}(\lifetime{c})};\\
    \codeline \hspace*{1em}*\detokenize{next} = \labeled{18}{\detokenize{n}}{\color{gray}: \rty{Rc}\ofRefCell};\label{line:rtypes-snippet:push-transforms-n}\\
    \codeline \hspace*{1em}\kw{drop}(\detokenize{next}, \textcolor{gray}{/-- snip --/} \detokenize{n}, \detokenize{x}, \detokenize{list}, );\label{line:rtypes-snippet:drop-in-push}\\
    \codeline \}\\
    \codeline $\langle \labeled{19}{\rty{void}} \rangle$ \detokenize{replace}($\langle \labeled{20}{\rty{mut}(\lifetime{d})}, \labeled{21}{\rty{i32}} \rangle$ \detokenize{dst},
        $\langle \labeled{22}{\rty{i32}} \rangle$ \detokenize{x},
    ) \{
    \textcolor{gray}{/-- snip --/}
    \}\\
    \codeline $\langle \labeled{23}{\rty{void}} \rangle$ \detokenize{replace_first}($\langle \labeled{24}{\rty{shared}(\lifetime{e})}, \labeled{25}{\rty{struct}(\detokenize{Node}, $\langle\rangle$)} \rangle$ \detokenize{list},
        $\langle \labeled{26}{\rty{i32}} \rangle$ \detokenize{x},
    ) \{
    \textcolor{gray}{/-- snip --/}
    \}\\
    \codeline $\langle \labeled{27}{\rty{i32}} \rangle$ \detokenize{first}($\langle \labeled{28}{\rty{shared}(\lifetime{f})}, \labeled{29}{\rty{struct}(\detokenize{Node}, $\langle\rangle$)} \rangle$ \detokenize{list},
    ) \{
    \textcolor{gray}{/-- snip --/}
    \}\\
    \codeline $\langle \labeled{30}{\rty{i32}} \rangle$ \detokenize{main}() \{\\
    \codeline \hspace*{1em}$\langle \labeled{31}{\rty{ghost}}, \labeled{32}{\rty{struct}(\detokenize{Node}, $\langle\rangle$)} \rangle$ \detokenize{list} = \detokenize{new}(); \\
    \codeline \hspace*{1em}\detokenize{push}(\labeled{33}{\detokenize{list}}{\color{gray}: \rty{mut}(\lifetime{h})}, \labeled{34}{$c_2$}{\color{gray}: \rty{i32}}, );\label{line:rtypes-snippet:main-calls-push}\\
    \codeline \hspace*{1em}$\langle \labeled{35}{\rty{shared}(\lifetime{i})}, \labeled{36}{\rty{Rc}\ofRefCell}, \labeled{37}{\rty{struct}(\detokenize{Node}, $\langle\rangle$)} \rangle$ \detokenize{next} = \labeled{38}{\&*(*\detokenize{list}).next}{\color{gray}: \rty{shared}(\lifetime{j})};\\
    \codeline \hspace*{1em}$\langle \labeled{39}{\rty{shared}\ofRefCell(\lifetime{k})}, \labeled{40}{\rty{struct}(\detokenize{Node}, $\langle\rangle$)} \rangle$ \detokenize{sublist} = \labeled{41}{*\detokenize{next}}{\color{gray}: \rty{shared}\ofRefCell(\lifetime{l})};\label{line:rtypes-snippet:next-is-live}\\
    \codeline \hspace*{1em}\detokenize{push}(\labeled{42}{\detokenize{sublist}}{\color{gray}: \rty{mut}(\lifetime{m})}, \labeled{43}{$c_3$}{\color{gray}: \rty{i32}}, );\\
    \codeline \hspace*{1em}\detokenize{replace_first}(\labeled{44}{\detokenize{list}}{\color{gray}: \rty{shared}(\lifetime{n})}, \labeled{45}{$c_4$}{\color{gray}: \rty{i32}}, );\label{line:rtypes-snippet:main-calls-replace-first}\\
    \codeline \hspace*{1em}\detokenize{replace_first}(\labeled{46}{\detokenize{sublist}}{\color{gray}: \rty{shared}(\lifetime{o})}, \labeled{47}{$c_5$}{\color{gray}: \rty{i32}}, );\\
    \codeline \hspace*{1em}\kw{drop}(\detokenize{sublist}, \detokenize{next}, \detokenize{list}, );\\
    \codeline \}
    \end{numberedtext}

    \end{minipage}
    \caption{\label{fig:rtypes-snippet}The program from \autoref{fig:source-snippet}, but with Rust types inferred by \refinator shown in place of source types. Rust types are also shown for the outermost fragment of each rvalue expression.}
\end{figure}

\section{Overview of Constraints Generated by \Refinator}
\label{sec:constraints-overview}

\Refinator generates SMT constraints on inferred Rust types that enforce semantic equivalence with the input source program and adherence to Rust's type-checking rules. It generates constraints to:
\begin{enumerate}
    \item[(\autoref{sec:consistency})] Ensure the inferred types are consistent with the source program and self-consistent. The constraints encode standard Rust typing rules---e.g., structs are well-formed, expressions are well-typed, and only mutable-typed pointers can be used mutably---according to the rules described in the Rust Reference~\cite{rust-reference} and the Rust Standard Library API~\cite{rust-std}.
    \item[(\autoref{sec:semantic})] Ensure the inferred types account for Rust's copy and move semantics, which differ from the source language's (C-like) copy semantics, and affect aliasing relationships and when memory locations are invalidated.
    \item[(\autoref{sec:borrowing})] Ensure Rust's borrowing-checking rules are satisfied, based on the operational description in Rust's non-lexical lifetimes RFC~\cite{nll-rfc} and compiler development guide~\cite{rustc-dev-guide}.
    \item[(\autoref{sec:precision})] Ensure the inferred Rust types are precise.
    These constraints are based on a principled model of the relative complexity and cost of Rust pointer types.
\end{enumerate}
The main paper
describes the generated constraints in an intuitive, example-driven way.
The formal details are in \supplement{appendices,
with \autoref{appendix:notation} introducing common notation and definitions}.

\section{Type Consistency Constraints}
\label{sec:consistency}

Here, we describe constraints \refinator generates to ensure that (1) the inferred Rust types are consistent with the source program's types and (2) a Rust translation using these types would adhere to Rust's type-checking rules, \emph{excluding borrow-checking rules} (which are handled in \autoref{sec:borrowing}).
The formal constraints generated by \refinator are presented in \supplement{\autoref{appendix:consistency}}.

\subsection{Type Parity}
\label{sec:consistency-parity}

The first set of constraints \refinator generates ensures that the inferred Rust types are consistent with the source types in the source program.
\Refinator emits constraints that, for each type label $\ell$ in the program, the \RTYPEFRAG assigned to $\ell$ matches the \STYPEFRAG assigned to $\ell$.

The general rule is that each primitive source type fragment matches the corresponding primitive Rust type fragment (e.g., \sty{i32} matches \rty{i32}), struct source type fragments match corresponding struct Rust type fragments with any lifetimes (i.e., \texttt{\sty{struct}($s$)} matches \texttt{\rty{struct}($s$, \textunderscore)}), and the pointer source type fragment matches any pointer Rust type fragment (i.e., \sty{ptr} matches $\rty{ghost}^\texttt{\textunderscore}$, $\rty{Box}^\texttt{\textunderscore}$ $\rty{Rc}^\texttt{\textunderscore}$, \texttt{$\rty{shared}^\texttt{\textunderscore}$(\textunderscore)}, \texttt{$\rty{mut}^\texttt{\textunderscore}$(\textunderscore)}).
The challenge in choosing correct, precise Rust types thus lies in choosing a pointer Rust type fragment for each pointer source type fragment.

For example, in \autoref{fig:source-snippet} and \autoref{fig:rtypes-snippet}, \texttt{Node::data} has source type $\langle \labeled{0}{\sty{ptr}}, \labeled{1}{\sty{i32}} \rangle$ and Rust type $\langle \labeled{0}{\rty{ghost}}, \labeled{1}{\rty{i32}} \rangle$.
This Rust type satisfies the type parity constraints based on the source type since the source type fragment \labeled{0}{\sty{ptr}} matches the Rust type fragment \labeled{0}{\rty{ghost}}, and likewise for the source type fragment \labeled{1}{\sty{i32}} and Rust type fragment \labeled{1}{\rty{i32}}.

\subsection{Struct Consistency}
\label{sec:consistency-struct}

\Refinator generates constraints to ensure that struct definitions are well-formed; that is, each lifetime parameter in the Rust type of a struct field appears in the list of parameters declared in the Rust struct definition.
For example, in \autoref{fig:rtypes-snippet}, the Rust type fragments of fields \lstinline{Node::data} and \lstinline{Node::next} must only be parameterized on lifetime variables declared in the definition of the struct \lstinline{Node}.
The inferred Rust types in
\autoref{fig:rtypes-snippet} satisfy this constraint because none of the type fragments of \lstinline{Node::data} and \lstinline{Node::next} are parameterized on any lifetime variables, and there are no lifetime variables declared in the definition of struct \lstinline{Node}.
If any of these type fragments contained a lifetime variable, the constraints would not be satisfied.

\subsection{Recursive Struct Definitions}
\label{sec:consistency-recursive}

Rust requires recursive structs to be defined with finite size~\cite{rust-reference}.
Consider \autoref{fig:rtypes-snippet}, where the Rust type of \lstinline{Node::next} is $\langle \labeled{2}{\rty{ghost}}, \labeled{3}{$\rty{Rc}\ofRefCell$}, \labeled{4}{\rty{struct}(Node, $\langle\rangle$)} \rangle$.
Despite being recursively defined, struct \lstinline{Node} is finitely sized due to the indirection in the Rust type of \lstinline{Node::next}.

Now, if the Rust type of \lstinline{Node::next} were $\langle \labeled{2}{\rty{ghost}}, \labeled{3}{\rty{ghost}}, \labeled{4}{\rty{struct}(Node, $\langle\rangle$)} \rangle$, which has no indirection, then the struct \lstinline{Node} has infinite size, and cannot compile.
\savespace{as in \autoref{fig:rtype-struct-recursive-rtypes}. This type cannot compile, as a mechanical translation to Rust in \autoref{fig:rtype-struct-recursive-translation} helps to show.}

\savespace{
\begin{figure}
    \small
    \centering
    \begin{subfigure}[t]{0.50\linewidth}
        \ttfamily
        \kw{struct} Node \{\\
        \hspace*{1em} \labeled{0}{\rty{i32}} data;\\
        \hspace*{1em} \labeled{1}{\rty{ghost}(\labeled{2}{\rty{struct}(Node, \kw{nil})})} next;\\
        \};
        \caption{\label{fig:rtype-struct-recursive-rtypes}A definition of struct \texttt{Node} in the source language with Rust types.}
    \end{subfigure}
    \hfill
    \begin{subfigure}[t]{0.40\linewidth}
        \lstset{basicstyle=\ttfamily\small}
        \lstinline|struct Node {|\\
        \lstinline|    data: i32,|\\
        \lstinline|    next: Node,|\\
        \lstinline|}|
        \caption{\label{fig:rtype-struct-recursive-translation}Mechanical translation of (a) into Rust. The result cannot compile.}
    \end{subfigure}
    \caption{\label{fig:rtype-struct-recursive}A recursive struct definition with infinite size.}
\end{figure}
}

\Refinator emits constraints to ensure structs are defined with finite size by introducing the notion of an integer \emph{rank} for each struct. \Refinator generates constraints requiring that a struct's rank be greater than the rank of any struct it contains directly as a field (i.e., without indirection through pointers).
Inferring $\langle \labeled{2}{\rty{ghost}}, \labeled{3}{\rty{ghost}}, \labeled{4}{\rty{struct}(Node, $\langle\rangle$)} \rangle$ as the Rust type of \lstinline{Node::next} requires $\StructRank(\texttt{Node}) > \StructRank(\texttt{Node})$, which is unsatisfiable.
\savespace{For example, \autoref{fig:rtype-struct-recursive-rtypes} does not satisfy the constraints because it requires $\Rank(\texttt{Node}) > \Rank(\texttt{Node})$.}

\subsection{Array Consistency}
\label{sec:consistency-array}

Since array types are not encoded in source types, \refinator generates constraints to insert array qualifiers in the appropriate Rust type fragments when a pointer is used as a pointer to an array of elements.
Namely, for each \texttt{\labeled{\ell}{\&$v$[$e$]}} rvalue expression, \refinator generates a constraint that the outermost type fragment of $v$ has an array qualifier.

A $\rty{ghost}^\texttt{[]}$ or $\rty{ghost}^\texttt{[\rty{Cell}]}$ type fragment indicates an array that is stored on the stack, which is only possible in Rust if the array has compile-time constant size.
Since \refinator does not determine the sizes of arrays, \refinator conservatively disallows any array-qualified \rty{ghost} pointer.
\Refinator thus emits constraints to disallow array qualifiers on \rty{ghost} pointers.

\subsection{Type Equivalence}
\label{sec:consistency-equiv}

\Refinator generates constraints that for each assignment,\footnote{Assignments occur at any instruction in \autoref{fig:source-lang} that includes the \texttt{=} symbol, at call sites when passing parameters, and at \texttt{\kw{return} \labeled{\ell}{$e$};} instructions (see \supplement{\autoref{appendix:notation-stypes} and \autoref{appendix:consistency-equiv}}).} the Rust types of the expressions on either side of the assignment are equal, except for lifetime variables.
This requirement matches the Rust type checker's rules.
For example, if \lstinline|x| is a variable of type \lstinline|i32|, then \lstinline|let p: &'a i32 = &'b x;| is allowed by the Rust type checker,\footnote{Assuming that \lstinline{'b} outlives \lstinline{'a} (\autoref{sec:borrowing-lifetimes}).} even though \lstinline|p| and \lstinline|&'b x| have different lifetimes.
But, \lstinline|let p: &'a i8 = &'b x;| is not allowed because the types \lstinline|&'a i8| and \lstinline|&'b i32| differ by more than their lifetimes.

\subsection{Pointer Transformations}
\label{sec:consistency-pointer}

\begin{figure}
    \centering
    \small
    \begin{tikzpicture}
        \node[anchor=mid] (ghost) at (0, 0) {$\rty{ghost}^\mu$};
        \node[anchor=mid] (Box) at (-3, 0) {$\rty{Box}^\mu$};
        \node[anchor=mid] (mut) at (-1.5, -1.75) {\texttt{$\rty{mut}^\mu$(\textunderscore)}};
        \node[anchor=mid] (Rc) at (1.5, -1.75) {$\rty{Rc}^\mu$};
        \node[anchor=mid] (shared) at (0, -3.5) {\texttt{$\rty{shared}^\mu$(\textunderscore)}};
        
        \path[->, transform canvas={yshift=-2}] (ghost.mid west) edge[dashed] node[below] {\lstinline[mathescape]|Box::new($p$)|} (Box.mid east);
        \path[->] (ghost) edge[dashed] node[above right] {\lstinline[mathescape]|Rc::new($p$)|} (Rc);
        \path[->] (ghost) edge node[below right] {\lstinline[mathescape]|&mut $\ p$|} (mut);
        \path[->] (Box) edge node[below left] {\lstinline[mathescape]|&mut *$p$|} (mut);
        \path[->, transform canvas={yshift=2}] (Box.mid east) edge[dashed] node[above] {\lstinline[mathescape]|*$p$|} (ghost.mid west);
        \path[->] (mut) edge[out=-120, in=-150, loop] node[below left] {\lstinline[mathescape]|&mut *$p$|} ();
        \path[->] (mut) edge node[below left] {\lstinline[mathescape]|&*$p$|} (shared);
        \path[->] (Rc) edge[out=-30, in=-60, loop] node[below right] {\lstinline[mathescape]|Rc::clone(&$p$)|} ();
        \path[->] (Rc) edge node[below right] {\lstinline[mathescape]|&*$p$|} (shared);
        \path[->] (shared) edge[loop below] node[below] {\lstinline[mathescape]|&*$p$|} ();

        \node[anchor=mid] (rule7-lhs) at (-3, 1.25) {$\rty{Box}^\texttt{[]}$};
        \node[anchor=mid east] (rule7-rhs) at (10, 1.25) {$\rty{Box}^\texttt{[\rty{Cell}]}$};
        \path[->, transform canvas={yshift=2}] (rule7-lhs.mid east) edge[dashed] node[above] {\lstinline[mathescape]|Vec::from_iter($p$.into_iter().map(Cell::new)).into_boxed_slice()|} (rule7-rhs.mid west);
        \path[->, transform canvas={yshift=-2}] (rule7-rhs.mid west) edge[dashed] node[below] {\lstinline[mathescape]|Vec::from_iter($p$.into_iter().map(Cell::into_inner)).into_boxed_slice()|} (rule7-lhs.mid east);

        \node[anchor=mid west] (rule6-lhs) at (3.25, 0) {$\rty{Box}$};
        \node[anchor=mid east] (rule6-rhs) at (10, 0) {$\rty{Box}^\rty{Cell}$};
        \path[->, transform canvas={yshift=2}] (rule6-lhs.mid east) edge[dashed] node[above] {\lstinline[mathescape]|Box::new(Cell::new(*$p$))|} (rule6-rhs.mid west);
        \path[->, transform canvas={yshift=-2}] (rule6-rhs.mid west) edge[dashed] node[below] {\lstinline[mathescape]|Box::new((*$p$).into_inner())|} (rule6-lhs.mid east);

        \node[anchor=mid west] (rule5-lhs) at (4.25, -1) {$\rty{Box}^\texttt{[\rty{Cell}]}$};
        \node[anchor=mid east] (rule5-rhs) at (10, -1) {$\rty{Box}^\rty{Cell}$};
        \path[->] (rule5-lhs.mid east) edge[dashed] node[above] {\lstinline[mathescape]|Box::new($p$[0])|} (rule5-rhs.mid west);

        \node[anchor=mid west] (rule4-lhs) at (5, -1.75) {$\rty{Box}^\texttt{[]}$};
        \node[anchor=mid east] (rule4-rhs) at (10, -1.75) {\rty{Box}};
        \path[->] (rule4-lhs.mid east) edge[dashed] node[above] {\lstinline[mathescape]|Box::new($p$[0])|} (rule4-rhs.mid west);

        \node[anchor=mid west] (rule2-lhs) at (5.75, -2.5) {\texttt{$R^\texttt{[]}$(\textunderscore)}};
        \node[anchor=mid east] (rule2-rhs) at (10, -2.5) {\texttt{$R$(\textunderscore)}};
        \path[->] (rule2-lhs.mid east) edge node[above] {\lstinline[mathescape]|&$p$[0]|} node[below] {\lstinline[mathescape]|&mut\ $p$[0]|} (rule2-rhs.mid west);

        \node[anchor=mid west] (rule3-lhs) at (4.875, -3.375){\texttt{$R^\texttt{[\rty{Cell}]}$(\textunderscore)}};
        \node[anchor=mid east] (rule3-rhs) at (10, -3.375) {\texttt{$R^\rty{Cell}$(\textunderscore)}};
        \path[->] (rule3-lhs.mid east) edge node[above] {\lstinline[mathescape]|&$p$[0]|} node[below] {\lstinline[mathescape]|&mut\ $p$[0]|} (rule3-rhs.mid west);

        \node[anchor=mid west] (rule1-lhs) at (4, -4.25) {\texttt{$\rty{shared}^\rty{Cell}$(\textunderscore)}};
        \node[anchor=mid east] (rule1-rhs) at (10, -4.25) {\texttt{$\rty{mut}$(\textunderscore)}};
        \path[->] (rule1-lhs.mid east) edge node[above] {\lstinline[mathescape]|$p$.borrow_mut()|} (rule1-rhs.mid west);
    \end{tikzpicture}

    \caption{Pointer transformations considered by \refinator. Each node is a set of Rust type fragments, where $\mu \in \PTRMOD$ and $R \in \{ \rty{shared}, \rty{mut} \}$.
    Each edge from $T_1$ to $T_2$ is labeled with a Rust expression that transforms a pointer $p$ of type $T_1$ to a pointer of type $T_2$. Expressions on dashed edges invalidate $p$, while expressions on solid edges do not. $T_1$ is \emph{transformable} (\autoref{sec:consistency-pointer}) to $T_2$ if $T_2$ is reachable from $T_1$ by traversing any edges. $T_1$ is \emph{nondestructively transformable} (\autoref{sec:semantic-pointer}) to $T_2$ if $T_2$ is reachable from $T_1$ by traversing only solid edges.}
    \label{fig:transform-pointers}
\end{figure}

To increase flexibility in type assignment, \Refinator takes advantage of \emph{pointer transformations}.
A pointer transformation from $T_1$ to $T_2$ is an operation that takes a pointer $p$ of Rust pointer type $T_1$ and produces a pointer to the same location as $p$, but has Rust pointer type $T_2$.

\refinator's constraints consider implicit pointer transformations at non-nested rvalue expressions.
To support these transformations, the source language tags each non-nested rvalue expression with a fresh type label.
Since each non-nested rvalue expression has a fresh type label, \refinator may choose a Rust type for the rvalue expression that differs from the Rust type of its path.
This difference in Rust types is reconciled by the pointer transformation.

For example, at line~\ref*{line:rtypes-snippet:push-transforms-n} in \autoref{fig:rtypes-snippet}, the path \texttt{n} has type $\langle \labeled{11}{\rty{ghost}}, \labeled{12}{\rty{struct}(Node, $\langle\rangle$)} \rangle$, but the rvalue expression \labeled{18}{n} has type $\langle \labeled{18}{\rty{Rc}\ofRefCell}, \labeled{12}{\rty{struct}(Node, $\langle\rangle$)} \rangle$.
So, there is an implicit pointer transformation, from $\rty{ghost}$ to $\rty{Rc}\ofRefCell$, at the rvalue expression \labeled{18}{n}. Specifically, the Rust expression \lstinline[mathescape]|Rc::new(RefCell::new(n))| would perform such a transformation.

\Refinator generates constraints that, for each pointer-typed non-nested rvalue expression, the type of its path must be transformable to the type of the rvalue expression in Rust using a combination of safe Rust language features (e.g., \lstinline{&mut *}) and the Rust standard library (e.g., \lstinline{RefCell::borrow_mut}), as shown in \autoref{fig:transform-pointers}.
(The figure uses both dashed and solid arrows for pointer transformations, but the distinction is not important in this section.)

\subsection{Immutability Preservation}
\label{sec:consistency-immutable}

In Rust, it is not permissible to obtain a mutable reference to data behind an immutable pointer, except when the referent is interior mutable.
For example, if $p$ is a pointer with (immutable) type \lstinline|&Node| or \lstinline|Rc<Node>|, then it is not possible to get a \lstinline|&mut| reference into the \lstinline|Node::data| field of \lstinline[mathescape]|*$p$|, %using the expression \lstinline[mathescape]|&mut *(*$p$).data|
so \refinator does not allow any rvalue expression with path \texttt{\&*(*$p$).data} to be a \rty{mut} reference.
\Refinator thus generates constraints that an rvalue expression cannot have a mutable pointer type if it takes a pointer to data behind an immutable pointer to a non-interior mutable type.

\subsection{Mutability}
\label{sec:consistency-mutability}

In Rust, only certain pointer types allow their referent to be mutated: the singly-owning \lstinline{Box<T>} pointer and the mutable reference \lstinline{&mut T}.
It is also possible to mutate a referent whose type provides interior mutability. For example, the referent of a \lstinline{&Cell<T>} or \lstinline{Rc<Cell<T>>} can be mutated even though shared references and \lstinline{Rc}s are immutable pointers.

In the source language, the only instruction that mutates the referent of a pointer is \texttt{*$v$ = \labeled{\ell}{$e$};}.
So, for each \texttt{*$v$ = \labeled{\ell}{$e$};} instruction, \Refinator generates a constraint that the outermost Rust type fragment of the variable $v$ allows mutation of its referent.

\subsection{Ownership}
\label{sec:consistency-ownership}

In Rust, all allocated objects must have an owner.
So, \refinator generates constraints that all instructions that allocate memory on the stack or heap and yield a pointer to the new allocation (e.g., \texttt{$\vec{\tau}$ $v$ = \kw{alloca}($c$);}) must yield an owning pointer (i.e., $\rty{ghost}^\texttt{\textunderscore}$, $\rty{Box}^\texttt{\textunderscore}$, or $\rty{Rc}^\texttt{\textunderscore}$).
\savespace{This includes the \texttt{$\vec{\tau}$ $v$ = \kw{alloca}($c$);} instruction and calls to certain \textsf{libc} functions, such as \lstinline[language=C]{malloc}.}

\subsection{External Functions}
\label{sec:consistency-external}

\Refinator generates constraints to ensure that the inferred types are consistent with the signatures of external functions whose bodies are not available---most importantly, for mutable pointer arguments to \textsf{libc} functions.
For example, \refinator generates a constraint that the outermost type fragment of \texttt{dst} in the function declaration \lstinline[language=C]{memcpy(void *dst, void *src, size_t n)} is \texttt{$\rty{mut}^\texttt{\textunderscore}$(\texttt{\textunderscore})}.
We derived these constraints manually for all \textsf{libc} functions called from our evaluated programs.

\section{Semantic Preservation Constraints}
\label{sec:semantic}

This section describes at a high level the constraints that \refinator generates to ensure that a Rust translation using its inferred types is semantically equivalent to the source program.
The formal constraints generated by \refinator are in \supplement{\autoref{appendix:semantic}}.

To preserve the semantics of the source program, it suffices to ensure that using \refinator's inferred types, there is a Rust translation of each source language instruction that does not illegally \emph{invalidate} any paths.
In the source language, using an expression as an rvalue never invalidates its path, e.g., in the source language, after the instruction \texttt{*$v_1$ = \labeled{\ell}{$v_2$};}, $v_2$ may always be used.
However, the same is not true in Rust due to its \emph{move semantics}.
In Rust, if $v_2$'s type is not \rty{Copy}, then $v_2$ cannot be used after the statement \lstinline[mathescape]{*$v_1$ = $v_2$;}, so $v_2$ is invalidated.\footnote{More precisely, this depends on whether it is possible to emulate a shallow copy of $v_2$, not whether $v_2$'s type is \rty{Copy}.}

When directly translating instructions in the source language to Rust, invalidation may be illegal for the following reasons: only variables and the dereference of a \lstinline{Box<T>} may be invalidated (addressed by \autoref{sec:semantic-referent}); an invalid path may not be read from (addressed by \autoref{sec:semantic-pointer}); and a borrowed path may not be invalidated (addressed by \autoref{sec:borrowing-conflict-move}).
Since the only Rust type fragments that may not be \rty{Copy} are (1) the pointer types and (2) struct and \rty{unknown} types accessed behind pointer types, \refinator only needs to constrain pointer type fragments to preserve the semantics of the source program.

\subsection{Referent Preservation}
\label{sec:semantic-referent}

In Rust, it is illegal to invalidate a path derived from a borrowed reference.
For example, if $p$ is a pointer with type \lstinline{&mut Box<T>}, then the statement \lstinline[mathescape]{let $q$ = *$p$;} invalidates the path \lstinline[mathescape]{*$p$}, but the Rust compiler prohibits this because $p$ is a mutable reference.
Although it is sometimes possible to invalidate a path derived from an \lstinline{Rc<T>} pointer via \lstinline{Rc::into_inner}, \refinator cannot statically guarantee that doing so will not introduce panics, so \refinator considers it illegal to invalidate any path derived from an \rty{Rc} pointer.
So, invalidating a path derived from a pointer may be legal only when the pointer is singly-owning, i.e., one that corresponds to $\rty{ghost}^\texttt{\textunderscore}$ or $\rty{Box}^\texttt{\textunderscore}$.
Therefore, for pointer-typed \labeled{\ell}{*$v$} and \labeled{\ell}{\&*(*$v$).$f$} rvalue expressions, \refinator generates constraints that ensure the path of the rvalue expression is not invalidated unless $v$ is a singly-owning pointer.

\subsection{Nondestructive Pointer Transformations}
\label{sec:semantic-pointer}

The constraints presented in this section prevent pointer transformations from invalidating live paths.
This is sufficient to prevent all reads of invalid paths.
To achieve this, we identify a subset of pointer transformations as \emph{nondestructive transformations}.
Nondestructive transformations are pointer transformations that do not invalidate any path in their operand, and are shown in \autoref{fig:transform-pointers} by solid edges.
\Refinator generates constraints that, for each pointer-typed non-nested rvalue expression, the type of the rvalue expression nondestructively transforms from the type of its path, unless the pointer-typed variable in the rvalue expression dies.

For example, consider line~\ref*{line:rtypes-snippet:main-calls-push} in \autoref{fig:rtypes-snippet}, where the variable \texttt{list} remains live because the rvalue expression \labeled{44}{list} is used later in line~\ref*{line:rtypes-snippet:main-calls-replace-first}.
One of \refinator's nondestructive pointer transformationation constraints is that the type of \texttt{list} is nondestructively transformable to the type of \labeled{44}{list}.
This constraint is satisfied since $\rty{ghost}$ to $\texttt{\rty{shared}(\lifetime{n})}$ is a nondestructive transformation.

\section{Borrow-Checking Constraints}
\label{sec:borrowing}

\Refinator's borrow-checking constraints are based on Rust's borrow-checking rules as described primarily in Rust's non-lexical lifetimes (NLL) RFC~\cite{nll-rfc}.
We based \refinator's borrow-checking constraints on the NLL RFC since that is what the Rust compiler currently uses for borrow checking. In contrast, the Polonius borrow checker~\cite{polonius}
accepts a superset of the programs accepted by NLL.
(Thus \refinator would technically not need to change if Polonius becomes Rust's borrow checker.)

In this section, we use the following sets.
\begin{gather*}
    \nt{RExprPath} \ni E \ ::= e \mid \texttt{*$v$} \mid \texttt{\&*(*$v$).$f$} \mid \texttt{\&$v$[$e$]} \quad
    \PATH \ni a \ ::= \ \texttt{*$v$}
        \mid \texttt{**$v$}
        \mid \texttt{*(*$v$).$f$} \\
    \POINT_M \ni p \ ::= \ (L)_\text{in}
        \mid (L)_\text{out} \quad
    \LOAN := \LIFETIMEVAR \times \PATH \quad
    L \in \INSTRLAB_M \quad
    M \in \nt{Program} \quad
\end{gather*}
The borrow-checking rules solve for the \emph{lifetime} of each lifetime inference variable.
A lifetime is a set of program points ($\POINT_M$), and we say the lifetime inference variable $\rho \in \LIFETIMEVAR$ is live at program point $p \in \POINT_M$ if $p$ is in the lifetime of $\rho$.
Intuitively, the lifetime of a $\rho \in \LIFETIMEVAR$ is the set of program points where a reference or struct containing $\rho$ in its type is live or aliased.

Using the solved-for lifetimes, the borrow-checking rules check that no use of a path \emph{conflicts} with the set of \emph{in-scope loans} at the program point immediately preceding it.
A \emph{loan} $(\rho, a) \in \LOAN$ is a pair comprised of a lifetime inference variable $\rho \in \LIFETIMEVAR$ and a path $a \in \PATH$.
We say the loan $(\rho, a)$ is \emph{in scope} at the program point $p \in \POINT_M$ if $a$ is borrowed by a reference with lifetime inference variable $\rho$ and $\rho$ is live at $p$.
A use of a path \emph{conflicts} with an in-scope loan if the use accesses a mutably borrowed path, writes to a borrowed path, or invalidates a borrowed path.
A drop of a variable also conflicts with in-scope loans of paths involving that variable, if that variable is not a reference.

\Refinator generates constraints that (1) compute the in-scope loans at each program point (\autoref{sec:borrowing-lifetimes} and \autoref{sec:borrowing-loans}) and (2) disallow borrowing conflicts (\autoref{sec:borrowing-conflict}).
\Supplement{\autoref{appendix:borrowing}} presents the constraints in detail.

\subsection{Lifetimes}
\label{sec:borrowing-lifetimes}

\Refinator generates constraints to infer (1) the lifetime of each lifetime inference variable based on \emph{liveness}, \emph{subtyping}, and \emph{reborrowing}, following the NLL RFC~\cite{nll-rfc}, and (2) the lifetime of each function \emph{lifetime parameter}, following the Rust Compiler Development Guide~\cite{rustc-dev-guide}.

\subsubsection*{Liveness}

A lifetime inference variable $\rho \in \LIFETIMEVAR$ is live at every program point where a variable with a type that mentions $\rho$ is live.
For example, in \autoref{fig:rtypes-snippet}, the variable \texttt{next} is live at $(\ref*{line:rtypes-snippet:next-is-live})_\text{in},$\footnote{$(\ref*{line:rtypes-snippet:next-is-live})_\text{in}$ is the program point immediately before the instruction at line~\ref*{line:rtypes-snippet:next-is-live}.} and the lifetime inference variable \lifetime{i} appears in the type of \texttt{next}, so \refinator generates a constraint that ensures \lifetime{i} is live at $(\ref*{line:rtypes-snippet:next-is-live})_\text{in}$.%
\footnote{Throughout \autoref{sec:borrowing}, we refer to Rust types in \autoref{fig:rtypes-snippet} to explain how each constraint works intuitively. Since Rust types are unknown during constraint generation, we describe what each constraint \emph{ensures} in terms of \autoref{fig:rtypes-snippet}, as opposed to the actual assertion emitted by \refinator.}

\subsubsection*{Subtyping}

The borrow-checking rules infer that for an assignment \texttt{$d$ = $s$}, the type of $s$ must be a subtype of the type of $d$.
Each lifetime inference variable in $s$ must outlive the corresponding lifetime inference variable in $d$, and depending on the types of $s$ and $d$, some outlives relationships in the opposite direction must hold.

According to the Rust Reference~\cite{rust-reference}, the shared reference type \lstinline{&'a T} is covariant in both \lstinline{'a} and \lstinline{T}; that is, the Rust type \lstinline{&'a T} is a subtype of \lstinline{&'b U} if \lstinline{'a} outlives \lstinline{'b} and \lstinline{T} is a subtype of \lstinline{U}.
However, the mutable reference type \lstinline{&'a mut T} is covariant in \lstinline{'a} but invariant in \lstinline{T}; that is, the Rust type \lstinline{&'a mut T} is a subtype of \lstinline{&'b mut U} if \lstinline{'a} outlives \lstinline{'b} and \lstinline{T} is exactly \lstinline{U}, including lifetimes.
Similarly, \lstinline{Cell<T>} is invariant in \lstinline{T}.

Consider the assignment \texttt{\detokenize{sublist} = \labeled{41}{*next}} at line~\ref*{line:rtypes-snippet:next-is-live} in \autoref{fig:rtypes-snippet}.
The types of the right- and left-hand sides are $\langle \labeled{41}{\rty{shared}(\lifetime{l})}, \labeled{37}{\rty{struct}(Node, $\langle\rangle$)} \rangle$ and $\langle \labeled{39}{\rty{shared}(\lifetime{k})}, \labeled{40}{\rty{struct}(Node, $\langle\rangle$)} \rangle$, respectively.
\Refinator generates a constraint based on the subtyping rules to ensure \lifetime{l} outlives \lifetime{k}.

\subsubsection*{Reborrowing}

According to the borrow-checking rules, for each borrow expression \lstinline{&$\rho \; a$} (which borrows the path $a$ with lifetime variable $\rho$), the lifetime of each reference that is dereferenced in $a$ outlives $\rho$.
The reborrowing constraints ensure that loans of paths by references in $a$ are in scope as long as references derived from $a$ are in scope.

When generating its reborrowing constraints, \Refinator treats the non-nested rvalue expressions \labeled{\ell}{*$v$} and \labeled{\ell}{$e$} that copy pointers as their equivalent reborrow expressions, \texttt{\&**$v$} and \texttt{\&*$e$}, respectively.
For example, the rvalue expression \labeled{41}{*next} in \autoref{fig:rtypes-snippet}, when treated as \texttt{\&**next}, dereferences \texttt{*next} and \texttt{next}.
Since \texttt{next} is a \texttt{\rty{shared}(\lifetime{i})} and \labeled{41}{*next} is a \texttt{\rty{shared}\ofRefCell(\lifetime{l})}, \refinator generates a constraint that ensures \lifetime{i} outlives \lifetime{l}.

\subsubsection*{Lifetime parameters}

According to the Rust Compiler Development Guide~\cite{rustc-dev-guide}, the lifetimes of a function's lifetime parameters (which in \refinator are the lifetime inference variables that appear in the function's signature) include the entire function body.
In \autoref{fig:rtypes-snippet}, \lifetime{a} appears in the signature of \texttt{push}, so \Refinator generates a constraint that \lifetime{a} is live at every point in the body of \texttt{push}.

\Refinator also generates constraints to propagate the relationships between the lifetime parameters of a function into relationships between the lifetimes of arguments and return values at the call sites for that function.
So, taking the relationships between lifetime parameters of a function as annotations in the function's signature, as required according to the Rust Compiler Development Guide~\cite{rustc-dev-guide}, the lifetime constraints expressed by these annotations are satisfied at each call site.

\subsection{In-Scope Loans}
\label{sec:borrowing-loans}

\Refinator generates constraints that, in effect, compute the in-scope loans at every program point using data-flow analysis,
based on the data-flow analysis described in the NLL RFC~\cite{nll-rfc}.
Data-flow facts are the set of in-scope loans, and control-flow merges take the union of facts because the analysis is a \emph{may} analysis.

Let $\Loans: \POINT_M \to \mathcal{P}(\LOAN)$ be a function that gives the data-flow facts at each program point.
The transfer function at an instruction at line $L \in \INSTRLAB_M$ is defined as follows using one \emph{gen} set, $\gen_\text{borrow}(L) \subseteq \LOAN$, and two \emph{kill} sets, $\dfkill_\text{lifetime}(L) \subseteq \LIFETIMEVAR$ and $\dfkill_\text{assign}(L) \subseteq \PATH$.
\begin{gather*}
    \Loans((L)_\text{out}) = ((\Loans((L)_\text{in}) \setminus \dfkill_\text{lifetime}(L) \times \PATH) \cup \gen_\text{borrow}(L)) \setminus \LIFETIMEVAR \times \dfkill_\text{assign}(L)
\end{gather*}
\label{sec:borrowing-transfer-function-overview}
The gen and kill sets are defined at each label $L \in \INSTRLAB_M$ as follows.
\begin{enumerate}
    \item For each $\rho \in \LIFETIMEVAR$, $\rho \in \dfkill_\text{lifetime}(L)$ if and only if $\rho$ is dead at $(L)_\text{out}$.
    \item $(\rho, a) \in \gen_\text{borrow}(L)$ if and only if an rvalue expression that borrows the path $a$ appears in the instruction at $L$ and $\rho$ is live at $(L)_\text{out}$.
    \item $\dfkill_\text{assign}(L) = \{ v, \texttt{*$v$} \}$ if the instruction at $L$ is \texttt{*$v$ = \labeled{\ell}{$e$};}. Otherwise, $\dfkill_\text{assign}(L) = \emptyset$.
\end{enumerate}

\subsection{Borrowing Conflicts}
\label{sec:borrowing-conflict}

Borrowing conflicts can occur when a path is \emph{borrowed} by an rvalue expression, \emph{moved} by an rvalue expression, \emph{assigned into}, or \emph{dropped}.

\subsubsection{Borrowing rvalue expressions}
\label{sec:borrowing-conflict-borrow}

A pointer-typed non-nested rvalue expression \labeled{\ell}{$E$} that appears in the instruction at $L \in \INSTRLAB_M$ and borrows $a \in \PATH$ conflicts with $(\rho', a') \in \LOAN$ if
\begin{enumerate}
    \item $(\rho', a')$ is in scope at $(L)_\text{in}$; and
    \item $a'$ is a prefix of $a$, or $a$ is obtained from $a'$ by stripping field accesses and dereferences;\footnote{These paths correspond to the paths of relevant loans for ``deep'' accesses of a path in the NLL RFC~\cite{nll-rfc}.} and
    \item \labeled{\ell}{$E$} or the rvalue expression that produced $(\rho', a')$ is a mutable reference.\footnote{A conflict only occurs if at least one of these accesses is a ``write'' access, which includes taking a mutable reference, according to the NLL RFC~\cite{nll-rfc}.}
\end{enumerate}
To avoid borrowing conflicts, \Refinator emits constraints that if conditions (1) and (2) are met, then (3) must not hold.
Specifically, if a pointer-typed non-nested rvalue expression \labeled{\ell}{$E$} that borrows $a \in \PATH$ appears in the instruction at $L \in \INSTRLAB_M$, and the rvalue expression \labeled{\ell'}{$E'$} produces $(\rho', a') \in \LOAN$ where $(\rho', a')$ is in scope at $(L)_\text{in}$, and $a'$ relates to $a$ as in (2), then both \labeled{\ell}{$E$} and \labeled{\ell'}{$E'$} must be shared references.

For example, consider the rvalue expression \labeled{44}{call}, which borrows \texttt{*call} and appears at line~\ref*{line:rtypes-snippet:main-calls-replace-first} in \autoref{fig:rtypes-snippet}.
At $(\ref*{line:rtypes-snippet:main-calls-replace-first})_\text{in}$, the loan $(\lifetime{j}, \texttt{*(*list).next})$, produced by \labeled{38}{\&*(*list).next}, is in scope and satisfies conditions (1) and (2).
So, \refinator generates a constraint that ensures that both \labeled{44}{call} and \labeled{38}{\&(*list).next} are shared references.

\subsubsection{Moving rvalue expressions}
\label{sec:borrowing-conflict-move}

A pointer-typed non-nested rvalue expression \labeled{\ell}{$E$} that appears at an instruction at $L \in \INSTRLAB_M$ that moves $a \in \PATH$ conflicts with $(\rho', a') \in \LOAN$ if
\begin{enumerate}
    \item $(\rho', a')$ is in scope at $(L)_\text{in}$; and
    \item $a'$ is a prefix of $a$, or $a$ is obtained from $a'$ by stripping field accesses and dereferences.
\end{enumerate}
To avoid borrowing conflicts, \Refinator emits constraints that if conditions (1) and (2) are met, then the rvalue expression's path must not be moved; that is, the type of \labeled{\ell}{$E$} must nondestructively transform from the type of its path $E$ (\autoref{sec:semantic-pointer}).

Consider again the rvalue expression \labeled{44}{list} in \autoref{fig:rtypes-snippet}.
As discussed in \autoref{sec:borrowing-conflict-borrow}, there is a loan that satisfies conditions (1) and (2).
So, \refinator generates a constraint that ensures that the type of \texttt{list} is nondestructively transformable to the type of \labeled{44}{list}.

\subsubsection{Assignments}
\label{sec:borrowing-conflict-assign}

An assignment into $a \in \PATH$ at $L \in \INSTRLAB_M$ conflicts with $(\rho', a') \in \LOAN$ if (1) $(\rho', a')$ is in scope at $(L)_\text{in}$, and (2) $a'$ is a prefix of $a$.\footnote{These paths correspond to the paths of relevant loans for ``shallow'' accesses of a path in the NLL RFC~\cite{nll-rfc}.}
To avoid borrowing conflicts, \refinator emits constraints that if conditions (1) and (2) hold, then $a$ has interior mutable type, and the rvalue expression that produced loan $(\rho', a')$ is a shared reference.

\subsubsection{Variable drops}
\label{sec:borrowing-conflict-drop}

When an owning pointer is dropped, the borrow-checking rules require that no references derived from it are live.
For a variable $v$ that is dropped at $L \in \INSTRLAB_M$, a loan $(\rho, a) \in \LOAN$ that is in scope at $(L)_\text{in}$ conflicts with the drop of $v$ if
\begin{enumerate}
    \item $v$ is an owning pointer, and $a = \texttt{*$v$}$ or $a = \texttt{*(*$v$).$f$}$; or
    \item \texttt{*$v$} is an owning pointer and $a = \texttt{**$v$}$.
\end{enumerate}
To avoid borrowing conflicts, \Refinator emits constraints that prohibit loans that conflict with a drop of a variable $v$ from being in scope at the point where $v$ is dropped.
For example, in \autoref{fig:rtypes-snippet}, because \texttt{n} in \texttt{push} is an owning pointer, \refinator generates a constraint that ensures that there are no loans that borrow paths derived from \texttt{*n} in scope at $(\ref*{line:rtypes-snippet:drop-in-push})_\text{in}$, where \texttt{n} is dropped.
Since \texttt{n} escapes the \texttt{push} function into \texttt{list}, if \lstinline{Node::next} were a borrowed reference, then a loan of \texttt{*n} would be in scope when \texttt{n} is dropped, so this constraint forces \lstinline{Node::next} to be an owning pointer.

\section{Precision}
\label{sec:precision}

The constraints presented so far ensure that Rust types are \emph{correct}, but the types may not be \emph{precise} (i.e., simplest and least permissive), which is desirable for minimizing run time and memory usage, as well as for avoiding run-time borrowing panics in translated function bodies using the inferred interface types (\autoref{sec:refinator-overview:limitations}).
To achieve precise types, \refinator minimizes the following objectives in descending order of priority:
(1) the number of \rty{Cell} qualifiers,
(2) the sum of the \emph{transformation cost} of each pointer Rust type fragment,
(3) the sum of the \emph{run-time cost} of each pointer Rust type fragment, and
(4) the number of array qualifiers.%
\footnote{In (1), \rty{Cell} qualifiers refer to both \rty{Cell} and \texttt{[\rty{Cell}]}. In (4), array qualifiers refer to both \texttt{[]} and \texttt{[\rty{Cell}]}.}
For these objectives, \refinator only considers type fragments that appear in interfaces.

The transformation cost of a pointer Rust type fragment roughly corresponds to the number of type fragments to or from which it is transformable, depending on its location.
The outermost type fragments of struct fields and a global variables are singled out because they correspond to extra indirection introduced by compilation to LLVM IR or conversion to the source language, and are likely redundant.
For these type fragments, \refinator considers owning pointer type fragments ($\rty{ghost}^\texttt{\textunderscore}$, $\rty{Box}^\texttt{\textunderscore}$, $\rty{Rc}^\texttt{\textunderscore}$) to be less costly than borrowed pointer type fragments since owning pointer type fragments are transformable \emph{to} more type fragments than borrowed pointer type fragments.
This cost function encourages redundant indirection to be removed via $\rty{ghost}^\texttt{\textunderscore}$ type fragments.
For other Rust type fragments, \refinator considers borrowed pointers (\texttt{$\rty{shared}^\texttt{\textunderscore}$(\textunderscore)}, followed by \texttt{$\rty{mut}^\texttt{\textunderscore}$(\textunderscore)}) to be less costly than owning pointer type fragments since borrowed pointers are transformable \emph{from} more type fragments than owning pointer type fragments.
This cost function encourages \refinator to choose the least permissive types whenever possible, e.g., using a \texttt{$\rty{shared}^\texttt{\textunderscore}$(\textunderscore)} instead of an $\rty{Rc}^\texttt{\textunderscore}$.

The run-time cost of a pointer Rust type fragment corresponds to the overhead introduced by its concrete Rust types.
\refinator considers $\rty{Rc}^\texttt{\textunderscore}$ to be the most costly, followed by $\rty{Box}^\texttt{\textunderscore}$, $\rty{ghost}^\texttt{\textunderscore}$, \texttt{$\rty{mut}^\texttt{\textunderscore}$(\textunderscore)}, and \texttt{$\rty{shared}^\texttt{\textunderscore}$(\textunderscore)}.
This ordering is strict in order to break ties in case multiple type assignments have the same cumulative transformation cost.
However, there is not necessarily a unique assignment of interface types that minimizes \refinator's optimality objectives.

\Supplement{\autoref{appendix:precision}} presents the precision constraints in detail.

\section{Evaluation}

We evaluate \refinator on three metrics: correctness and precision of inferred Rust interfaces and run-time performance of \refinator.

\subsection{Implementation}
\label{sec:impl}

We implemented \refinator's two top-level components: conversion to source language and constraint generation analysis (\autoref{fig:refinator-overview}).

\subsubsection*{Conversion to source language}

\Refinator's conversion from the input C program is implemented as an LLVM compiler pass preceded by LLVM's \textsf{mem2reg} optimization to reduce the size and complexity of the IR~\cite{llvm}.
Note that the implementation's source language supports more features, such as LLVM IR \texttt{switch} and \texttt{unreachable} instructions, than the version presented in \autoref{fig:source-lang}.
However, it does not support function pointers, \lstinline[language=C]{union} types, variadic functions, or ternary operators.

\Refinator infers source types from LLVM IR, including monomorphic \lstinline[language=C]{void} pointers (i.e., \lstinline[language=C]{void} pointers cast to a single other pointer type).
The implementation cannot handle polymorphic pointers (i.e., \lstinline[language=C]{void} pointers cast to multiple other pointer types) in general (\autoref{sec:refinator-overview:limitations}).
When encountering polymorphic pointers or unsupported type casts (\autoref{sec:refinator-overview:limitations}), the implementation infers \sty{unknown} for some source type fragments, but this may cause constraint generation to fail.

\subsubsection*{Constraint generation analysis}

We implemented \refinator's constraint generation analysis as a standalone program
written in Rust that analyzes the source program and generates SMT constraints.
The analysis uses the Rust \textsf{z3} crate to generate and solve constraints with Z3~\cite{z3}.

To improve solving performance, the analysis includes the following optimizations:
\begin{itemize}
    \item Rust type fragments are factored into separate, independently constrained components.
    \item Lists of lifetime variables are represented implicitly, without using an SMT datatype.
    \item Each \LOAN is represented by the rvalue expression that generates it---there is a one-to-one correspondence between possible loans and pointer-typed non-nested rvalue expressions.
    \item Every function and relation the paper presents as defined SMT functions is either inlined, or declared as an uninterpreted function in SMT and constrained by a set of axioms.
    \item Superfluous constraints are not generated, e.g., loans from different functions cannot conflict.
    \later{\item It emits constraints that fix known parts of \textsf{libc} function signatures, e.g. the format string argument of \lstinline|printf| is known to be compatible with Rust type \lstinline|&Vec<u8>|, which is optimal with respect to the Rust types considered by \Refinator.
    \mike{Should this be part of Design?}
    \victor{I'm unsure about this. This doesn't impact the results of \Refinator, but it may impact performance by reducing how many types Z3 needs to solve.}
    \mike{Oh I see, what you're describing here is more specific than \refinator's knowing the borrowing semantics of \textsf{libc} functions (which I think should be in Design).}}
\end{itemize}

\subsection{Methodology}
\label{sec:meth}

Our evaluation uses the following C programs as input:
the paper's running example (\autoref{fig:running-example:c});
the four smallest programs (in lines of C code\footnote{We inadvertently selected the smallest programs based on the size of the program \emph{including testing code}, which is not part of the input to \refinator.
Our reported lines of C code do not however include the testing code.})
in CRUST-Bench, \bench{ulidgen}, \bench{leftpad}, \bench{amp}, and \bench{nandc}; and another CRUST-Bench program, \bench{rbtree}, because it uses a recursive datatype extensively.

We modified each program by combining multiple files into one source file to simplify inputting it to \refinator.
We additionally modified \bench{nandc} and \bench{rbtree} slightly to conform to \refinator's current restrictions without meaningfully changing its functionality, as described later for each program.

To confirm \refinator's translated interfaces were correct, we manually translated each C program to a (safe) Rust program that used \refinator's interface.
To assess precision, we manually translated each C program to Rust line-by-line and attempted to use the most precise types,\footnote{By performing translation line-by-line, all aliasing present in the C program is preserved in the Rust translation.} and compared this program's interface to \refinator's interface.
For \bench{amp} and \bench{rbtree}, we manually translated the interfaces ``blind'' (i.e., without looking at \refinator's interface). For the other programs, which arguably have simpler ownership and borrowing choices, we looked at \refinator's interface first.
Although \refinator is correct and precise (according to its cost function) by design, this manual evaluation of \refinator's inferred interface provides an empirical assessment of these claims.

In addition to the above fully evaluated programs, we evaluated three larger programs from CRUST-Bench solely for the purpose of measuring \refinator's performance scalability. We initially selected programs that were closest in size (lines of C code excluding testing code) to $2^k$ times larger than \bench{rbtree} for integers $k\ge1$. However, we encountered technical issues---source language conversion fails amid heavy use of polymorphic pointers and function pointers---with several of these programs (\bench{recordmanager}, \bench{megalania}, and \bench{libfor}), so we replaced them with the next-closest-in-size programs and ended up with the following programs: \bench{carrays}, \bench{jccc}, and \bench{tisp}. We modified these programs to not use the ternary operator, \lstinline[language=C]{union}s, function pointers, polymorphic pointers, pointer-to-integer casts, and a few unsupported C library functions that resulted in instructions that \refinator does not currently handle (\lstinline{isdigit}, \lstinline{isalnum}, and \lstinline{calloc}). These changes did not significantly affect the size of these programs or the complexity of inferring interfaces for them.

To measure \refinator's analysis performance, we summed the wall-clock times for source language translation and constraint generation analysis.
We measured \refinator's constraint-solving performance as the wall-clock time used by Z3 to solve the constraints.
To account for running time variability, we ran \refinator on each program five times and reported the average run time.

The experiments ran as jobs on nodes of a Linux cluster with Xeon Platinum 8268 CPUs at 2.9\,GHz with 170\,GB of RAM. Each run of \refinator, which is single-threaded, ran on a dedicated socket. The experiments used LLVM 17.0.6 and Z3 4.13.3.

\subsection{Correctness, Precision, and Performance Results}
\label{sec:results}

\autoref{tab:stats-perf} reports characteristics of each source program and run-time performance of \refinator.
The first column reports non-empty, non-comment lines of the C code \refinator translated to its source language.
The next three columns count instructions, type fragments (i.e., type labels), and interface variables in the source program.
To count interface variables, we excluded functions that are external to the source program (e.g., \textsf{libc} and LLVM intrinsics) and \lstinline{main} (whose interface variables' types \refinator leaves unconstrained, since C and Rust have different signatures for \lstinline{main}). The table reports all instructions and type labels since they represent the source program \refinator analyzed. The \emph{\Refinator performance} columns report the number of SMT constraints generated, average \refinator analysis time, and average Z3 solving time.
The \emph{Inferred interface} columns report whether \refinator's inferred interfaces are correct and precise.

\begin{table}[t]
\caption{Benchmark characteristics, \refinator performance, and inferred interface correctness and precision. Constraints are rounded to three significant figures, and average run times are rounded to two significant figures and the nearest 0.1\,s. The last three rows show the ``scalability'' benchmarks, for which we did not manually analyze the inferred interfaces.}
\label{tab:stats-perf}

\newcommand\aiv[1]{{\footnotesize ?}} 

\small
\centering
\begin{tabular}{@{}l|rrrr|rrr|c@{\;}c@{}}
              & \mc{4}{@{}c|}{\bf Program characteristics}             & \mc{3}{c|}{\bf \Refinator performance} & \mc{2}{c@{}}{\bf Inferred interface} \\
            & C LoC      & Source & Type  & Interface & Constraints & Analysis & Solving & Correct & Precise \\
\bf Program &            & instrs & frags & vars      &             & time     & time    &         &         \\\hline
\autoref{fig:running-example:c} & \stat{41} & \stat{34} & \stat{114} & \stat{14} & \assertions{1080} & $<\,$\tsec{0.1} & \tsec{0.573629093} & \checkmark & \checkmark \\ 
\bench{ulidgen} & \stat{69} & \stat{184} & \stat{479} & \stat{3} & \assertions{9370} & \tsec{0.126378952} & \tsec{32.720790864} & \checkmark & \checkmark \\
\bench{leftpad} & \stat{72} & \stat{131} & \stat{344} & \stat{6} & \assertions{5897} & \tsec{0.084292143} & \tsec{7.80041016} & \checkmark & \checkmark \\
\bench{amp} & \stat{99} & \stat{197} & \stat{536} & \stat{17} & \assertions{8855} & \tsec{0.144421344} & \tsec{16.9997365} & \checkmark & \checkmark \\
\bench{nandc} & \stat{122} & \stat{383} & \stat{971} & \stat{42} & \assertions{6456} & \tsec{0.09272542} & \tsec{9.934883869} & \checkmark & \checkmark \\
\bench{rbtree} & \stat{346} & \stat{767} & \stat{2081} & \stat{43} & \assertions{111877} & \tsec{2.100333544} & \tsec{1126.70732631} & \checkmark & \checkmark \\\hline
\bench{carrays} & \stat{618} & \stat{740} & \stat{1799} & \aiv{165} & \assertions{32706} & \tsec{0.667987229} & \tsec{148.121643313} & ? & ? \\
\bench{jccc} & \stat{1394} & \stat{1983} & \stat{5016} & \aiv{432} & \assertions{188208} & \tsec{4.221009811} & \tsec{16718.967761750} & ? & ? \\
\bench{tisp} & \stat{3656} & \stat{2908} & \stat{7540} & \aiv{401} & \assertions{283289} & \tsec{7.071361051} & \tsec{25026.26454} & ? & ? \\
\end{tabular}
\end{table}

These results show that \refinator analyzed programs with dozens to thousands of lines of C code, corresponding to up to thousands of instructions in the source language, and hundreds to thousands of Rust type fragments to solve for.
\recentsavespace{The number of interface variables
varies across the programs and does not track closely with program size.}
\Refinator's conversion to its source language and constraint generation are fast. Constraint-solving time dominates, scaling superlinearly with program size and number of generated constraints.
\label{sec:eval-scalability}%
Interestingly, although \bench{carrays} is about the same size as \bench{rbtree} in terms of instructions, \refinator generates significantly fewer constraints and solves them significantly faster. We believe the main reason is that \bench{carrays}'s functions tend to be smaller than \bench{rbtree}'s, and few large functions lead to more constraints than many small functions do. The solving times for the two largest programs are significantly higher than for \bench{rbtree}, although we note that solving time increases linearly with program size from \bench{jccc} to \bench{tisp}, suggesting that solving time may scale reasonably well with program size if not function size. Future work is needed to measure performance for larger programs and to develop optimizations to improve scalability.

We manually assessed
the correctness and precision of the interfaces produced by \refinator, using the methodology described in \autoref{sec:meth}.
For every (non-scalability) program, we were able to validate \refinator's inferred interface by writing a safe Rust program that used the interface and preserved semantics, as the \emph{Correct} column reports.
We also determined that \refinator's interfaces were precise---we were unable to discover an interface for any program that we considered to be more precise than \refinator's inferred interface, as the \emph{Precise} column reports.

In the following, we discuss each non-scalability benchmark in turn.

\subsubsection*{Running example}

The running example in \autoref{fig:running-example:c} creates and updates a linked list with interesting aliasing patterns.
The source program, shown in abbreviated form in \autoref{fig:source-snippet},
has 14 interface variables, for which \refinator chose the types shown in \autoref{fig:rtypes-snippet},
which are correct as evidenced by the correct function bodies in \autoref{fig:rtypes-snippet}.
We were unable to create an interface more precise than the inferred one.

\subsubsection*{\bench{ulidgen}}

\bench{ulidgen} is a library from CRUST-Bench that generates probabilistically unique identifiers.
\later{\victor{This is actually possible in safe Rust using slices, but \refinator doesn't (have the machinery to) support this in full generality.}}
The program consists of functions \lstinline{ulidgen_r} and \lstinline{main} and a global variable \lstinline{b32alphabet}.

Our manually translated interface differs from \refinator's interface only in choosing \lstinline{&str} instead of \lstinline{&[u8]} for \lstinline{b32alphabet}.
Since \lstinline{&str} is a more idiomatic version of \lstinline{&[u8]}, we consider the inferred interface to be correct and precise.

\subsubsection*{\bench{nandc}}

\bench{nandc} is a program from CRUST-Bench that simulates digital circuits.
It consists of functions that simulate logic gates and
adders and subtractors, some of which have a \lstinline[language=C]{bool*} output parameter.
We replaced a ternary operator expression with an equivalent branching conditional statement since \refinator's source language translation does not currently support the ternary operator.

\bench{nandc} consists of (relatively many) small functions, making its interface large compared to those of similarly sized programs.
\Refinator inferred all occurrences of the \lstinline[language=C]{bool*} type in output parameters to be
\lstinline{&mut i8},
as a result of LLVM IR not supporting \lstinline[language=C]{bool*} types.
Since the concrete Rust types \lstinline{bool}, \lstinline{i8}, and \lstinline{u8}
have identical cost, we consider
\lstinline{&mut i8}
to be correct and precise.

\subsubsection*{\bench{leftpad}}

\bench{leftpad} is a library from CRUST-Bench for padding a string on the left to a fixed length. The program's functionality consists of a single \lstinline[language=C]{leftpad} function and a \lstinline[language=C]{main} function allowing one to use the library as a command-line interface.
There are only six interface variables in this program, since the \lstinline[language=C]{main} function is not considered.

Our manual implementation uses \lstinline{&String}, \lstinline{Option<&String>}, and \lstinline{Option<&mut String>} for the string parameters of the \lstinline{leftpad} function.
\Refinator infers types $\langle \texttt{$\rty{shared}^\texttt{[]}$(\textunderscore)}, \rty{i8} \rangle$, which maps to \lstinline{&[u8]} and \lstinline{Option<&[u8]>}, and $\langle \texttt{$\rty{mut}^\texttt{[]}$(\textunderscore)}, \rty{i8} \rangle$, which maps to \lstinline{Option<&mut [u8]>}.
Since \lstinline{String} is an idiomatic wrapper around \lstinline{Vec<u8>}, and thus \lstinline{&String} is an idiomatic wrapper around \lstinline{&[u8]}, we consider \refinator's chosen types to be precise.

\subsubsection*{\bench{amp}}

\bench{amp} is a library from CRUST-Bench that implements a message encoding and decoding protocol.
\bench{amp}'s interface includes one struct and five functions excluding \lstinline{main}, and has a total of 17 interface variables.

We wrote a manual translation of \bench{amp} before looking at \refinator's inferred interface---which we found was more precise than ours.
Specifically, the function \lstinline{amp_encode} can take a borrowed \lstinline{&[T]} instead of an owned \lstinline{[T; N]} as a parameter, which is arguably better since passing by reference copies only one word to the stack, while passing by value copies three words.
We consider \refinator's interface to be correct and precise.

\subsubsection*{\bench{rbtree}}

\bench{rbtree} is a library from CRUST-Bench that implements a red--black tree.
The library consists of two \lstinline[language=C]{int} global variables, \lstinline[language=C]{RBTREE_RED} and \lstinline[language=C]{RBTREE_BLACK}; two structs, \lstinline[language=C]{rbtree} and \lstinline[language=C]{node_t}; and 13 functions
that use \lstinline[language=C]{rbtree*} or \lstinline[language=C]{node_t*} parameters and return values.
Since our \refinator implementation cannot handle polymorphic \lstinline[language=C]{void} pointers in general, we manually monomorphized \lstinline[language=C]{malloc} and \lstinline[language=C]{free} from \textsf{libc} (but did not include them in our count of interface variables).


As for \bench{amp}, we initially wrote a manual translation of \bench{rbtree} before looking at \refinator's interface, which turned out to be significantly different from ours.
In our manual translation, many functions that modify the fields of \lstinline{node_t} structs take \lstinline{Rc<RefCell<node_t>>} parameters, where the \lstinline{RefCell} is necessary in order to modify the fields of \lstinline{node_t}s behind aliased pointers.
However, in \refinator's interface, all the fields of \lstinline{node_t} are interior mutable, removing the need to use \lstinline{RefCell} to translate most \lstinline[language=C]{node_t*} function parameters.
\Refinator's interface also identifies \lstinline[language=C]{node_t*} function parameters that can use \lstinline{&node_t} instead of the more costly \lstinline{Rc<RefCell<node_t>>}.
We were still unable to produce an implementation of \bench{rbtree} using an interface more precise than the one produced by \refinator, so we consider \refinator's interface to be precise.

\subsection{Comparison with CRUST-Bench's Hand-Translated Interfaces}
\label{sec:crust-bench-comparison}

CRUST-Bench provides a hand-translated Rust interface for each benchmark~\cite{CRUST-Bench}, but an apples-to-apples comparison with \refinator's interfaces is hard because CRUST-Bench's interfaces do not adhere to the same constraints as \refinator's interfaces. Here we provide a qualitative comparison.

Overall, the CRUST-Bench and \refinator interfaces are largely the same. They differ in the following ways, which we group into four categories:


\subsubsection*{CRUST-Bench's interfaces are not one-to-one with the C program in a few cases.}

    \begin{itemize}
    \item CRUST-Bench's \bench{leftpad} interface omits a parameter \lstinline{dest_sz} from the \lstinline{leftpad} API function.
    \item CRUST-Bench's \bench{rbtree} interface omits from the \lstinline{rbtree} struct a sentinel \lstinline{nil} field, which the C program uses to represent node absence, and instead uses \lstinline{Option} to allow for node absence. In contrast, \refinator's interface retains the \lstinline{nil} field.
    \end{itemize}

\subsubsection*{CRUST-Bench's interfaces are imprecise in a few cases.}
    CRUST-Bench's \bench{rbtree} interface uses the type \lstinline{Rc<RefCell<Node>>} for all parameters that were \lstinline[language=C]{node_t*} in C. However, it is possible to correctly use the less costly \lstinline{&Node} or \lstinline{Rc<Node>} types for all parameters in exchange for making all and only the fields of \lstinline{node_t} interior mutable, as mentioned in \autoref{sec:results}.
    \refinator chooses the latter, which makes the interface, when considered as a whole, more precise.

\subsubsection*{\Refinator's interfaces use less idiomatic types than CRUST-Bench's types.}

    \begin{itemize}
    \item \Refinator's interface for \bench{ulidgen} uses \lstinline{u8} instead of \lstinline{char}.
    \item \Refinator's interfaces for \bench{leftpad} and \bench{amp} use \lstinline{[u8]} instead of \lstinline{str}.
    \item In a similar vein, \refinator's interface for \bench{nandc} uses \lstinline{&mut i8} instead of \lstinline{&mut bool}, due to LLVM IR not supporting pointers to Booleans (\autoref{sec:results}).
    \end{itemize}

\subsubsection*{\Refinator's definition of interface correctness is more strict than CRUST-Bench's.}
    Specifically, CRUST-Bench's \bench{amp} interface uses reference types in places where \refinator's interface uses owned types because \refinator preserves copy semantics (\autoref{sec:semantic}). More specifically:
    \begin{itemize}
    \item The return value of \lstinline{decode_arg} has reference type \lstinline{&str} in CRUST-Bench's interface, but has owned type \lstinline{Rc<[Cell<u8>]>} in \refinator's interface.
    The C implementation of this function copies a string out of the \lstinline{self} reference parameter and returns that copy; since \refinator preserves copy semantics, the return type must be owned. Using a reference for the return value still admits a correct implementation, but that implementation borrows from \lstinline{self} rather than copying out of it, involving different copy semantics.
    \item Similarly, the \lstinline{buf} parameter of \lstinline{decode} has reference type \lstinline{&str} in CRUST-Bench's interface, but has owned type \lstinline{Rc<[Cell<u8>]>} in \refinator's interface.
    The C program assigns \lstinline{self.buf} to an alias of \lstinline{buf}; as \lstinline{self.buf} is a \lstinline{String} in the CRUST-Bench struct interface and \lstinline{buf} is a \lstinline{&str}, such an assignment would require copying instead of aliasing. For the same reason as for \lstinline{decode_arg}, a correct implementation of the CRUST-Bench interface is possible, but would have different copy semantics.
    \end{itemize}


\subsection{Summary}

The results show that \refinator achieves its goals of correctness and precision, modulo its current limitations. \Refinator's performance scales to C programs with thousands of lines of code, but more work is needed to scale to much larger programs.

\section{Conclusion}

\Refinator is a novel constraint-based static program analysis that chooses Rust types for an input C program that simultaneously satisfy Rust type constraints and preserve semantics. \Refinator generates correct, precise interfaces, as our evaluation demonstrates on moderately sized C programs. Scalability to larger programs, relaxing adherence to the C program's aliasing patterns, and supporting more C language features remain challenges for future work. This work advances the state of the art by demonstrating the feasibility of automatic C-to-Rust interface translation, which can help make C-to-Rust program translation modular, incremental, and tractable.

\begin{acks}
We thank the anonymous reviewers, Anirudh Khatry, and Zhao Fu for valuable feedback.
Vincent Beardsley, Chuyang Chen, Zhiqiang Lin, Melih Sirlanci, Chris Xiong, and Carter Yagemann influenced this project in its early stages.
This work is supported by NSF grants CSR-2106117 and SaTC-2348754.
\end{acks}

\section*{Data Availability Statement}

The following are publicly available:
\begin{itemize}
\item \refinator's source code and evaluated benchmarks.\footnote{\url{https://github.com/PLaSSticity/refinator-impl}}
\item An artifact reproducing the paper's results~\cite{refinator-artifact}.
\end{itemize}

\bibliographystyle{ACM-Reference-Format}
\bibliography{paper}

\iftoggle{includes-appendix}{
\appendix
\allowdisplaybreaks

\section{Definitions and Notation}
\label{appendix:notation}

This section defines definitions and notation used throughout our formulation of the constraints.

\subsection{Notational Conventions}
\label{appendix:notation-conventions}

We write most sets, including sets of strings described by a grammar and sets derived from an input program, in boldface font, e.g., \BASETYPE and $\TYPELAB_M$.
We write functions evaluated at constraint generation time in snake case, e.g., \borrowpath and \deeppaths, and write SMT functions and relations in Pascal case, e.g., \BaseType and \Array.
For adjustable parameters used during constraint generation, we use Fraktur script letters, e.g., $\mathfrak{B}_\text{generics}$ and $\mathfrak{F}_\text{heap}$.
For a sequence $\vec{x} \in X^n$ for some $n \in \NN$, we write $(\vec{x})_i$ to mean the $i$th element of $\vec{x}$.
In SMT formulas, relation symbols ($=$, $\leq$ $\in$, $\subseteq$, $\vartriangleright$, $\blacktriangleright$) have higher precedence than conjunction ($\land$) and disjunction ($\lor$), which have higher precedence than implication ($\Rightarrow$) and the biconditional ($\Leftrightarrow$), and $\Rightarrow$ is right-associative.

For a source language program $M \in \nt{Program}$, we define the following sets of unique identifiers (along with the metavariables we use):
\begin{enumerate}
    \item $\TYPELAB_M$ is the set of type labels ($\ell$) that appear in $M$.
    \item $\INSTRLAB_M$ is the set of identifiers for instructions ($L$) in $M$.
    \item $\STRUCT_M$ is the set of struct identifiers ($s$) in $M$.
    \item $\FUNCTION_M$ is the set of function identifiers ($F$) in $M$.
    \item $\VAR_M$ is the set of variable identifiers ($v$) in $M$.
\end{enumerate}
In our inference rules, we use the following judgments:
\begin{enumerate}
    \item $M \vdash_\textsf{instr} I \ @ \ L \in F$, read as: ``In program $M$, the instruction labeled $L$ in function $F$ is $I$.''
    \item $M \vdash_\textsf{src} \ell : \tau$, read as: ``In program $M$, type label $\ell$ has source type fragment $\tau$.''
    \item $M \vdash_\textsf{fld} s, f : \vec{\tau}$, read as: ``In program $M$, field $f$ of struct $s$ is defined with source type $\vec{\tau}$.''
    \item $M \vdash_\textsf{func} F : (\vec{\tau}_1, ..., \vec{\tau_n}) \rightarrow \vec{\tau}_\text{ret}$, read as: ``In program $M$, the function $F$ takes parameters of source types $\vec{\tau_1}, ..., \vec{\tau_n}$ and has return source type $\vec{\tau}_\text{ret}$.''
    \item $M \vdash_\textsf{var} v : \vec{\tau}$, read as: ``In program $M$, variable $v$ is declared with source type $\vec{\tau}$.''
    \item $M \vdash_\textsf{rval} \labeled{\ell}{$E$} : \vec{\tau} \ @ \ L \in F$, read as: ``In program $M$, the rvalue expression $E$ with type label $\ell$ and source type $\vec{\tau}$ appears in the instruction labeled $L$ in function $F$.''
    \item $M \vdash_\textsf{asn} \labeled{\ell_1}{$\tau_1$} \leftarrow \labeled{\ell_2}{$\tau_2$} \ @ \ \labeled{\ell_\text{rval}}{$E$}, i$, read as: ``In program $M$, rvalue expression \labeled{\ell_\text{rval}}{$E$}'s $i$th type fragment, \labeled{\ell_2}{$\tau_2$}, \emph{corresponds by assignment} to \labeled{\ell_1}{$\tau_1$}.'' We define correspondence by assignment in \autoref{appendix:notation-stypes}.
    \item $M \vdash_\textsf{cfg} L_1 \to L_2$, read as: ``In program $M$'s control-flow graph, $L_1$ is a direct predecessor of $L_2$.''
    \item $M \vdash_\textsf{live} v \ @ \ p$, read as: ``In program $M$, variable $v$ is live at program point $p$.''
    \item $M \vdash \boxed{\phi}$, read as: ``For program $M$, \refinator emits the SMT constraint $\phi$.''
    \item $M \vDash \phi$, read as: ``For program $M$, the SMT formula $\phi$ holds in \refinator's constraint system.''
    \item $M \vDash_\textsf{rs} \ell : T$, read as: ``In program $M$, \refinator chooses Rust type fragment $T$ for type label $\ell$.''
\end{enumerate}

\subsection{Source Types}
\label{appendix:notation-stypes}

The source type of a non-nested rvalue expression is defined in terms of the source type fragments of its type label and its path.
In general, the first source type fragment of a non-nested rvalue expression is the type fragment associated with the type label of the rvalue expression, and the remaining type fragments come from the source type of the path of the rvalue expression.
For example, in \autoref{fig:source-snippet}, the source type \labeled{18}{n} is $\langle \labeled{18}{\sty{ptr}}, \labeled{12}{\sty{struct}(Node)} \rangle$, where \sty{ptr} is the type fragment associated with the type label of the rvalue expression (18), and \texttt{\sty{struct}(Node)} comes from the source type of \texttt{n}.

We can describe the source types of non-nested rvalue expressions using the following rules:
{
\small
\begin{gather*}
    \irule{D-Rval-Load}{
        \begin{gathered}
            M \vdash_\textsf{instr} \texttt{\textunderscore \ = \labeled{\ell}{*$v$};} \ @ \ \texttt{\textunderscore}
            \quad M \vdash_\textsf{src} \ell : \tau
            \\ M \vdash_\textsf{var} v : \langle \labeled{\ell_1}{$\tau_1$}, \labeled{\ell_2}{$\tau_2$}, \labeled{\ell_3}{$\tau_3$} ..., \labeled{\ell_n}{$\tau_n$} \rangle
        \end{gathered}
    }{
        M \vdash_\textsf{rval} \labeled{\ell}{*$v$} : \langle \labeled{\ell}{$\tau$}, \labeled{\ell_3}{$\tau_3$}, ..., \labeled{\ell_n}{$\tau_n$} \rangle
    }{}
    \quad
    \irule{D-Rval-Store}{
        \begin{gathered}
            M \vdash_\textsf{instr} \texttt{\textunderscore \ = \labeled{\ell}{$e$};} \ @ \ \texttt{\textunderscore}
            \quad M \vdash_\textsf{src} \ell : \tau
            \\ M \vdash_\textsf{var} e : \langle \labeled{\ell_1}{$\tau_1$}, \labeled{\ell_2}{$\tau_2$}, ..., \labeled{\ell_n}{$\tau_n$} \rangle
        \end{gathered}
    }{
        M \vdash_\textsf{rval} \labeled{\ell}{$e$} : \langle \labeled{\ell}{$\tau$}, \labeled{\ell_2}{$\tau_2$}, ..., \labeled{\ell_n}{$\tau_n$} \rangle
    }{}
    \\[0.25em]
    \irule{D-Rval-Elt}{
        \begin{gathered}
            M \vdash_\textsf{instr} \texttt{\textunderscore \ = \labeled{\ell}{\&$v$[\textunderscore]};} \ @ \ \texttt{\textunderscore}
            \quad M \vdash_\textsf{src} \ell : \tau
            \\ M \vdash_\textsf{var} v : \langle \labeled{\ell_1}{$\tau_1$}, \labeled{\ell_2}{$\tau_2$} ..., \labeled{\ell_n}{$\tau_n$} \rangle
        \end{gathered}
    }{
        M \vdash_\textsf{rval} \labeled{\ell}{\&$v$[\textunderscore]} : \langle \labeled{\ell}{$\tau$}, \labeled{\ell_2}{$\tau_2$}, ..., \labeled{\ell_n}{$\tau_n$} \rangle
    }{}
    \quad
    \irule{D-Rval-Call}{
        \begin{gathered}
            M \vdash_\textsf{instr} \texttt{\textunderscore \ = $F$(\labeled{\ell'_1}{$e_1$}, ..., \labeled{\ell'_n}{$e_n$});} \ @ \ \texttt{\textunderscore}
            \\ M \vdash_\textsf{src} \ell : \tau
            \quad M \vdash_\textsf{var} e_i : \langle \labeled{\ell_1}{$\tau_1$}, \labeled{\ell_2}{$\tau_2$} ..., \labeled{\ell_n}{$\tau_n$} \rangle
        \end{gathered}
    }{
        M \vdash_\textsf{rval} \labeled{\ell}{$e_i$} : \langle \labeled{\ell}{$\tau$}, \labeled{\ell_2}{$\tau_2$}, ..., \labeled{\ell_n}{$\tau_n$} \rangle
    }{}
    \\[0.25em]
    \irule{D-Rval-Phi}{
        \begin{gathered}
            M \vdash_\textsf{instr} \texttt{\textunderscore \ = \kw{phi}(\labeled{\ell'_1}{$e_1$}: \textunderscore, $...$, \labeled{\ell'_n}{$e_n$},): \textunderscore;} \ @ \ \texttt{\textunderscore}
            \\ M \vdash_\textsf{src} \ell : \tau
            \quad M \vdash_\textsf{var} e_i : \langle \labeled{\ell_1}{$\tau_1$}, \labeled{\ell_2}{$\tau_2$}, ..., \labeled{\ell_n}{$\tau_n$} \rangle
        \end{gathered}
    }{
        M \vdash_\textsf{rval} \labeled{\ell}{$e_i$} : \langle \labeled{\ell}{$\tau$}, \labeled{\ell_2}{$\tau_2$}, ..., \labeled{\ell_n}{$\tau_n$} \rangle
    }{}
    \quad
    \irule{D-Rval-Use}{
        \begin{gathered}
            M \vdash_\textsf{instr} \texttt{\textunderscore \ = \kw{use}(\labeled{\ell'_1}{$v_1$}, $...$, \labeled{\ell'_n}{$v_n$},);} \ @ \ \texttt{\textunderscore}
            \\ M \vdash_\textsf{src} \ell : \tau
            \quad M \vdash_\textsf{var} v_i : \langle \labeled{\ell_1}{$\tau_1$}, \labeled{\ell_2}{$\tau_2$}, ..., \labeled{\ell_n}{$\tau_n$} \rangle
        \end{gathered}
    }{
        M \vdash_\textsf{rval} \labeled{\ell}{$v_i$} : \langle \labeled{\ell}{$\tau$}, \labeled{\ell_2}{$\tau_2$}, ..., \labeled{\ell_n}{$\tau_n$} \rangle
    }{}
    \\[0.25em]
    \irule{D-Rval-CondBr}{
        \begin{gathered}
            M \vdash_\textsf{instr} \texttt{\kw{if} (\labeled{\ell}{$e$}) \kw{goto} \textunderscore; \kw{else} \kw{goto} \textunderscore;} \ @ \ \texttt{\textunderscore}
            \\ M \vdash_\textsf{src} \ell : \tau
            \quad M \vdash_\textsf{var} e : \langle \labeled{\ell_1}{$\tau_1$}, \labeled{\ell_2}{$\tau_2$}, ..., \labeled{\ell_n}{$\tau_n$} \rangle
        \end{gathered}
    }{
        M \vdash_\textsf{rval} \labeled{\ell}{$e$} : \langle \labeled{\ell}{$\tau$}, \labeled{\ell_2}{$\tau_2$}, ..., \labeled{\ell_n}{$\tau_n$} \rangle
    }{}
    \quad
    \irule{D-Rval-Ret}{
        \begin{gathered}
            M \vdash_\textsf{instr} \texttt{\kw{return} \labeled{\ell}{$e$};} \ @ \ \texttt{\textunderscore}
            \quad M \vdash_\textsf{src} \ell : \tau
            \\ M \vdash_\textsf{var} v_i : \langle \labeled{\ell_1}{$\tau_1$}, \labeled{\ell_2}{$\tau_2$}, ..., \labeled{\ell_n}{$\tau_n$} \rangle
        \end{gathered}
    }{
        M \vdash_\textsf{rval} \labeled{\ell}{$e$} : \langle \labeled{\ell}{$\tau$}, \labeled{\ell_2}{$\tau_2$}, ..., \labeled{\ell_n}{$\tau_n$} \rangle
    }{}
    \\[0.25em]
    \irule{D-Rval-Fld}{
        \begin{gathered}
            M \vdash_\textsf{instr} \texttt{\textunderscore \ = \labeled{\ell}{\&*(*$v$).$f$};} \ @ \ \texttt{\textunderscore}
            \quad M \vdash_\textsf{src} \ell : \tau
            \quad M \vdash_\textsf{var} v : \langle \texttt{\textunderscore}, \labeled{\texttt{\textunderscore}}{\sty{struct}($s$)} \rangle
            \\ M \vdash_\textsf{fld} s, f : \langle \labeled{\ell_1}{$\tau_1$}, \labeled{\ell_2}{$\tau$}, ..., \labeled{\ell_n}{$\tau_n$} \rangle 
        \end{gathered}
    }{
        M \vdash_\textsf{rval} \labeled{\ell}{\&*(*$v$).$f$} : \langle \labeled{\ell}{$\tau$}, \labeled{\ell_2}{$\tau_2$}, ..., \labeled{\ell_n}{$\tau_n$} \rangle
    }{}
\end{gather*}
}%
Note that the omission of $\ell_2$ from the conclusion of [{\scshape D-Rval-Load}] is correct---it accounts for the dereference.

At assignments that do not cross function boundaries (i.e., all instructions in \autoref{fig:source-lang} that include the \texttt{=} symbol, except for \texttt{$\vec{\tau}$ $v$ = $F$(\textunderscore);}, plus \texttt{\kw{return} \labeled{\ell}{$e$};}), we say that pairs of type labels in the types of the source and destination \emph{correspond by assignment}.
We formalize this correspondence using the following rules:
{
\small
\begin{gather*}
    \irule{D-Cor-Gen}{
        \begin{gathered}
            M \vdash_\textsf{instr} \texttt{$\langle \labeled{\ell'_1}{$\tau'_1$}, ..., \labeled{\ell'_n}{$\tau'_n$} \rangle$ \texttt{\textunderscore} = \labeled{\ell}{$E$};} \ @ \ \texttt{\textunderscore}
            \\ M \vdash_\textsf{rval} \labeled{\ell}{$E$} : \langle \labeled{\ell_1}{$\tau_1$}, ..., \labeled{\ell_n}{$\tau_n$} \rangle \ @ \ \texttt{\textunderscore}
        \end{gathered}
    }{
        M \vdash_\textsf{asn} \labeled{\ell'_i}{$\tau'_i$} \leftarrow \labeled{\ell_i}{$\tau_i$} \ @ \ \labeled{\ell}{$E$}, i
    }{}
    \quad
    \irule{D-Cor-Store}{
            \begin{gathered}
            M \vdash_\textsf{instr} \texttt{*$v$ = \labeled{\ell}{$E$};} \ @ \ \texttt{\textunderscore}
            \quad M \vdash_\textsf{var} v : \langle \labeled{\ell'_1}{$\tau'_1$}, ..., \labeled{\ell'_n}{$\tau'_{n+1}$} \rangle
            \\ M \vdash_\textsf{rval} \labeled{\ell}{$E$} : \langle \labeled{\ell_1}{$\tau_1$}, ..., \labeled{\ell_n}{$\tau_n$} \rangle \ @ \ \texttt{\textunderscore}
        \end{gathered}
    }{
        M \vdash_\textsf{asn} \labeled{\ell'_{i+1}}{$\tau'_{i+1}$} \leftarrow \labeled{\ell_i}{$\tau_i$} \ @ \ \labeled{\ell}{$E$}, i
    }{}
    \\[0.25em]
    \irule{D-Cor-Phi}{
        \begin{gathered}
            M \vdash_\textsf{instr} \texttt{$\langle \labeled{\ell_1}{$\tau_1$}, ..., \labeled{\ell_n}{$\tau_n$} \rangle$ \textunderscore \ = \kw{phi}(\labeled{\ell''_1}{$e_1$}: \textunderscore, $...$, \labeled{\ell''_m}{$e_m$}: \textunderscore,);} \ @ \ \texttt{\textunderscore}
            \\ M \vdash_\textsf{rval} \labeled{\ell''_j}{$e_i$} : \langle \labeled{\ell'_1}{$\tau'_1$}, ..., \labeled{\ell'_n}{$\tau'_n$} \rangle \ @ \ \texttt{\textunderscore}
        \end{gathered}
    }{
        M \vdash_\textsf{asn} \labeled{\ell_i}{$\tau_i$} \leftarrow \labeled{\ell'_i}{$\tau'_i$} \ @ \ \labeled{\ell''_j}{$e_j$}, i
    }{}
    \\[0.25em]
    \irule{D-Cor-Ret}{
        M \vdash_\textsf{instr} \texttt{\kw{return} \labeled{\ell}{$e$};} \ @ \ \texttt{\textunderscore} \in F
        \quad M \vdash_\textsf{rval} \labeled{\ell}{$e$} : \langle \labeled{\ell_1}{$\tau_1$}, ..., \labeled{\ell_n}{$\tau_n$} \rangle \ @ \ \texttt{\textunderscore}
        \quad M \vdash_\textsf{func} F: \texttt{\textunderscore} \rightarrow \langle \labeled{\ell'_1}{$\tau'_1$}, ..., \labeled{\ell'_n}{$\tau'_n$} \rangle
    }{
        M \vdash_\textsf{asn} \labeled{\ell'_i}{$\tau'_i$} \leftarrow \labeled{\ell_i}{$\tau_i$} \ @ \ \labeled{\ell}{$e$}, i
    }{}
\end{gather*}
}

\subsection{Rust Types}
\label{appendix:notation-rust}

\Refinator constrains \RTYPEFRAG{}s by their components: base type (Rust type fragment ignoring qualifiers and lifetimes), array qualifiers, \rty{Cell} qualifiers, and lifetime variables.
The base types are enumerated in the \BASETYPE SMT datatype, defined below.
\begin{align*}
    \BASETYPE \ni \beta \ &{::=} \
             \rty{unknown}
        \mid \rty{void}
        \mid \rty{bool}
        \mid \rty{i8}
        \mid \rty{i16}
        \mid \rty{i32}
        \mid \rty{i64}
        \mid \rty{f32}
        \mid \rty{f64}
        \mid \texttt{\rty{struct}($s$)} \\
        &\mid \rty{ghost}
        \mid \rty{Box}
        \mid \rty{Rc}
        \mid \rty{shared}
        \mid \rty{mut}
\end{align*}
In the SMT constraints, each \LIFETIMEVAR is represented by a natural number, and a \LIFETIMEVAR may be used where a natural number is expected (e.g., as operands of $\leq$ or as $\NN$).
We write \LIFETIMEVAR instead of $\NN$ to clarify when a natural number should be interpreted as a lifetime variable.
\Refinator's constraints represent the \RTYPEFRAG of each $\TYPELAB_M$ in the following SMT functions and relations:
\begin{enumerate}
    \item $\BaseType: \TYPELAB_M \to \BASETYPE$ maps each type label to its base type (i.e., Rust type fragment ignoring qualifiers and lifetimes).
    \item $\Array \subseteq \TYPELAB_M$ is the set of type labels of Rust type fragments with array qualifiers.
    \item $\Cell \subseteq \TYPELAB_M$ is the set of type labels of Rust type fragments with \rty{Cell} qualifiers.
    \item $\RefLifetime: \TYPELAB_M \to \LIFETIMEVAR$ maps each type label of a reference Rust type fragment to its lifetime variable.
    \item $\StructLifetime: \TYPELAB_M \times \NN \to \LIFETIMEVAR$ maps each type label of a struct Rust type fragment and index $i$ to the $i$th lifetime variable of the Rust type fragment.
\end{enumerate}
\Refinator can recover the \RTYPEFRAG of a type label from the interpretations of the functions introduced above produced by the SMT solver according to the following rules:
{
\small
\begin{gather*}
    \irule{R-Primitive}{
        M \vDash \BaseType(\ell) = \beta
    }{
        M \vDash_\textsf{rs} \ell : \beta
    }{
        $\beta \in \{ \rty{void}, \rty{bool}, \rty{i8}, \rty{i16}, \rty{i32}, \rty{i64}, \rty{f32}, \rty{f64}, \rty{unknown} \}$
    }
    \\[0.25em]
    \irule{R-Struct}{
        \begin{gathered}
            M \vDash \BaseType(\ell) = \texttt{\rty{struct}($s$)}
            \quad M \vDash \StructGenerics(s) = n
            \\ M \vDash \StructLifetime(\ell, 1) = \rho_1 \; \cdots \; M \vDash \StructLifetime(\ell, n) = \rho_n
        \end{gathered}
    }{
        M \vDash_\textsf{rs} \ell : \texttt{\rty{struct}($s$, $\langle \rho_1, ..., \rho_n \rangle$)}
    }{}
    \\[0.25em]
    \irule{R-Ptr-Bare}{
        \begin{gathered}
            M \vDash \BaseType(\ell) = \beta
            \\ M \vDash \ell \notin \Array
            \quad M \vDash \ell \notin \Cell
        \end{gathered}
    }{
        M \vDash_\textsf{rs} \ell : \beta
    }{
        $\beta \in \{ \rty{ghost}, \rty{Box}, \rty{Rc}\}$
    }
    \quad
    \irule{R-Ptr-Slice}{
        \begin{gathered}
            M \vDash \BaseType(\ell) = \beta
            \\ M \vDash \ell \in \Array
            \quad M \vDash \ell \notin \Cell
        \end{gathered}
    }{
        M \vDash_\textsf{rs} \ell : \beta^\texttt{[]}
    }{
        $\beta \in \{ \rty{ghost}, \rty{Box}, \rty{Rc}\}$
    }
    \\[0.25em]
    \irule{R-Ptr-Cell}{
        \begin{gathered}
            M \vDash \BaseType(\ell) = \beta
            \\ M \vDash \ell \notin \Array
            \quad M \vDash \ell \in \Cell
        \end{gathered}
    }{
        M \vDash_\textsf{rs} \ell : \beta^\rty{Cell}
    }{
        $\beta \in \{ \rty{ghost}, \rty{Box}, \rty{Rc}\}$
    }
    \quad
    \irule{R-Ptr-SliceCell}{
        \begin{gathered}
            M \vDash \BaseType(\ell) = \beta
            \\ M \vDash \ell \in \Array
            \quad M \vDash \ell \in \Cell
        \end{gathered}
    }{
        M \vDash_\textsf{rs} \ell : \beta^\texttt{[\rty{Cell}]}
    }{
        $\beta \in \{ \rty{ghost}, \rty{Box}, \rty{Rc}\}$
    }
    \\[0.25em]
    \irule{R-Ref-Bare}{
        \begin{gathered}
            M \vDash \BaseType(\ell) = \beta
            \\ M \vDash \RefLifetime(\ell) = \rho
            \\ M \vDash \ell \notin \Array
            \quad M \vDash \ell \notin \Cell
        \end{gathered}
    }{
        M \vDash_\textsf{rs} \ell : \texttt{$\beta$($\rho$)}
    }{
        $\beta \in \{ \rty{shared}, \rty{mut} \}$
    }
    \quad
    \irule{R-Ref-Slice}{
        \begin{gathered}
            M \vDash \BaseType(\ell) = \beta
            \\ M \vDash \RefLifetime(\ell) = \rho
            \\ M \vDash \ell \in \Array
            \quad M \vDash \ell \notin \Cell
        \end{gathered}
    }{
        M \vDash_\textsf{rs} \ell : \texttt{$\beta^\texttt{[]}$($\rho$)}
    }{
        $\beta \in \{ \rty{shared}, \rty{mut} \}$
    }
    \\[0.25em]
    \irule{R-Ref-Cell}{
        \begin{gathered}
            M \vDash \BaseType(\ell) = \beta
            \\ M \vDash \RefLifetime(\ell) = \rho
            \\ M \vDash \ell \notin \Array
            \quad M \vDash \ell \in \Cell
        \end{gathered}
    }{
        M \vDash_\textsf{rs} \ell : \texttt{$\beta^\rty{Cell}$($\rho$)}
    }{
        $\beta \in \{ \rty{shared}, \rty{mut} \}$
    }
    \quad
    \irule{R-Ref-SliceCell}{
        \begin{gathered}
            M \vDash \BaseType(\ell) = \beta
            \\ M \vDash \RefLifetime(\ell) = \rho
            \\ M \vDash \ell \in \Array
            \quad M \vDash \ell \in \Cell
        \end{gathered}
    }{
        M \vDash_\textsf{rs} \ell : \texttt{$\beta^\texttt{[\rty{Cell}]}$($\rho$)}
    }{
        $\beta \in \{ \rty{shared}, \rty{mut} \}$
    }
\end{gather*}
}

\Refinator emits constraints involving the lifetime variables in the Rust type of $\labeled{\ell}{\&*(*$v$.$f$)}$ rvalue expressions.
The lifetime variables in the types of these expressions depend on the lifetime variables in both the type of $v$ and the definition of the struct field $f$ of $s$, where $v$ is a pointer to an $s$ struct.
So, we define the following SMT functions to resolve the correct lifetime variables:
\begin{enumerate}
    \item $\refbind: \TYPELAB_M \times \NN \to \LIFETIMEVAR$ maps each type label of a non-nested rvalue expression and an index $i$, where the $i$th type fragment of the rvalue expression's Rust type is a reference type, to the type fragment's lifetime variable.
    \item $\structbind: \TYPELAB_M \times \NN \times \NN \to \LIFETIMEVAR$ maps each type label of a non-nested rvalue expression and an index $i$, where the $i$th type fragment of the rvalue expression's Rust type is a struct type, and an index $j$, to the type fragment's $j$th lifetime variable.
\end{enumerate}
We define $\refbind$ and $\structbind$ according to the following rules:
{
\small
\begin{gather*}
    \irule{D-RefBind-Gen}{
        M \vdash_\textsf{rval} \labeled{\ell}{$E$} : \vec{\tau} \ @ \ \texttt{\textunderscore}
        \quad (\vec{\tau})_i = \labeled{\ell'}{\sty{ptr}}
    }{
        \refbind(\ell, i) = \RefLifetime(\ell')
    }{$E \neq \texttt{\&*(*$v$).$f$} \lor i = 1$}
    \quad
    \irule{D-StructBind-Gen}{
        M \vdash_\textsf{rval} \labeled{\ell}{$E$} : \vec{\tau} \ @ \ \texttt{\textunderscore}
        \quad (\vec{\tau})_i = \labeled{\ell'}{\sty{struct}(\textunderscore)}
    }{
        \structbind(\ell, i, j) = \StructLifetime(\ell', j)
    }{}
    \\[0.25em]
    \irule{D-RefBind-Fld}{
        M \vdash_\textsf{rval} \labeled{\ell}{\&*(*$v$).$f$} : \vec{\tau} \ @ \ \texttt{\textunderscore}
        \quad (\vec{\tau})_i = \labeled{\ell'}{\sty{ptr}}
        \quad M \vdash_\textsf{var} v : \vec{\tau}' \ @ \ \texttt{\textunderscore}
        \quad (\vec{\tau}')_2 = \labeled{\ell''}{\textunderscore}
    }{
        \refbind(\ell, i) = \StructLifetime(\ell'', \RefLifetime(\ell'))
    }{$i > 1$}
    \\[0.25em]
    \irule{D-StructBind-Fld}{
        M \vdash_\textsf{rval} \labeled{\ell}{\&*(*$v$).$f$} : \vec{\tau} \ @ \ \texttt{\textunderscore}
        \quad (\vec{\tau})_i = \labeled{\ell'}{\sty{struct}(\textunderscore)}
        \quad M \vdash_\textsf{var} v : \vec{\tau}' \ @ \ \texttt{\textunderscore}
        \quad (\vec{\tau}')_2 = \labeled{\ell''}{\textunderscore}
    }{
        \structbind(\ell, i, j) = \StructLifetime(\ell'', \StructLifetime(\ell', j))
    }{}
\end{gather*}
}

\section{Type Consistency Constraints}
\label{appendix:consistency}

Here, we show the constraints \refinator generates for type consistency introduced in \autoref{sec:consistency}.

\subsection{Type Parity}
\label{appendix:consistency-parity}

\Refinator generates the following constraints for \autoref{sec:consistency-parity}:
{
\small
\begin{gather*}
    \irule{C-Parity-Void}{
        M \vdash_\textsf{src} \ell : \sty{void}
    }{
        M \vdash \boxed{\BaseType(\ell) = \rty{void}}
    }{}
    \quad
    \irule{C-Parity-Bool}{
        M \vdash_\textsf{src} \ell : \sty{bool}
    }{
        \boxed{\BaseType(\ell) = \rty{bool}}
    }{}
    \quad
    \irule{C-Parity-Int8}{
        M \vdash_\textsf{src} \ell : \sty{i8}
    }{
        M \vdash \boxed{\BaseType(\ell) = \rty{i8}}
    }{}
    \\[0.25em]
    \irule{C-Parity-Int16}{
        M \vdash_\textsf{src} \ell : \sty{i16}
    }{
        M \vdash \boxed{\BaseType(\ell) = \rty{i16}}
    }{}
    \quad
    \irule{C-Parity-Int32}{
        M \vdash_\textsf{src} \ell : \sty{i32}
    }{
        M \vdash \boxed{\BaseType(\ell) = \rty{i32}}
    }{}
    \quad
    \irule{C-Parity-Int64}{
        M \vdash_\textsf{src} \ell : \sty{i64}
    }{
        M \vdash \boxed{\BaseType(\ell) = \rty{i64}}
    }{}
    \\[0.25em]
    \irule{C-Parity-Float32}{
        M \vdash_\textsf{src} \ell : \sty{f32}
    }{
        M \vdash \boxed{\BaseType(\ell) = \rty{f32}}
    }{}
    \quad
    \irule{C-Parity-Float64}{
        M \vdash_\textsf{src} \ell : \sty{f64}
    }{
        M \vdash \boxed{\BaseType(\ell) = \rty{f64}}
    }{}
    \quad
    \irule{C-Parity-Struct}{
        M \vdash_\textsf{src} \ell : \texttt{\sty{struct}($s$)}
    }{
        M \vdash \boxed{\BaseType(\ell) = \texttt{\rty{struct}($s$)}}
    }{}
    \\[0.25em]
    \irule{C-Parity-Ptr}{
        M \vdash_\textsf{src} \ell : \sty{ptr}
    }{
        M \vdash \boxed{\BaseType(\ell) \in \{ \rty{ghost}, \rty{Box}, \rty{Rc}, \rty{shared}, \rty{mut} \}}
    }{}
    \quad
    \irule{C-Parity-Unknown}{
        M \vdash_\textsf{src} \ell : \sty{unknown}
    }{
        M \vdash \boxed{\BaseType(\ell) = \rty{unknown}}
    }{}
\end{gather*}
}

\subsection{Struct Consistency}
\label{appendix:consistency-struct}

This section describes the constraints for \autoref{sec:consistency-struct}.
\Refinator represents the number of generic lifetimes in a struct definition using the following SMT function, which maps each struct identifier to the number of generic lifetimes in its definition.
\begin{gather*}
    \StructGenerics: \STRUCT_M \to \NN
\end{gather*}
\Refinator generates the following constraint to ensure that the number of generic lifetimes for each struct definition is in the interval $[0, \mathfrak{B}_\text{generics}]$, where $\mathfrak{B}_\text{generics}$ is the desired maximum number of generic lifetimes for any struct definition, chosen by the user of \refinator before constraint generation:
{
\small
\begin{gather*}
    \irule{C-Struct-Bound}{}{
        M \vdash \boxed{0 \leq \StructGenerics(s) \leq \mathfrak{B}_\text{generics}}
    }{$s \in \STRUCT_M$}
\end{gather*}
}%
\Refinator generates the following constraints to ensure that each lifetime variable used in a Rust type fragment inside the definition of struct $s$ corresponds to one of the struct's generic lifetimes, i.e., it is in the interval $[1, \StructGenerics(s)]$:
{
\small
\begin{gather*}
    \irule{C-LifeParam-Ptr}{
        M \vdash_\textsf{fld} s, f : \vec{\tau}
        \quad (\vec{\tau})_i = \labeled{\ell}{\sty{ptr}}
    }{
        M \vdash \boxed{1 \leq \RefLifetime(\ell) \leq \StructGenerics(s)}
    }{}
    \\[0.25em]
    \irule{C-LifeParam-Struct}{
        M \vdash_\textsf{fld} s, f : \vec{\tau}
        \quad (\vec{\tau})_i = \labeled{\ell}{\sty{struct}(\textunderscore)}
    }{
      M \vdash \boxed{1 \leq \StructLifetime(\ell, j) \leq \StructGenerics(s)}  
    }{$1 \leq j \leq \mathfrak{B}_\text{generics}$}
\end{gather*}
}

\subsection{Recursive Struct Definitions}
\label{appendix:consistency-recursive}

This section shows the constraints that \autoref{sec:consistency-recursive} introduced.
\Refinator represents the rank of each struct using the SMT function
\begin{gather*}
    \StructRank: \STRUCT_M \to \NN,
\end{gather*}
and generates the following constraint:
{
\small
\begin{gather*}
    \irule{C-Rank}{
        M \vdash_\textsf{fld} s, f : \langle \labeled{\ell_1}{\sty{ptr}}, ..., \labeled{\ell_{n-1}}{\sty{ptr}}, \labeled{\ell_n}{\sty{struct}($s'$)} \rangle
    }{
        M \vdash \boxed{
            \textstyle
            \left(\bigwedge_{i=1}^{n-1}{\BaseType(\ell_i) = \rty{ghost}}\right) \Rightarrow \StructRank(s') < \StructRank(s)
        }
    }{}
\end{gather*}
}

\subsection{Array Consistency}
\label{appendix:consistency-array}

This section shows the constraints that \autoref{sec:consistency-array} introduced.
\Refinator generates the following constraints:
{
\small
\begin{gather*}
    \irule{C-Array}{
        M \vdash_\textsf{rval} \labeled{\ell}{\&$v$[$e$]} \ @ \ \texttt{\textunderscore}
    }{
        M \vdash \boxed{\ell \in \Array}
    }{}
    \quad
    \irule{C-Slice}{
        M \vdash_\textsf{src} \ell : \sty{ptr}
    }{
        M \vdash \boxed{\ell \in \Array \Rightarrow \BaseType(\ell) \neq \rty{ghost}}
    }{}
\end{gather*}
}%
\textsc{C-Array} is more strict than the first constraint described in \autoref{sec:consistency-array}.
Due to the pointer transformation constraints, if the outermost type fragment of \labeled{\ell}{\&$v$[$e$]} has an array qualifier, then the outermost type fragment of $v$ also has an array qualifier.

\subsection{Type Equivalence}
\label{appendix:consistency-equiv}

This section shows the constraints that \autoref{sec:consistency-equiv} introduced.
\Refinator generates the following constraints:
{
\small
\begin{gather*}
    \irule{C-Equiv-Cor}{
        M \vdash_\textsf{asn} \labeled{\ell_1}{\textunderscore} \leftarrow \labeled{\ell_2}{\textunderscore} \ @ \ \texttt{\textunderscore}
    }{
        M \vdash \boxed{ \BaseType(\ell_1) = \BaseType(\ell_2) \land (\ell_1 \in \Array \Leftrightarrow \ell_2 \in \Array) \land (\ell_1 \in \Cell \Leftrightarrow \ell_2 \in \Cell) }
    }{}
    \\[0.25em]
    \irule{C-Equiv-Args}{
        \begin{gathered}
            M \vdash_\textsf{instr} \texttt{\textunderscore \ = $F$(\labeled{\ell''_1}{$e_1$}, ..., \labeled{\ell''_n}{$e_n$});} \ @ \ \texttt{\textunderscore}
            \quad M \vdash_\textsf{func} F : (\vec{\tau}_1, ..., \vec{\tau}_n) \rightarrow \texttt{\textunderscore}
            \\ M \vdash_\textsf{rval} \labeled{\ell''_j}{$e_j$} : \langle \labeled{\ell_1}{\textunderscore}, ..., \labeled{\ell_m}{\textunderscore} \rangle \ @ \ \texttt{\textunderscore}
            \quad \vec{\tau}_j = \langle \labeled{\ell'_1}{\textunderscore}, ..., \labeled{\ell'_m}{\textunderscore} \rangle
        \end{gathered}
    }{
        M \vdash \boxed{\BaseType(\ell_i) = \BaseType(\ell'_i) \land (\ell_i \in \Array \Leftrightarrow \ell'_i \in \Array) \land (\ell_i \in \Cell \Leftrightarrow \ell'_i \in \Cell)}
    }{}
    \\[0.25em]
    \irule{C-Equiv-Ret}{
        M \vdash_\textsf{instr} \texttt{$\langle \labeled{\ell_1}{\textunderscore}, ..., \labeled{\ell_n}{\textunderscore} \rangle$ \textunderscore \ = $F$(\textunderscore);} \ @ \ \texttt{\textunderscore}
        \quad M \vdash_\textsf{func} F : \texttt{\textunderscore} \rightarrow \langle \labeled{\ell'_1}{\textunderscore}, ..., \labeled{\ell'_n}{\textunderscore} \rangle
    }{
        M \vdash \boxed{\BaseType(\ell_i) = \BaseType(\ell'_i) \land (\ell_i \in \Array \Leftrightarrow \ell'_i \in \Array) \land (\ell_i \in \Cell \Leftrightarrow \ell'_i \in \Cell)}
    }{}
\end{gather*}
}%

\subsection{Pointer Transformations}
\label{appendix:consistency-pointer}

This section shows the constraints that \autoref{sec:consistency-pointer} introduced.
We write $\RustType(\ell_1) \vartriangleright \RustType(\ell_2)$ to mean that the Rust type fragment for type label $\ell_1$ is transformable to the Rust type fragment for type label $\ell_2$, as in \autoref{fig:transform-pointers}.
We can write $\RustType(\ell_1) \vartriangleright \RustType(\ell_2)$ as an SMT formula by noting that $\RustType(\ell_1) \vartriangleright \RustType(\ell_2)$ if there exists a transformation in \autoref{fig:transform-pointers} from $T_1$ to $T_2$ such that using the inference rules in \autoref{appendix:notation-rust}, (1) the SMT formulas in the premises and conditions to infer $M \vDash_\textsf{rs} \ell_1 : T_1$ hold, and (2) the SMT formulas in the premises and conditions to infer $M \vDash_\textsf{rs} \ell_2: T_2$ hold.
\Refinator generates the following constraints:
{
\small
\begin{gather*}
    \irule{C-Transform-Var}{
        M \vdash_\textsf{rval} \labeled{\ell}{$v$} : \texttt{\textunderscore} \ @ \ \texttt{\textunderscore}
        \quad
        M \vdash_\textsf{var} v : \langle \labeled{\ell'}{\sty{ptr}}, ...  \rangle
    }{
        M \vdash \boxed{\RustType(\ell') \vartriangleright \RustType(\ell)}
    }{}
    \quad
    \irule{C-Transform-Deref}{
        M \vdash_\textsf{rval} \labeled{\ell}{*$v$} : \texttt{\textunderscore} \ @ \ \texttt{\textunderscore}
        \quad
        M \vdash_\textsf{var} v : \langle \labeled{\texttt{\textunderscore}}{\textunderscore}, \labeled{\ell'}{\sty{ptr}}, ...  \rangle
    }{
        M \vdash \boxed{\RustType(\ell') \vartriangleright \RustType(\ell)}
    }{}
    \\[0.25em]
    \irule{C-Transform-Fld}{
        M \vdash_\textsf{rval} \labeled{\ell}{\&*(*$v$).$f$} : \texttt{\textunderscore} \ @ \ \texttt{\textunderscore}
        \quad M \vdash_\textsf{var} v : \langle \texttt{\textunderscore}, \labeled{\texttt{\textunderscore}}{\sty{struct}($s$)} \rangle
        \quad M \vdash_\textsf{fld} s, f : \langle \labeled{\ell'}{\texttt{\textunderscore}}, ... \rangle
    }{
        M \vdash \boxed{\RustType(\ell') \vartriangleright \RustType(\ell)}
    }{}
    \\[0.25em]
    \irule{C-Transform-Elt}{
        M \vdash_\textsf{rval} \labeled{\ell}{\&$v$[\textunderscore]} : \texttt{\textunderscore} \ @ \ \texttt{\textunderscore}
        \quad
        M \vdash_\textsf{var} v : \langle \labeled{\ell'}{\textunderscore}, ... \rangle
    }{
        M \vdash \boxed{\RustType(\ell') \vartriangleright \RustType(\ell)}
    }{}
\end{gather*}
}

\subsection{Immutability Preservation}
\label{appendix:consistency-immutable}

This section shows the constraints that \autoref{sec:consistency-immutable} introduced.
\Refinator generates the following constraints:
{
\small
\begin{gather*}
    \irule{C-Immut-Deref}{
        M \vdash_\textsf{rval} \labeled{\ell}{*$v$} : \texttt{\textunderscore} \ @ \ \texttt{\textunderscore}
        \quad
        M \vdash_\textsf{var} v : \langle \labeled{\ell'}{\sty{ptr}}, ...  \rangle
    }{
        M \vdash \boxed{\BaseType(\ell') \in \{ \rty{shared}, \rty{Rc} \} \land \ell' \notin \Cell \Rightarrow \BaseType(\ell) \neq \rty{mut} }
    }{}
    \\[0.25em]
    \irule{C-Immut-Fld}{
        M \vdash_\textsf{rval} \labeled{\ell}{\&*(*$v$).$f$} : \texttt{\textunderscore} \ @ \ \texttt{\textunderscore}
        \quad
        M \vdash_\textsf{var} v : \langle \labeled{\ell'}{\sty{ptr}}, ...  \rangle
    }{
        M \vdash \boxed{\BaseType(\ell') \in \{ \rty{shared}, \rty{Rc} \} \land \ell' \notin \Cell \Rightarrow \BaseType(\ell) \neq \rty{mut} }
    }{}
\end{gather*}
}

\subsection{Mutability}
\label{appendix:consistency-mutability}

This section shows the constraint that \autoref{sec:consistency-mutability} introduced.
\Refinator generates the following constraint:
{
\small
\begin{gather*}
    \irule{C-Mut}{
        M \vdash_\textsf{instr} \texttt{*$v$ = \textunderscore;} \ @ \ \texttt{\textunderscore}
        \quad M \vdash_\textsf{var} v : \langle \labeled{\ell}{\textunderscore}, ... \rangle
    }{
        M \vdash \boxed{\BaseType(\ell) \in \{ \rty{ghost}, \rty{Box}, \rty{mut} \} \lor \ell \in \Cell}
    }{}
\end{gather*}
}

\subsection{Ownership}
\label{appendix:consistency-ownership}

This section shows the constraints that \autoref{sec:consistency-ownership} introduced.
\Refinator generates the following constraints:
{
\small
\begin{gather*}
    \irule{C-Own-Stack}{
        M \vdash_\textsf{instr} \texttt{$\langle \labeled{\ell}{\textunderscore}, ... \rangle$ \textunderscore \ = \kw{alloca}(\textunderscore);} \ @ \ \texttt{\textunderscore}
    }{
        M \vdash \boxed{\BaseType(\ell) \in \{ \rty{ghost}, \rty{Box}, \rty{Rc} \}}
    }{}
    \qquad
    \irule{C-Own-Heap}{
        M \vdash_\textsf{instr} \texttt{$\langle \labeled{\ell}{\textunderscore}, ... \rangle$ \textunderscore \ = F(\textunderscore);} \ @ \ \texttt{\textunderscore}
    }{
        M \vdash \boxed{\BaseType(\ell) \in \{ \rty{ghost}, \rty{Box}, \rty{Rc} \}}
    }{$F \in \mathfrak{F}_\text{heap}$}
\end{gather*}
}%
By default, \refinator uses $\mathfrak{F}_\text{heap} = \{ \texttt{malloc}, \texttt{calloc}, \texttt{valloc}, \texttt{realloc}, \texttt{aligned\_alloc} \}$.

\subsection{External Functions}
\label{appendix:consistency-extenral}

This section shows the constraints that \autoref{sec:consistency-external} introduced.
We specify additional constraints on the signatures of external functions using $\mathfrak{F}_\text{external} \subseteq \nt{SigSpec}$, where \nt{SigSpec} is defined as follows.
\begin{align*}
    \nt{SigSpec} \ni \sigma \ {::=} \ F : (\{\vec{\gamma},\}^\ast) \to \vec{\gamma} \quad
    \nt{TypeSpec} \ni \vec{\gamma} \ {::=} \ \langle \{\gamma,\}^+ \rangle \quad
    \nt{BaseSpec} \ni \gamma \ {::=} \ \beta \mid \rty{any}
\end{align*}
\Refinator generates the following constraint:
{
\small
\begin{gather*}
    \irule{C-External}{
        M \vdash_\textsf{func} F : (\vec{\tau}_1, ..., \vec{\tau}_n) \rightarrow \vec{\tau}_0
        \quad (\vec{\gamma}_1, ..., \vec{\gamma}_m) \to \vec{\gamma}_0 \in \mathfrak{F}_\text{external}
        \quad (\vec{\tau}_i)_j = \labeled{\ell}{\textunderscore}
    }{
        M \vdash \boxed{\BaseType(\ell) = (\vec{\gamma}_i)_j}
    }{$(\vec{\gamma}_i)_j \neq \rty{any}$}
\end{gather*}
}%
By default, \refinator puts the following entries in $\mathfrak{F}_\text{external}$, where \texttt{*} is a wildcard.
Note that any parameters not present in a \nt{SigSpec} are left unconstrained.
\begin{align*}
    \texttt{llvm.memcpy.*}&: (\langle \rty{mut}, \rangle, ) \to \langle \rty{mut}, \rangle \\
    \texttt{llvm.memmove.*}&: (\langle \rty{mut}, \rangle, ) \to \langle \rty{mut}, \rangle \\
    \texttt{llvm.memset.*}&: (\langle \rty{mut}, \rangle, ) \to \langle \rty{mut}, \rangle \\
\end{align*}

\section{Semantic Preservation Constraints}
\label{appendix:semantic}

Here, we show the constraints \refinator generates for semantic preservation introduced in \autoref{sec:semantic}.

\subsection{Referent Preservation}
\label{appendix:semantic-referent}

This section shows the constraints that \autoref{sec:semantic-referent} introduced.
Similar to \autoref{appendix:consistency-pointer}, we write $\RustType(\ell_1) \blacktriangleright \RustType(\ell_2)$ to mean that the Rust type fragment for type label $\ell_1$ is nondestructively transformable to the Rust type fragment for type label $\ell_2$, as in \autoref{fig:transform-pointers}.
\Refinator generates the following constraints:
{
\small
\begin{gather*}
    \irule{S-Referent-Deref}{
        M \vdash_\textsf{rval} \labeled{\ell}{*$v$} : \texttt{\textunderscore} \ @ \ \texttt{\textunderscore}
        \quad M \vdash_\textsf{var} v : \langle \labeled{\ell'}{\textunderscore}, \labeled{\ell''}{\sty{ptr}}, ... \rangle
    }{
        M \vdash \boxed{\BaseType(\ell') \notin \{ \rty{ghost}, \rty{Box} \} \Rightarrow \RustType(\ell'') \blacktriangleright \RustType(\ell) }
    }{}
    \\[0.25em]
    \irule{S-Referent-Fld}{
        M \vdash_\textsf{rval} \labeled{\ell}{\&*(*$v$).$f$} : \texttt{\textunderscore} \ @ \texttt{\textunderscore}
        \quad M \vdash_\textsf{var} v : \langle \labeled{\ell'}{\textunderscore}, \labeled{\texttt{\textunderscore}}{\sty{struct}($s$)}, ... \rangle
        \quad M \vdash_\textsf{fld} s, f : \langle \labeled{\ell''}{\texttt{\textunderscore}}, ... \rangle
    }{
        M \vdash \boxed{\BaseType(\ell') \notin \{ \rty{ghost}, \rty{Box} \} \Rightarrow \RustType(\ell'') \blacktriangleright \RustType(\ell) }
    }{}
\end{gather*}
}%

\subsection{Nondestructive Pointer Transformations}
\label{appendix:semantic-pointer}

This section shows the constraints that \autoref{sec:semantic-pointer} introduced.
\refinator generates the following constraints:
{
\small
\begin{gather*}
    \irule{S-Transform-Var}{
        \begin{gathered}
            M \vdash_\textsf{rval} \labeled{\ell}{$v$} : \texttt{\textunderscore} \ @ \ L \in \texttt{\textunderscore}
            \\
            M \vdash_\textsf{var} v : \langle \labeled{\ell'}{\sty{ptr}}, ...  \rangle
            \quad M \vdash_\textsf{live} v \ @ \ (L)_\text{out}
        \end{gathered}
    }{
        M \vdash \boxed{\RustType(\ell') \blacktriangleright \RustType(\ell)}
    }{}
    \quad
    \irule{S-Transform-Deref}{
        \begin{gathered}
            M \vdash_\textsf{rval} \labeled{\ell}{*$v$} : \texttt{\textunderscore} \ @ \ L \in \texttt{\textunderscore}
            \\
            M \vdash_\textsf{var} v : \langle \labeled{\texttt{\textunderscore}}{\textunderscore}, \labeled{\ell'}{\sty{ptr}}, ...  \rangle
            \quad M \vdash_\textsf{live} v \ @ \ (L)_\text{out}
        \end{gathered}
    }{
        M \vdash \boxed{\RustType(\ell') \blacktriangleright \RustType(\ell)}
    }{}
    \\[0.25em]
    \irule{S-Transform-Fld}{
        \begin{gathered}
            M \vdash_\textsf{rval} \labeled{\ell}{\&*(*$v$).$f$} : \texttt{\textunderscore} \ @ \ L \in \texttt{\textunderscore}
            \quad M \vdash_\textsf{var} v : \langle \texttt{\textunderscore}, \labeled{\texttt{\textunderscore}}{\sty{struct}($s$)} \rangle
            \quad M \vdash_\textsf{fld} s, f : \langle \labeled{\ell'}{\texttt{\textunderscore}}, ... \rangle
            \quad M \vdash_\textsf{live} v \ @ \ (L)_\text{out}
        \end{gathered}
    }{
        M \vdash \boxed{\RustType(\ell') \blacktriangleright \RustType(\ell)}
    }{}
    \\[0.25em]
    \irule{S-Transform-Elt}{
        M \vdash_\textsf{rval} \labeled{\ell}{\&$v$[\textunderscore]} : \texttt{\textunderscore} \ @ \ L \in \texttt{\textunderscore}
        \quad
        M \vdash_\textsf{var} v : \langle \labeled{\ell'}{\textunderscore}, ... \rangle
        \quad M \vdash_\textsf{live} v \ @ \ (L)_\text{out}
    }{
        M \vdash \boxed{\RustType(\ell') \blacktriangleright \RustType(\ell)}
    }{}
\end{gather*}
}

\section{Borrow-Checking Constraints}
\label{appendix:borrowing}

This section presents the formal constraints generated by \refinator for borrow-checking rules, introduced in \autoref{sec:borrowing}.

\subsection{Lifetimes}
\label{appendix:borrowing-lifetimes}

Here, we show the constraints that \autoref{sec:borrowing-lifetimes} introduced.
\Refinator's constraints represent the lifetime of each lifetime inference variable in the following SMT functions:
\begin{enumerate}
    \item $\Lifetime: \LIFETIMEVAR \to \mathcal{P}(\POINT_M)$ maps each lifetime inference variable to the set of program points contained by its lifetime.
    \item $\LifetimeEnd: \LIFETIMEVAR \to \mathcal{P}(\LIFETIMEVAR)$ maps each lifetime inference variable to the set of function lifetime parameters it outlives.\footnote{In the NLL RFC~\cite{nll-rfc}, these are called ``end regions.''}
\end{enumerate}

\subsubsection{Liveness}

\Refinator generates the following constraints to encode the liveness rules:
{
\small
\begin{gather*}
    \irule{L-Live-Ptr}{
        M \vdash_\textsf{var} v : \vec{\tau}
        \quad (\vec{\tau})_i : \labeled{\ell}{\sty{ptr}}
        \quad M \vdash_\textsf{live} v \ @ \ p
    }{
        M \vdash \boxed{\BaseType(\ell) \in \{ \rty{shared}, \rty{mut} \} \Rightarrow p \in \Lifetime(\RefLifetime(\ell))}
    }{}
    \\[0.25em]
    \irule{L-Live-Struct}{
        M \vdash_\textsf{var} v : \vec{\tau}
        \quad (\vec{\tau})_i : \labeled{\ell}{\sty{struct}($s$)}
        \quad M \vdash_\textsf{live} v \ @ \ p
    }{
        M \vdash \boxed{i \leq \StructGenerics(s) \Rightarrow p \in \Lifetime(\StructLifetime(\ell, i))}
    }{$1 \leq i \leq \mathfrak{B}_\text{generics}$}
\end{gather*}
}

\subsubsection{Subtyping}
\label{appendix:subtyping}

\Refinator generates the following constraints to encode the subtyping rules.
The [\textsc{L-Covariant-$\ast$}] rules enforce that the source outlives the destination in an assignment, and the [\textsc{L-Invariant-$\ast$}] rules enforce that the destination outlives the source in an assignment when required by mutable references or interior mutable types.
{
\small
\begin{gather*}
    \irule{L-Covariant-Ptr}{
        M \vdash_\textsf{asn} \labeled{\ell_1}{\sty{ptr}} \leftarrow \labeled{\ell_2}{\textunderscore} \ @ \ \labeled{\ell}{$E$}, i
    }{
        M \vdash \boxed{\textstyle\{ \BaseType(\ell_1), \BaseType(\ell_2) \} \subseteq \{ \rty{shared}, \rty{mut} \} \Rightarrow \Lifetime(\refbind(\ell, i)) \supseteq \Lifetime(\RefLifetime(\ell_1))}
    }{}
    \\[0.25em]
    \irule{L-Covariant-Struct}{
        M \vdash_\textsf{asn} \labeled{\ell_1}{\sty{struct}($s$)} \leftarrow \labeled{\ell_2}{\textunderscore} \ @ \ \labeled{\ell}{$E$}, i
    }{
        M \vdash \boxed{j \leq \StructGenerics(s) \Rightarrow \Lifetime(\structbind(\ell, i, j)) \supseteq \Lifetime(\StructLifetime(\ell_1, j))}
    }{$1 \leq j \leq \mathfrak{B}_\text{generics}$}
    \\[0.25em]
    \irule{L-Invariant-Ptr}{
        M \vdash_\textsf{asn} \labeled{\ell_1}{\sty{ptr}} \leftarrow \labeled{\ell_2}{\textunderscore} \ @ \ \labeled{\ell}{$E$}, i
        \quad M \vdash_\textsf{asn} \labeled{\ell_1'}{\sty{ptr}} \leftarrow \labeled{\ell'_2}{\textunderscore} \ @ \ \labeled{\ell}{$E$}, j
    }{
        M \vdash \boxed{\begin{aligned}
            &\BaseType(\ell_1) = \rty{mut} \lor \BaseType(\ell_2) = \rty{mut} \lor \ell_1 \in \Cell \lor \ell_2 \in \Cell
            \\
            \Rightarrow \ &\textstyle\{ \BaseType(\ell_1), \BaseType(\ell_2) \} \subseteq \{ \rty{shared}, \rty{mut} \} \Rightarrow \Lifetime(\refbind(\ell, j)) \supseteq \Lifetime(\RefLifetime(\ell_1'))
        \end{aligned}}
    }{}
    \\[0.25em]
    \irule{L-Invariant-Struct}{
        M \vdash_\textsf{asn} \labeled{\ell_1}{\sty{ptr}} \leftarrow \labeled{\ell_2}{\textunderscore} \ @ \ \labeled{\ell}{$E$}, i
        \quad M \vdash_\textsf{asn} \labeled{\ell_1'}{\sty{struct}($s$)} \leftarrow \labeled{\ell'_2}{\textunderscore} \ @ \ \labeled{\ell}{$E$}, j
    }{
        M \vdash \boxed{\begin{aligned}
            &\BaseType(\ell_1) = \rty{mut} \lor \BaseType(\ell_2) = \rty{mut} \lor \ell_1 \in \Cell \lor \ell_2 \in \Cell
            \\
            \Rightarrow \ &k \leq \StructGenerics(s)
            \Rightarrow \Lifetime(\structbind(\ell, j, k)) \supseteq \Lifetime(\StructLifetime(\ell_1', k))
        \end{aligned}}
    }{$1 \leq k \leq \mathfrak{B}_\text{generics}$}
\end{gather*}
}

\subsubsection{Reborrowing}

\Refinator generates the following constraints to encode the reborrowing rules:
{
\small
\begin{gather*}
    \irule{L-Reborrow-Var}{
        M \vdash_\textsf{rval} \labeled{\ell_1}{$v$}: \labeled{\ell_1}{\sty{ptr}} \ @ \ \texttt{\textunderscore}
        \quad M \vdash_\textsf{var} v : \langle \labeled{\ell_2}{\textunderscore}, ... \rangle
    }{
        M \vdash \boxed{\textstyle\{ \BaseType(\ell_1), \BaseType(\ell_2) \} \subseteq \{ \rty{shared}, \rty{mut} \} \Rightarrow \Lifetime(\RefLifetime(\ell_2)) \supseteq \Lifetime(\RefLifetime(\ell_1)) }
    }{}
    \\[0.25em]
    \irule{$\text{L-Reborrow-Deref}_1$}{
        M \vdash_\textsf{rval} \labeled{\ell_1}{*$v$}: \labeled{\ell_1}{\sty{ptr}} \ @ \ \texttt{\textunderscore}
        \quad M \vdash_\textsf{var} v : \langle \labeled{\ell_2}{\textunderscore}, ... \rangle
    }{
        M \vdash \boxed{\textstyle\{ \BaseType(\ell_1), \BaseType(\ell_2) \} \subseteq \{ \rty{shared}, \rty{mut} \} \Rightarrow \Lifetime(\RefLifetime(\ell_2)) \supseteq \Lifetime(\RefLifetime(\ell_1)) }
    }{}
    \\[0.25em]
    \irule{$\text{L-Reborrow-Deref}_2$}{
        M \vdash_\textsf{rval} \labeled{\ell_1}{*$v$}: \labeled{\ell_1}{\sty{ptr}} \ @ \ \texttt{\textunderscore}
        \quad M \vdash_\textsf{var} v : \langle \texttt{\textunderscore}, \labeled{\ell_2}{\textunderscore}, ... \rangle
    }{
        M \vdash \boxed{\textstyle\{ \BaseType(\ell_1), \BaseType(\ell_2) \} \subseteq \{ \rty{shared}, \rty{mut} \} \Rightarrow \Lifetime(\RefLifetime(\ell_2)) \supseteq \Lifetime(\RefLifetime(\ell_1)) }
    }{}
    \\[0.25em]
    \irule{$\text{L-Reborrow-Fld}_1$}{
        M \vdash_\textsf{rval} \labeled{\ell_1}{\&*(*$v$).$f$}: \labeled{\ell_1}{\textunderscore} \ @ \ \texttt{\textunderscore}
        \quad M \vdash_\textsf{var} v : \langle \labeled{\ell_2}{\textunderscore}, ... \rangle
    }{
        M \vdash \boxed{\textstyle\{ \BaseType(\ell_1), \BaseType(\ell_2) \} \subseteq \{ \rty{shared}, \rty{mut} \} \Rightarrow \Lifetime(\RefLifetime(\ell_2)) \supseteq \Lifetime(\RefLifetime(\ell_1)) }
    }{}
    \\[0.25em]
    \irule{$\text{L-Reborrow-Fld}_2$}{
        M \vdash_\textsf{rval} \labeled{\ell_1}{\&*(*$v$).$f$}: \labeled{\ell_1}{\textunderscore} \ @ \ \texttt{\textunderscore}
        \quad M \vdash_\textsf{var} v : \langle \texttt{\textunderscore}, \labeled{\ell}{\sty{struct}($s$)}, ... \rangle
        \quad M \vdash_\textsf{fld} s, f : \langle \labeled{\ell_2}{\textunderscore}, ... \rangle
    }{
        M \vdash \boxed{\textstyle\{ \BaseType(\ell_1), \BaseType(\ell_2) \} \subseteq \{ \rty{shared}, \rty{mut} \} \Rightarrow \Lifetime(\StructLifetime(\ell, \RefLifetime(\ell_2))) \supseteq \Lifetime(\RefLifetime(\ell_1)) }
    }{}
    \\[0.25em]
    \irule{L-Reborrow-Elt}{
        M \vdash_\textsf{rval} \labeled{\ell_1}{\&$v$[\textunderscore]}: \labeled{\ell_1}{\textunderscore} \ @ \ \texttt{\textunderscore}
        \quad M \vdash_\textsf{var} v : \langle \labeled{\ell_2}{\textunderscore}, ... \rangle
    }{
        M \vdash \boxed{\{ \BaseType(\ell_1), \BaseType(\ell_2) \} \subseteq \{ \rty{shared}, \rty{mut} \} \Rightarrow \Lifetime(\RefLifetime(\ell_2)) \supseteq \Lifetime(\RefLifetime(\ell_1)) }
   }{}
\end{gather*}
}

\subsubsection{Lifetime parameters}

Here, we formalize the constraints \refinator generates regarding function lifetime parameters.

The following constraints ensure that the lifetimes of lifetime variables that appear in a function's signature contain the entire function body:
{
\small
\begin{gather*}
    \irule{L-Param-Body-Ptr}{
        M \vdash_\textsf{func} F : (\vec{\tau}_1, ..., \vec{\tau_n}) \rightarrow \vec{\tau}_0
        \quad (\vec{\tau}_i)_j = \labeled{\ell}{\sty{ptr}}
        \quad M \vdash_\textsf{instr} \texttt{\textunderscore} \ @ \ L \in F
    }{
        M \vdash \boxed{\BaseType(\ell) \in \{ \rty{shared}, \rty{mut} \} \Rightarrow \textstyle\{ (L)_\text{in}, (L)_\text{out} \} \subseteq \Lifetime(\RefLifetime(\ell))}
    }{}
    \\[0.25em]
    \irule{L-Param-End-Ptr}{
        M \vdash_\textsf{func} F : (\vec{\tau}_1, ..., \vec{\tau_n}) \rightarrow \vec{\tau}_0
        \quad (\vec{\tau}_i)_j = \labeled{\ell}{\sty{ptr}}
    }{
        M \vdash \boxed{\BaseType(\ell) \in \{ \rty{shared}, \rty{mut} \} \Rightarrow \RefLifetime(\ell) \in \LifetimeEnd(\RefLifetime(\ell))}
    }{}
    \\[0.25em]
    \irule{L-Param-Body-Struct}{
        M \vdash_\textsf{func} F : (\vec{\tau}_1, ..., \vec{\tau_n}) \rightarrow \vec{\tau}_0
        \quad (\vec{\tau}_i)_j = \labeled{\ell}{\sty{struct}($s$)}
        \quad M \vdash_\textsf{instr} \texttt{\textunderscore} \ @ \ L \in F
    }{
        M \vdash \boxed{j \leq \StructGenerics(s) \Rightarrow \textstyle\{ (L)_\text{in}, (L)_\text{out} \} \subseteq \Lifetime(\StructLifetime(\ell, j))}
    }{$1 \leq j \leq \mathfrak{B}_\text{generics}$}
    \\[0.25em]
    \irule{L-Param-End-Struct}{
        M \vdash_\textsf{func} F : (\vec{\tau}_1, ..., \vec{\tau_n}) \rightarrow \vec{\tau}_0
        \quad (\vec{\tau}_i)_j = \labeled{\ell}{\sty{struct}($s$)}
    }{
        M \vdash \boxed{j \leq \StructGenerics(s) \Rightarrow \StructLifetime(\ell, j) \in \LifetimeEnd(\StructLifetime(\ell, j))}
    }{$1 \leq j \leq \mathfrak{B}_\text{generics}$}
\end{gather*}
}

To generate constraints that propagate relationships between lifetime variables in the signature of a function into the function's call sites, we define, for each call site $L$, a bijection $\Phi_L$ between SMT terms that maps an SMT term that evaluates to a lifetime variable at the call site to an SMT term that evaluates to the corresponding lifetime variable in the function signature:
{
\small
\begin{align*}
    \Phi_L := &\left\{ (\refbind(\ell_i, j), \RefLifetime(\ell)) :
    \begin{aligned}
        &M \vdash_\textsf{instr} \texttt{$\vec{\tau}_0$ \textunderscore \ = $F$(\labeled{\ell_1}{$e_1$}, $...$, \labeled{\ell_n}{$e_n$},);} \ @ \ L \in \texttt{\textunderscore} \text{ and } \\
        &M \vdash_\textsf{func} F : (\vec{\tau}'_1, ..., \vec{\tau}'_n) \to \vec{\tau}'_0 \text{ and } (\vec{\tau}_i)_j = \labeled{\ell}{\sty{ptr}}
    \end{aligned}
    \right\} \ \cup \\
    &\left\{ (\structbind(\ell_i, j, k), \StructLifetime(\ell, k)) :
    \begin{aligned}
        &M \vdash_\textsf{instr} \texttt{$\vec{\tau}_0$ \textunderscore \ = $F$(\labeled{\ell_1}{$e_1$}, $...$, \labeled{\ell_n}{$e_n$},);} \ @ \ L \in \texttt{\textunderscore} \text{ and } \\
        &M \vdash_\textsf{func} F : (\vec{\tau}'_1, ..., \vec{\tau}'_n) \to \vec{\tau}'_0 \text{ and } (\vec{\tau}_i)_j = \labeled{\ell}{\sty{struct}(\textunderscore)} \text{ and } \\
        &1 \leq k \leq \mathfrak{B}_\text{generics}
    \end{aligned}
    \right\}
\end{align*}
}%
\Refinator then generates the following constraint:
{
\small
\begin{gather*}
    \irule{L-Hom}{
        \Phi_L(t_1) = t_1'
        \quad \Phi_L(t_2) = t_2'
    }{
        M \vdash \boxed{\LifetimeEnd(t_1') \ni t_2' \Rightarrow \Lifetime(t_1) \supseteq \Lifetime(t_2) }
    }{}
\end{gather*}
}

\subsection{In-Scope Loans}
\label{appendix:borrowing-loans}

Here, we show the constraints that \autoref{sec:borrowing-loans} introduced.
\Refinator's constraints represent the in-scope loans at each program point in the SMT function defined as follows.
\begin{gather*}
    \Loans: \POINT_M \to \mathcal{P}(\TYPELAB_M)
\end{gather*}
Instead of representing a loan as a pair in $\LIFETIMEVAR \times \PATH$ as in \autoref{sec:borrowing}, \refinator uses the type label of the non-nested rvalue expression that produces the loan, as there is a one-to-one correspondence between pointer-typed non-nested rvalue expressions and $\LIFETIMEVAR \times \PATH$ pairs---the non-nested rvalue expression \labeled{\ell}{$E$} may only produce the loan with lifetime variable $\RefLifetime(\ell)$ and path $\borrowpath(E)$.
For a non-nested rvalue expression $E$, we denote the path it borrows as $\borrowpath(E)$, defined as follows.
{
\small
\begin{gather*}
    \borrowpath(E) := \begin{cases}
        \texttt{*$v$} &\text{if $E = v$} \\
        \texttt{**$v$} &\text{if $E = \texttt{*$v$}$} \\
        \texttt{*(*$v$).$f$} &\text{if $E = \texttt{\&*(*$v$).$f$}$} \\
        \texttt{*$v$} &\text{if $E = \texttt{\&$v$[$e$]}$}
    \end{cases}
\end{gather*}
}%
\Refinator generates the following constraints:
{
\small
\begin{gather*}
    \irule{L-Merge}{
        M \vdash_\textsf{cfg} L_1 \rightarrow L_2
    }{
        M \vdash \boxed{\Loans((L_2)_\text{in}) \supseteq \Loans((L_1)_\text{out})}
    }{}
    \quad
    \irule{L-Transfer-Assign}{
        M \vdash_\textsf{instr} \texttt{*$v$ = \textunderscore;} \ @ \ L \in F
        \quad M \vdash_\textsf{rval} \labeled{\ell}{$v$} : \texttt{\textunderscore} \ @ \ \texttt{\textunderscore} \in F
    }{
        M \vdash \boxed{\ell \notin \Loans((L)_\text{out})}
    }{}
    \\[0.25em]
    \irule{L-Transfer-Borrow}{
        M \vdash_\textsf{rval} \labeled{\ell}{$E$} : \langle \labeled{\ell}{\sty{ptr}}, ... \rangle \ @ \ L \in F
    }{
        M \vdash \boxed{(L)_\text{out} \in \Lifetime(\RefLifetime(\ell)) \Rightarrow \ \ell \in \Loans((L)_\text{out})}
    }{}
    \\[0.25em]
    \irule{L-Transfer-Lifetime}{
        M \vdash_\textsf{rval} \labeled{\ell}{$E$} : \langle \labeled{\ell}{\sty{ptr}}, ... \rangle \ @ \ \texttt{\textunderscore} \in F
        \quad M \vdash_\textsf{instr} \texttt{\textunderscore} \ @ \ L \in F
    }{
        M \vdash \boxed{
            \ell \in \Loans((L)_\text{in}) \land (L)_\text{out} \in \Lifetime(\RefLifetime(\ell)) \Leftrightarrow
            \ell \in \Loans((L)_\text{out})
        }
    }{}
\end{gather*}
}

\subsection{Borrowing Conflicts}
\label{appendix:borrowing-conflict}

Here, we show the constraints that \autoref{sec:borrowing-conflict} introduced.
For $a \in \PATH$, we denote the set of paths whose loans may conflict with deep accesses of $a$, as described in the NLL RFC~\cite{nll-rfc}, by $\deeppaths(a)$, defined as follows.
{
\small
\begin{align*}
    \deeppaths(a) := \begin{cases}
        \{ \texttt{*$v$} \} \cup \{ \texttt{*(*$v$).$f$} \mid f \in \IDENT \} &\text{if $a = \texttt{*$v$}$} \\
        \{ \texttt{*$v$}, \texttt{**$v$} \} &\text{if $a = \texttt{**$v$}$} \\
        \{ \texttt{*$v$}, \texttt{*(*$v$).$f$} \} &\text{if $a = \texttt{*(*$v$).$f$}$}
    \end{cases}
\end{align*}
}

\subsubsection{Borrowing rvalue expressions}
\label{appendix:borrowing-conflict-borrow}

\Refinator generates the following constraint for \autoref{sec:borrowing-conflict-borrow}:
{
\small
\begin{gather*}
    \irule{C-Borrow}{
        M \vdash_\textsf{rval} \labeled{\ell}{$E$} : \langle \labeled{\ell}{\sty{ptr}}, ... \rangle \ @ \ L \in F
        \quad M \vdash_\textsf{rval} \labeled{\ell'}{$E'$} : \langle \labeled{\ell'}{\sty{ptr}}, ... \rangle \ @ \ \texttt{\textunderscore} \in F
    }{
        M \vdash \boxed{\ell' \in \Loans((L)_\text{in}) \Rightarrow \BaseType(\ell) = \BaseType(\ell') = \rty{shared}}
    }{$\borrowpath(E') \in \deeppaths(\borrowpath(E))$}
\end{gather*}
}

\subsubsection{Moving rvalue expressions}
\label{appendix:borrowing-conflict-move}

\Refinator generates the following constraints for \autoref{sec:borrowing-conflict-move}:
{
\small
\begin{gather*}
    \irule{C-Move-Var}{
        \begin{gathered}
            M \vdash_\textsf{rval} \labeled{\ell}{$v$} : \langle \labeled{\ell}{\sty{ptr}}, ... \rangle \ @ \ L \in F
            \quad M \vdash_\textsf{var} v : \langle \labeled{\ell'}{\textunderscore}, ... \rangle
            \\ M \vdash_\textsf{rval} \labeled{\ell''}{$E$} : \langle \labeled{\ell''}{\sty{ptr}}, ... \rangle \ @ \ \texttt{\textunderscore} \in F
        \end{gathered}
    }{
        M \vdash \boxed{\ell'' \in \Loans((L)_\text{in}) \Rightarrow \RustType(\ell') \blacktriangleright \RustType(\ell))}
    }{$\borrowpath(E) \in \deeppaths(\texttt{*$v$})$}
    \\[0.25em]
    \irule{C-Move-Deref}{
        \begin{gathered}
            M \vdash_\textsf{rval} \labeled{\ell}{*$v$} : \langle \labeled{\ell}{\sty{ptr}}, ... \rangle \ @ \ L \in F
            \quad M \vdash_\textsf{var} v : \langle \texttt{\textunderscore}, \labeled{\ell'}{\textunderscore}, ... \rangle
            \\ M \vdash_\textsf{rval} \labeled{\ell''}{$E$} : \langle \labeled{\ell''}{\sty{ptr}}, ... \rangle \ @ \ \texttt{\textunderscore} \in F
        \end{gathered}
    }{
        M \vdash \boxed{\ell'' \in \Loans((L)_\text{in}) \Rightarrow \RustType(\ell') \blacktriangleright \RustType(\ell))}
    }{$\borrowpath(E) \in \deeppaths(\texttt{**$v$})$}
    \\[0.25em]
    \irule{C-Move-Elt}{
        \begin{gathered}
            M \vdash_\textsf{rval} \labeled{\ell}{\&$v$[\textunderscore]} : \langle \labeled{\ell}{\sty{ptr}}, ... \rangle \ @ \ L \in F
            \quad M \vdash_\textsf{var} v : \langle \labeled{\ell'}{\textunderscore}, ... \rangle
            \\ M \vdash_\textsf{rval} \labeled{\ell''}{$E$} : \langle \labeled{\ell''}{\sty{ptr}}, ... \rangle \ @ \ \texttt{\textunderscore} \in F
        \end{gathered}
    }{
        M \vdash \boxed{\ell'' \in \Loans((L)_\text{in}) \Rightarrow \RustType(\ell') \blacktriangleright \RustType(\ell))}
    }{$\borrowpath(E) \in \deeppaths(\texttt{*$v$})$}
\end{gather*}
}

\subsubsection{Assignments}
\label{appendix:borrow-conflict-assign}

\Refinator generates the following constraint for \autoref{sec:borrowing-conflict-assign}:
{
\small
\begin{gather*}
    \irule{C-Assign}{
        M \vdash_\textsf{instr} \texttt{*$v$ = \textunderscore;} \ @ \ L \in F
        \quad M \vdash_\textsf{var} v : \langle \labeled{\ell}{\textunderscore}, ... \rangle
        \quad M \vdash_\textsf{rval} \labeled{\ell'}{$E$} : \langle \labeled{\ell'}{\sty{ptr}}, ... \rangle \ @ \ \texttt{\textunderscore} \in F
    }{
        M \vdash \boxed{\ell' \in \Loans((L)_\text{in}) \Rightarrow \BaseType(\ell) = \BaseType(\ell') = \rty{shared}}
    }{$\borrowpath(E) = \texttt{*$v$}$}
\end{gather*}
}

\subsubsection{Variable drops}
\label{appendix:borrow-conflict-drop}

\Refinator generates the following constraints for \autoref{sec:borrowing-conflict-drop}:
{
\small
\begin{gather*}
    \irule{C-Drop-Var}{
        M \vdash_\textsf{instr} \texttt{\kw{drop}($v$);} \ @ \ L \in F
        \quad M \vdash_\textsf{rval} \labeled{\ell}{$v$} \ @ \ \texttt{\textunderscore} \in F
        \quad M \vdash_\textsf{var} v : \langle \labeled{\ell'}{\sty{ptr}}, ... \rangle
    }{
        M \vdash \boxed{\ell \in \Loans((L)_\text{in}) \Rightarrow \BaseType(\ell') = \rty{shared}}
    }{}
    \\[0.25em]
    \irule{C-Drop-Deref}{
        M \vdash_\textsf{instr} \texttt{\kw{drop}($v$);} \ @ \ L \in F
        \quad M \vdash_\textsf{rval} \labeled{\ell}{*$v$} \ @ \ \texttt{\textunderscore} \in F
        \quad M \vdash_\textsf{var} v : \langle \texttt{\textunderscore}, \labeled{\ell'}{\sty{ptr}}, ... \rangle
    }{
        M \vdash \boxed{\ell \in \Loans((L)_\text{in}) \Rightarrow \BaseType(\ell') = \rty{shared}}
    }{}
    \\[0.25em]
    \irule{C-Drop-Fld}{
        M \vdash_\textsf{instr} \texttt{\kw{drop}($v$);} \ @ \ L \in F
        \quad M \vdash_\textsf{rval} \labeled{\ell}{\&*(*$v$).$f$} \ @ \ \texttt{\textunderscore} \in F
        \quad M \vdash_\textsf{var} v : \langle \labeled{\ell'}{\sty{ptr}}, ... \rangle
    }{
        M \vdash \boxed{\ell \in \Loans((L)_\text{in}) \Rightarrow \BaseType(\ell') = \rty{shared}}
    }{}
    \\[0.25em]
    \irule{C-Drop-Elt}{
        M \vdash_\textsf{instr} \texttt{\kw{drop}($v$);} \ @ \ L \in F
        \quad M \vdash_\textsf{rval} \labeled{\ell}{\&$v$[\textunderscore]} \ @ \ \texttt{\textunderscore} \in F
        \quad M \vdash_\textsf{var} v : \langle \labeled{\ell'}{\sty{ptr}}, ... \rangle
    }{
        M \vdash \boxed{\ell \in \Loans((L)_\text{in}) \Rightarrow \BaseType(\ell') = \rty{shared}}
    }{}
\end{gather*}
}

\section{Precision}
\label{appendix:precision}
Here, we show the optimality objectives introduced in \autoref{sec:precision}.
We define three $\BASETYPE \to \NN$ cost functions: $\cost_\text{height}$, which is the transformation cost favoring owning pointers; $\cost_\text{depth}$, which is the transformation cost favoring borrowed pointers; and $\cost_\text{dynamic}$, which favors pointers with lower run-time overhead.
We show the definitions of these functions in \autoref{tab:cost-def}.

\begin{table}[t]
    \centering
    \small
    \caption{Definitions of cost functions $\cost_\text{height}$, $\cost_\text{depth}$, and $\cost_\text{dynamic}$.}
    \begin{tabular}{l|rrr}
        $\beta$ & $\cost_\text{height}(\beta)$ & $\cost_\text{depth}(\beta)$ & $\cost_\text{dynamic}(\beta)$ \\ \hline
        \rty{shared} & 3 & 1 & 1 \\
        \rty{mut} & 2 & 2 & 2 \\
        \rty{ghost} & 1 & 3 & 3 \\
        \rty{Box} & 1 & 3 & 4 \\
        \rty{Rc} & 1 & 3 & 5 \\
        other & 0 & 0 & 0 \\
    \end{tabular}
    \label{tab:cost-def}
\end{table}

Let $\mathscr{L} \subset \TYPELAB_M$ be the set of type labels in $M$ that appear in interface variables, and let $\mathscr{L}_\text{outer} \subseteq \mathscr{L}$ be those that are the outermost type labels for struct fields and global variables, and $\mathscr{L}_\text{inner} := \mathscr{L} \setminus \mathscr{L}_\text{inner}$.
Specifically,
{
\small
\begin{align*}
    \mathscr{L} := &\left\{ \ell : M \vdash_\textsf{fld} \texttt{\textunderscore} : \vec{\tau} \text{ and } (\vec{\tau})_i = \labeled{\ell}{\textunderscore} \right\} \cup
    \left\{ \ell : M \vdash_\textsf{func} \texttt{\textunderscore} : (\vec{\tau}_1, ..., \vec{\tau}_n) \rightarrow \vec{\tau}_0 \text{ and } (\vec{\tau}_i)_j = \labeled{\ell}{\textunderscore} \right\} \cup \\
    &\left\{ \ell : M \vdash_\textsf{var} v : \vec{\tau} \text{ and } (\vec{\tau})_i = \labeled{\ell}{\textunderscore} \text{ and $v$ is a global variable} \right\} \\
    \mathscr{L}_\text{outer} := &\left\{ \ell : M \vdash_\textsf{fld} \texttt{\textunderscore} : \langle \labeled{\ell}{\textunderscore}, ... \rangle \right\} \cup
    \left\{ \ell : M \vdash_\textsf{var} : \langle \labeled{\ell}{\textunderscore}, ... \rangle \text{ and $v$ is a global variable} \right\}
\end{align*}
}%
For an SMT formula $\phi$, let $\chi(\phi) = 1$ if $\phi$ is true, and $\chi(\phi) = 0$ otherwise.
Now, \refinator's optimality objectives, in the same order as presented in \autoref{sec:precision}, are as follows.
\begin{enumerate}
    \item $\sum_{\ell \in \mathscr{L}} \chi(\ell \in \Cell)$
    \item $\sum_{\ell \in \mathscr{L}_\text{outer}} \cost_\text{height}(\BaseType(\ell)) + \sum_{\ell \in \mathscr{L}_\text{inner}} \cost_\text{depth}(\BaseType(\ell))$
    \item $\sum_{\ell \in \mathscr{L}} \cost_\text{dynamic}(\BaseType(\ell))$
    \item $\sum_{\ell \in \mathscr{L}} \chi(\ell \in \Array)$
\end{enumerate}

\section{Computing Source Types from LLVM IR}
\label{appendix:computing-source-types}

In LLVM IR, pointer types are opaque; that is, they do not contain information about their referent type.
However, to encode Rust's type and borrow-checking rules, as well as the semantic preservation rules, \refinator needs to know the referent type for each pointer.
To recover this information, \refinator's source language conversion pass performs source type inference.

\Refinator's source type inference considers the type information that appears explicitly in LLVM IR, and uses the relationships between the types of variables, function returns, and struct fields implied by LLVM IR instructions to infer pointer type information.
For example, at the LLVM IR instruction \lstinline[language=LLVM]{%v = load i32, %p}, \refinator's source type inference determines that the type of \lstinline[language=LLVM]{

\Refinator's source type inference extracts explicit types from struct type definitions, function declarations, and variable declarations. It extracts equality relationships from LLVM IR instructions \lstinline[language=LLVM]{alloca}, \lstinline[language=LLVM]{load}, \lstinline[language=LLVM]{store}, \lstinline[language=LLVM]{getelementptr}, \lstinline[language=LLVM]{call}, \lstinline[language=LLVM]{phi}, and \lstinline[language=LLVM]{ret}.
Then, it solves for the source types for each struct field, function return, and variable by solving this system of equalities.
}{}

\end{document}